\documentclass{article}
\usepackage[utf8]{inputenc}
\usepackage{amssymb}
\usepackage{authblk}
\usepackage{amsfonts}
\usepackage{mathrsfs}
\usepackage{amsmath}
\usepackage{appendix}
\usepackage{graphicx}
\usepackage{dsfont}
\usepackage{float}
\usepackage{diagbox} 
\usepackage{subcaption}

\usepackage{graphics}
\usepackage{csquotes}
\usepackage{tabularx, booktabs, makecell, caption}
\usepackage[all]{xy}
\usepackage[margin=1.4in,marginparwidth=1in,marginparsep=.05in]{geometry}
\usepackage{siunitx}
\PassOptionsToPackage{hyphens}{url}
\usepackage{tikz-cd}
\usepackage{braket}
\usepackage{xcolor}
\usepackage{verbatim}
\usepackage{float}
\usepackage{mathrsfs}
\usepackage{extarrows}
\usepackage{tikz}
\usetikzlibrary{shapes,arrows,positioning}
\usepackage{mathrsfs}
\usepackage{mathtools}
\usepackage{cite}

\usepackage{hyperref}
\usepackage{wrapfig}

\usepackage{algorithm} 
\usepackage{algpseudocode} 
\usepackage{fancyhdr}

\title{\textbf{Forest density is more effective than tree rigidity at reducing the onshore energy flux of tsunamis: Evidence from Large Eddy Simulations with Fluid-Structure Interactions}}

\date{\today}
\begin{document}

\author[1]{Abhishek Mukherjee\thanks{Corresponding author: am2455@njit.edu}}
\author[2]{Juan Carlos Cajas}
\author[3]{Guillaume Houzeaux}
\author[3]{Oriol Lehmkuhl}
\author[4]{Jenny Suckale}
\author[1]{Simone Marras}

\affil[1]{Department of Mechanical Engineering, New Jersey Institute of Technology, Newark, NJ}
\affil[2]{ENES, Universidad Nacional Autonoma de Mexico,  M\'erida, M\'exico}
\affil[3]{CASE, Barcelona Supercomputing Center, Spain}
\affil[4]{Department of Geophysics, Stanford University, Palo Alto, CA}

\maketitle

\begin{abstract}
Communities around the world are increasingly interested in nature-based solutions to mitigation of coastal risks like coastal forests, but it remains unclear how much protective benefits vegetation provides, particularly in the limit of highly energetic flows after tsunami impact.
The current study, using a three-dimensional incompressible computational fluid dynamics model with a fluid-structure interaction approach, 
aims to quantify how energy reflection and dissipation vary with different degrees of rigidity and vegetation density of a coastal forest.
 We represent tree trunks as cylinders and use the elastic modulus of hardwood trees such as pine or oak to characterize the rigidity of these cylinders.
The numerical results show that energy reflection increases with rigidity only for a single cylinder. In the presence of multiple cylinders, the difference in energy reflection created by varying rigidity diminishes as the number of cylinders increases. Instead of rigidity, we find that the blockage area created by the presence of multiple tree trunks dominates energy reflection.  
As tree trunks are deformed by the hydrodynamic forces, they alter the flow field around them, causing turbulent kinetic energy generation in the wake region. As a consequence, trees dissipate flow energy, highlighting coastal forests reducing the onshore energy flux of tsunamis by means of both reflection and dissipation. 
\end{abstract}

\section{Introduction}
\label{intro}
The partial failure of a multi-billion dollar concrete sea wall upon the impact of the T\={o}hoku tsunami in 2011 invigorated the debate about possible nature-based approaches to mitigating tsunami risk (e.g., \cite{cochard2008,behrensEtAl2021, rabyEtal2015}). A few years before, damage assessments after the Indian Ocean Tsunami pointed to a reduction in wave impact near natural features \cite{danielsen2005asian,kathiresan2005coastal,dahdouh2005effective,tanaka2007coastal,tanaka2007effects,chatenoux2007impacts,iverson2007using,olwig2007using,bayas2011influence}, but the causal attribution of lessened damage to vegetation has remained contentious \cite{kerr2006comments,kerr2007natural,kerr2009reply}. Adding to the challenge of clearly identifying the potential protective benefit of vegetation are field observations from other contexts. For example, the Japan Tsunami annihilated a coastal forest with an estimated 70,000 pine trees in Rikuzentakata, Iwate Prefecture, highlighting that there are no fail-safe approaches to countering a tsunami's destructive power. However, while the Rikuzentakata coastal forest was not able to withstand the 2011 event, it might have mitigated the impacts of the 1933 Showa Sanriku tsunami and the Chilean tsunami in 1960 \cite{liu2013,mori2013overview}.

One step towards adding clarity to the ongoing debate about the role of vegetation in tsunami-risk mitigation is to distinguish between different ecosystems and the differing effect they have on tsunami runup. Contrary to sea walls, vegetation cannot completely stop a tsunami, and its effectiveness depends both on the magnitude of the tsunami and on the structure and response of the vegetation. To leverage vegetation in tsunami-risk mitigation, we need to quantitatively understand the mechanism through which the vegetation interacts with the wave, the limits of this mechanism, and the degree to which it leads to a reduction in the energy of the tsunami \cite{tanaka2009vegetation}. The differences can appear quite subtle: For example, a detailed field assessment after the 1998 tsunami in Papua New Guinea found that  \emph{Casuarina} trees were able to withstand wave impact better than palm trees that were uprooted or snapped off at mid trunk level, transforming into projectiles and increasing rather than decreasing damage \cite{dengler2003mitigation}.  

The lessons learned from both the Japan and the Papua New Guinea tsunami highlight that advancing our understanding of the protective benefit of coastal forests requires not only a quantification of the effect that the trees have on the tsunami but also an analysis of the effect that the tsunami has on the trees. One important difference between Casuarina trees and palm trees is the bendability of their trunks, with Casuarina trees having more flexible trunks. More bendability, however, is not necessarily better. In fact, field surveys in Sri Lanka and Thailand after the Indian Ocean Tsunami suggested that older and relatively sparse belts of \emph{Casuarina equisetifolia} withstood the tsunami but did not appear to provide much protection \cite{tanaka2009vegetation}, emphasizing the potential importance of forest density in addition to individual tree properties. 

The goal of this study is to quantify how the energy dissipation and partial wave reflection of individual coastal trees vary with different degrees of bendability and the number of trees. We approach this problem numerically by solving the Navier-Stokes equations for turbulent flow with a free surface and fluid-structure interactions (FSI) using a multi-physics software package developed by some of the authors and described in  \cite{houzeaux2009,houzeauxAubryVazquez20111}. The FSI module allows to capture the two-way coupling between the fluid flow and the mechanical bending of the trees due to wave forcing. While our work is motivated by prior field studies, we study the interactions between fluid flow and tree bendability in an idealized laboratory-like setting that is not intended to mimic any specific tree species or location. To enable transferability of our results, we consider a non-dimensional parameter range that is representative of tsunami flows and realistic hardwood properties \cite{green1999mechanical}.

Our approach of leveraging high-fidelity numerical methods complements existing approaches that adopted the depth-averaged shallow-water, Boussinesq or Serre--Green--Naghdi approximation in 1D or 2D (e.g. \cite{lunghino2020protective, okalSynolakis2015, borreroEtAl2015b, apotsos2011wave, lin2004numerical}). These models are powerful in capturing regional-scale inundation but are more limited in estimating forces on individual structures \cite{marrasMandli2021}. As demonstrated by comparing a 2D against a 3D model in \cite{qin2018comparison}, the solution of the 3D Navier–Stokes equations is required to quantify impact with the due precision. The limitation of not being able to capture 3D impact forces is less severe when the vegetation under consideration is relatively flat as compared to the water height, as might be the case with seagrass or a marshland, but is important in the context of bendable tree trunks experiencing tsunami impact. 

Our work is not the first attempt to leverage Navier-Stokes-based simulations to improve our understanding of tsunami impacts on vegetation. Maza et al. \cite{mazaEtAl2015} solved the Unsteady Reynolds Averaged Navier-Stokes (URANS) equations to study the interaction of a solitary wave with vegetation, where vegetation was modeled by rigid cylinders after the experiments by \cite{huangEtAl2011}. The same authors built  a 2D linear model for vegetation motion and simulated the swaying dynamics of submerged vegetation \cite{mazaEtAl2013}. An extension to 3D cylinders is presented in \cite{tsaiEtAl2017} where a rigid motion of the cylinders is allowed around a fixed point on the ground. Unlike what we show in this article, the above cited studies do not account for the cylinder bendability and elasticity along its length.

Navier-Stokes-based simulations of tsunami impacts come with their own challenges, such as the need for careful validation \cite{larsen2019full1}, sufficient resolution in boundary layers \cite{williamsFuhrman2016,larsen2019full2}, and the potential overproduction of turbulence \cite{larsenFuhrman2018}. Fortunately, there is an increasing number of laboratory experiments that are essential for the validation of numerical simulations \cite{tonkin2003tsunami,bayas2011influence,lakshmanan2012,irish2014laboratory,ali2019energy}. Similarly, increasing computational power and data storage have solved some issues, particularly those related to insufficient resolution, making scientific computing for large scale tsunami simulations increasingly accepted \cite{behrensDias2015, synolakis2006tsunami}. This increased interest in high-performance computing for tsunami runup has led to several inter-comparison efforts published in recent years \cite{watanabeEtAl2022, ghani2019numerical, yang2017impact, kundu2016numerical, mattisEtAl2015, mattisEtAl2012, iimura2012numerical, venayagamoorthy2006numerical} that add valuable transparency to high-fidelity computational approaches.

While high-fidelity simulations provide a high level of detail regarding the flow field around submerged vegetation, post-processing is necessary to distill these details into a simple metric that is indicative of protective benefit. In much of the literature about the role of vegetation in coastal risk reduction, this metric is inundation distance, defined as the horizontal distance that a tsunami reached inland. In fact, one of the criticisms of the field studies that identified associations between vegetation and damage from satellite data \cite{danielsen2005asian, chatenoux2007impacts} was that vegetation did not appear to have a mitigating effect on inundation distance \cite{kerr2007natural}. However, inundation distance as a metric of protective benefit might be more valuable in the storm-surge context for which it was originally introduced \cite{USACE1963, fosberg1971mangroves} than for tsunamis. In contrast to storm surges, the main destructive potential of a tsunami is related to the energy transported not to the presence of water per se. We hence adopt reduction in energy flux as a metric of potential benefit, implying that there might be a significant potential benefit even if the inundation distance remains relatively unchanged, as seen in \cite{lunghino2020protective}.

In this work, we implement a two-way coupling fluid-structure interaction model to investigate tsunami flow energy damping performance for different vegetation parameters. We use circular base cylinders as an idealized approximation for tree trunks following previous work on emergent, flexible plants in flow \cite{nepf2012flow}. The remainder of this paper is organized as follows: \S~\ref{methodology} describes the methodology followed by the model validation tests described in  \S~\ref{benchmarking}. The numerical setup and the different scaling parameters are described in \S~\ref{model setup}. In \S~\ref{results} we present the energy budget comparisons for different vegetation parameters, including velocity and turbulent kinetic energy distribution and wave profile evolution. In \S~\ref{sct:discussion}, we discuss how our results could inform the ongoing debate on nature-based approaches to tsunami-risk reduction and end with a brief conclusion in \S~\ref{sct:conclusions}.
 
\section{Numerical model description}
\label{methodology}

\subsection{Mathematical model}
The interaction of a two phase fluid flow (water and air) with different sets of deformable structures is studied using a two-way-coupled fluid-structure interaction (FSI) model for multi-phase turbulent flows. In order to track the moving interface between the solid body and the two phase fluid flow, referred to as the wet surface hereafter, the Navier-Stokes equations for large eddy simulation (LES) cast in arbitrary Lagrangian Eulerian (ALE) formulation are used \cite{cajasEtAl2018}. The underlining idea of LES is the separation of scales in turbulent flows. This separation is usually achieved by a filtering operation that, for any vector or scalar quantity $\gamma(x)$, is such that 
\[
\gamma(x) = \overline{\gamma}(x) + \gamma'(x),
\]
where the barred values are the filtered, resolved values which are separated from the un-resolved primed quantities at the sub-grid scale. The filtered Navier-Stokes equations of incompressible flows that account for the coupled deformation of solid structures are:

\begin{equation}
\label{eq:NSles}
    \begin{aligned}
        \rho_{i} \frac{\partial \mathbf{\overline{u}_{i}}}{\partial t} + \rho_{i} [(\mathbf{\overline{u}}_{i} - \mathbf{u}_{m}) \cdot  \nabla]\mathbf{\overline{u}}_{i} - \mu_{i} \nabla^2\mathbf{\overline{u}}_{i} +
        \nabla \tau^{sgs}_i + \nabla \overline{p}_i = -\rho_{i} g {\bf k},\\
        \nabla  \cdot  \mathbf{\overline{u}}_{i} = 0,
    \end{aligned}
\end{equation}
where $\mathbf{\overline{u}}_{i}$ is the velocity vector of fluid $i$, $p_i$ is its pressure, $\rho_{i}$ is the fluid $i$'s density, $\mu_i$ is its dynamic viscosity, $\tau^{sgs}_i=(\overline{\bf u u})_i - (\overline{\bf u}~\overline{\bf u})_i$ is the subgrid-scale stress tensor, $g$ is the magnitude of the acceleration of gravity which points downwards in the direction of ${\bf k}=[0, 0, 1]$, and $\mathbf{u}_{m}$ corresponds to the domain velocity which is obtained from the domain displacement given by the structure's deformation. The $i$ subscript indicates the phase under consideration $i=1$ for water and $i=2$ for air. We model sub-grid scales using the approach of Vreman \cite{vreman2004}, described also in \cite{sagautBook}, with a model constant $c_{\rm vr}=0.1$ used in all of the simulations.

At the free surface, the two fluids interact by means of a shear stress imposed via the two conditions:
\begin{eqnarray*}
    \mathbf{u_1}=\mathbf{u_2}\\
    \mathbf{t_1} = \mathbf{t_2},
\end{eqnarray*}
where $\mathbf{t_i}$ represent the traction forces. We model the dynamics and deformation of the solid bodies representing the vegetation by the Euler equations as derived in \cite{casoni2015}:
\begin{equation}
\label{eqs:solids}
\rho_{s}\frac{\partial^2 \mathbf{d}_{s}}{\partial t^2} = \nabla \cdot \mathbf{P} + \mathbf{b}
\end{equation}
where $\rho_{s}$ is the solid's density, $\mathbf{d}_{s}$ represents the displacement field, $\mathbf{P}$ is the first Piola-Kirchhoff stress tensor, and $\mathbf{b}$ is the load distribution. At the wet surface $\Gamma_s$, the kinematic and dynamic coupling conditions are imposed as:
\begin{equation*}
    \begin{aligned}
        \mathbf{d}_{f} = \mathbf{d}_{s}\\
        \mathbf{t}_{f} = -\mathbf{t}_{s}\\
        \mathbf{u}_{f} = \frac{\partial \mathbf{d}_{f}}{\partial t}
    \end{aligned}
\end{equation*}
where $\mathbf{d}_{f}$ is the displacement of the interface seen from the fluid perspective, $\mathbf{t}_{f}$ and $\mathbf{t}_{s}$ correspond to the traction forces exerted on the fluid-solid interface respectively, and $\mathbf{u}_{f}$ is the velocity of the fluid. These conditions express continuity of the displacements, stresses, and velocities at the wet surface. 
We non-dimensionalize equation \ref{eqs:solids} using the length scale $h_{avg}$, velocity scale $u_{avg}$, and the stress scale $\rho_{f}u_{avg}^{2}$ \cite{thekkethil2019level} yielding:
\begin{equation}
\label{eqs:solids-nondim}
\rho_{r}\frac{\partial^2 \mathbf{d^{*}}}{\partial \tau^2} = \nabla \cdot \mathbf{P^{*}} + \mathbf{b^{*}}
\end{equation}
where $\mathbf{d^{*}} = \frac{\mathbf{d_{s}}}{h_{avg}}$, $\mathbf{b^{*}} = \frac{\mathbf{b}h_{avg}}{u_{avg}^{2}}, \mathbf{P^{*}}= \frac{\mathbf{P}}{\rho_{f}u_{avg}^{2}} $ \\
In our case, the length scale $h_{avg}$ represents the average dam bore height, and $u_{avg}$ corresponds to the average flow velocity of the bore before hitting the cylinder.
This non-dimensionalization results in non-dimensional parameters such as the density ratio ($\rho_{r}=\rho_{s}/\rho_{f}$), and non-dimensional modulus of elasticity $E^{*} = E/\rho_{f}u_{avg}^{2}$.

\subsection{Numerical implementations}
We discretize equations \ref{eq:NSles} and \ref{eqs:solids} in space via linear finite elements \cite{vazquezEtAlALYA2016, houzeauxAubryVazquez20111} and advanced in time via a backward differentiation formula of second order (BDF2) \cite{codina2006some}. The fluid-structure interaction algorithm is detailed in \cite{cajasEtAl2018}.
For the solid mechanics problem, we discretize the Euler equations (\ref{eqs:solids}) using a standard Galerkin method for large deformations with a Newmark time integration scheme, details of the numerical strategy can be found in \cite{belytschko2000w} and \cite{casoni2015alya}. The FSI problem is solved with an iterative staggered multi-code approach. We use a strong Gauss-Seidel coupling algorithm with an Aitken under-relaxation factor. In each time step, we solve the fluid flow first to calculate the loads on the deformable structure required in the solid mechanics solver. We then solve the solid mechanics problem, and the new position of the body is sent back to the fluid mechanics solver. Convergence is reached when the residual of successive coupling iterations is less than the residual of coupling parameters. 
Details of the multi-code coupling strategy can be found in \cite{cajasEtAl2018}.

We use a conservative level set model to track the motion of the water-air interface. The level set equations are approximated by a stabilized finite element method using orthogonal subgrid scales for numerical stability as detailed in \cite{owenCodina2007}.

\subsection{Boundary conditions}
All boundaries (lateral, left, right and top walls) are specified as no-flux with free slip. We apply a no-slip boundary condition at the cylinder wall and at the bottom surface through the wall-modeled LES approach described in \cite{owen2020wall} with the following Reichardt's law of the wall: \cite{reichardt1951vollstandige}
\begin{equation}
    u^{+} = \frac{1}{\kappa} \ln(1+\kappa y^{+}) + 7.8 (1-e^{-\frac{y^{+}}{11}}-\frac{y^{+}}{11}e^{-0.33y^{+}}),
    \label{reichard-wall}
\end{equation}\\
where $u^{+} = \frac{\overline{u}}{u_{\tau}}$ is the dimensionless velocity, $y^{+} = \frac{yu_{\tau}}{\nu}$ is the dimensionless wall distance, $u_{\tau} = \sqrt{\frac{\tau_{w}}{\rho}}$ is the friction velocity, $\tau_{w}$ is the wall shear stress, $\nu$ is the kinematic viscosity and $\kappa=0.41$ is the von K\'arman constant. 



\section{Model validation}
\label{benchmarking}

In the current study, we have selected two experimental analyses to validate the numerical model. We use Benchmark I mainly to validate our simple dam break computation without any obstacles. In contrast, Benchmark II is valuable to further assess our model in the presence of obstacles.

\subsection{Benchmark I} 
\begin{figure}[H]
\centering
	\includegraphics[width=\textwidth]{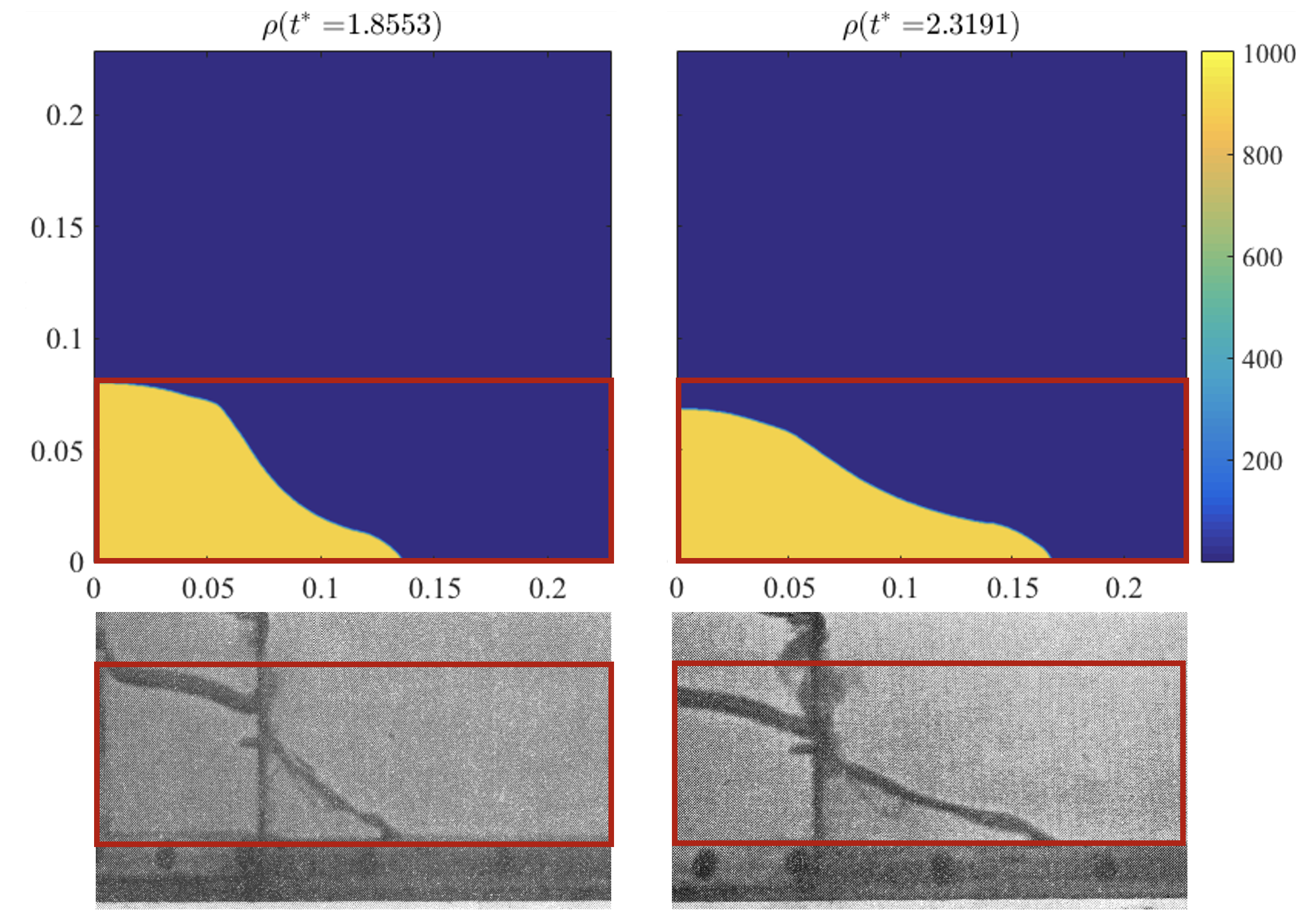}
 	\caption{Benchmark I: Qualitative comparison of the numerical results of the collapsing column against the lab photographs of \cite{martinMoyce1952}.}
\label{fig:MoyceVsAlya}
\end{figure}

\begin{figure}
\centering
	\includegraphics[width=0.5\textwidth]{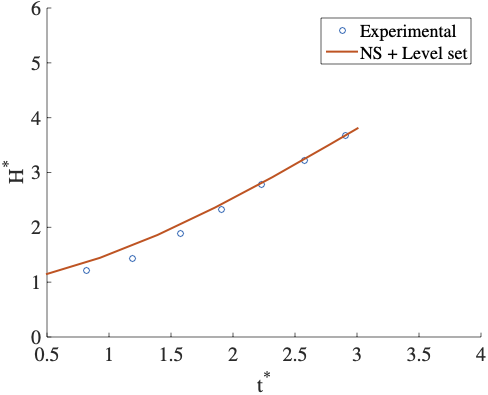}
 		\caption{Benchmark I: Computed front position against the measurements by \cite{martinMoyce1952}.}
	      \label{fig:computedVSmeasureMoyce}
\end{figure}

In 1952, Martin and Moyce \cite{martinMoyce1952} executed a comprehensive set of measurements to characterize the dynamics of a collapsing liquid column on a rigid horizontal plane. We use the non-dimensional data reported therein to validate our flow solver in the absence of obstacles. The choice of Benchmark I  is particularly suited to validate a numerical model because, as observed by Martin and Moyce \cite{martinMoyce1952}, the measurements were not influenced by either viscosity or surface tension and can hence be generalized across scales for both viscous and inviscid flow solvers. 
To compare the numerical results against the measurements, the following non-dimensional quantities are used:

\begin{align}
    H^* = \eta/(2a)\\
    t^* = t\sqrt{g/a}
\end{align}
where $\eta$ is the water depth and $a$ is the initial base. The time evolution of the collapsing column at $t^* = 1.8553$ and $t^*=2.3191$ is qualitatively compared in figure~\ref{fig:MoyceVsAlya} against the experiments by Martin and Moyce whereas figure~\ref{fig:computedVSmeasureMoyce} shows the quantitative comparison of the front position versus the non-dimensional time. 

\subsection{Benchmark II}
\begin{figure}[t]
\centering
\centerline{\includegraphics[trim={0 23cm 5cm 15cm},clip,scale=0.25]{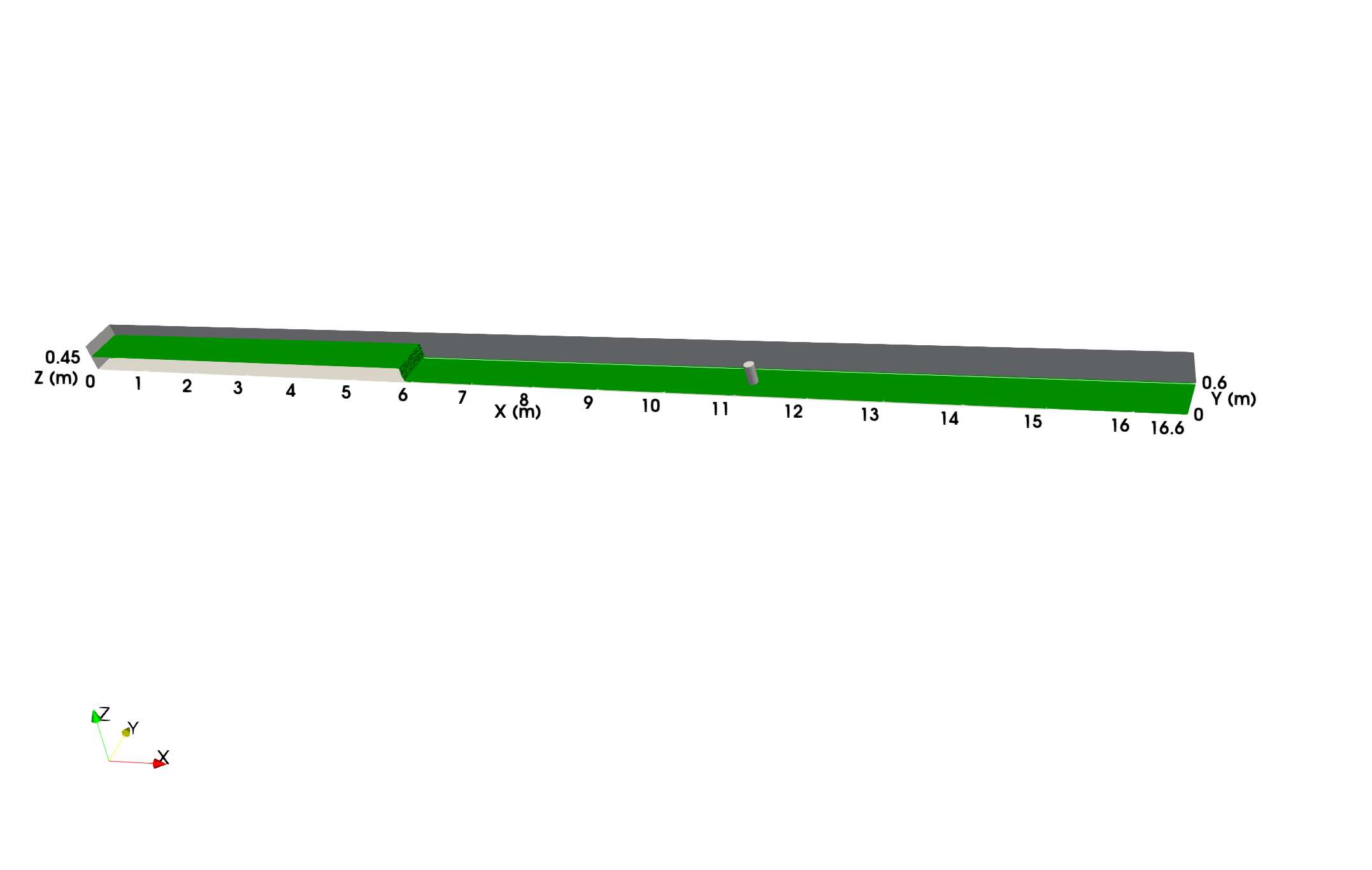}}
\centering
\caption{Benchmark II: Schematic illustration of numerical testing flume with cylindrical column for model validation used in the physical experiment by Arnason et al. \cite{arnason2009tsunami} 
}
\label{fig:arnason_case_setup}
\end{figure}

For Benchmark II, we compare the output of our code against the experimental data of \cite{arnason2009tsunami}. The experiment consists of a flow generated by a collapsing column of water over a uniform, thin water layer. The interaction of the collapsing water column and the thin water layer is responsible for the formation of a bore propagating downstream. 
The domain is $16.6\,{\rm m}$ long, $0.6\,{\rm m}$ wide, and $0.45\,{\rm m}$ deep, and the water flume is separated in two parts. In the upstream region, the water is $0.25\,{\rm m}$ deep, and downstream the water depth is $0.02\,{\rm m}$. The gate is situated at $5.9\,{\rm m}$ from the upstream wall, and the cylinder is placed $5.2\,{\rm m}$ from the dam. The diameter of the cylinder is $0.14\,{\rm m}$. The schematic of the model is shown in figure \ref{fig:arnason_case_setup}.

\begin{figure}[t]
\centering
\begin{subfigure}[t]{0.49\textwidth}
\includegraphics[trim={190 0 210 50},clip,scale=0.4]{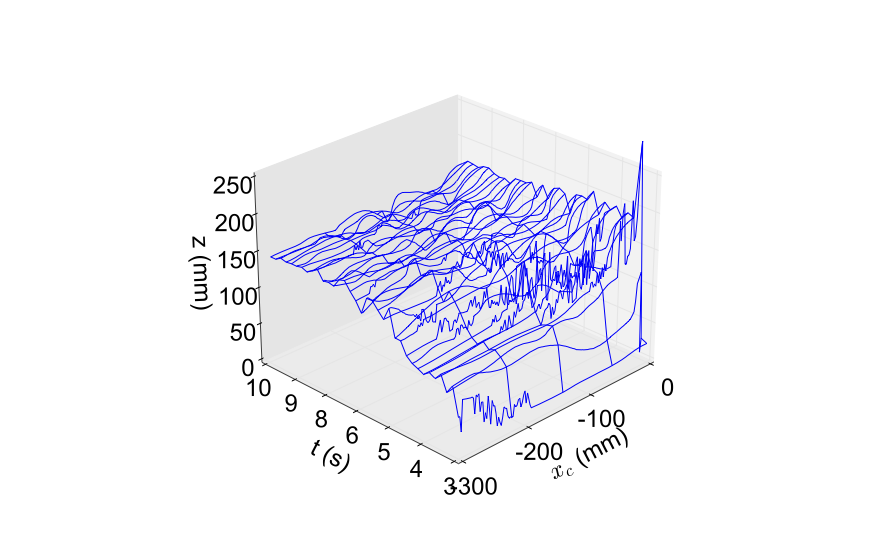}
\end{subfigure}%
\begin{subfigure}[t]{0.49\textwidth}
\includegraphics[trim={280 0 250 50},clip,scale=0.38]{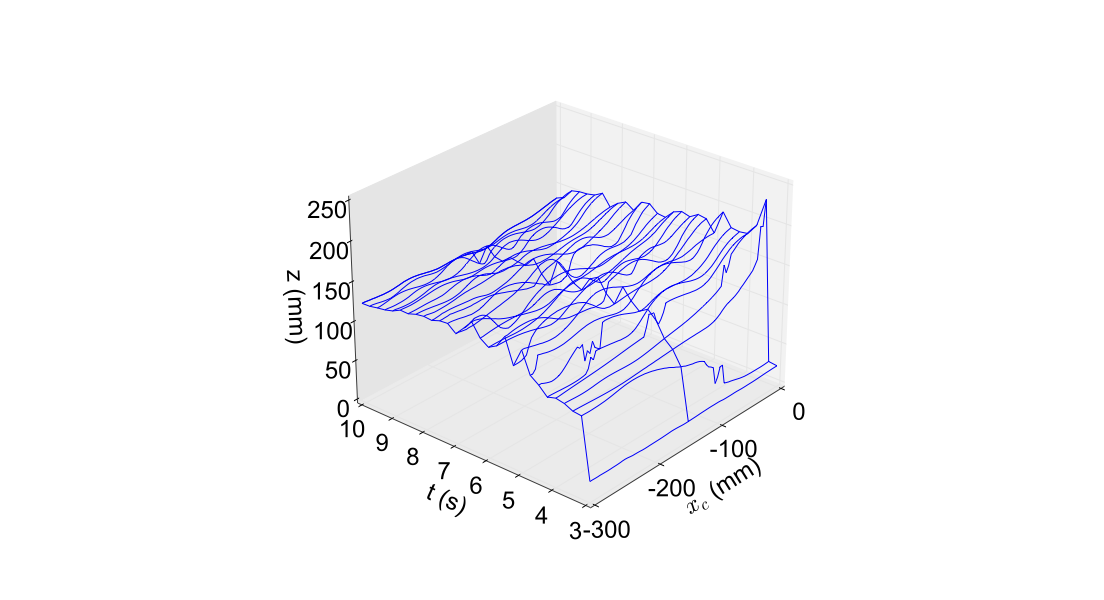}
\end{subfigure}
\caption{
Comparison of the evolution of water surface elevation along the streamwise direction from the cylinder leading edge. Left: Fine mesh scheme. Right: Coarse mesh scheme}
\label{fig:benchmarking1}
\end{figure}
We have conducted our tests for different grid resolutions. Table \ref{table1} shows the details of grid sizes for fine and coarse resolution grid setup.
The comparison of water depth evolution between the fine grid and coarse grid setup along the streamwise direction from the cylinder leading edge is given in figure \ref{fig:benchmarking1}. The location $x_{c} = 0$ corresponds to the position of the leading edge of the cylinder, and $x_{c}$ is negative towards the upstream direction. 
At coarse grid setup, the maximum bore depth at cylinder leading edge reaches around $0.225\,{\rm m}$ at $3.6\,{\rm s}$, whereas as reported by Arnason et al. \cite{arnason2009tsunami}, the maximum bore depth reached $0.27\,{\rm m}$ at $3.5\,{\rm s}$, which indicates that our computation underestimates the water depth at this coarse grid resolution and is unable to capture the water level rise after the bore impact due to numerical dissipation.

\begin{figure}[t]
\centering
	\includegraphics[width=0.6\textwidth]{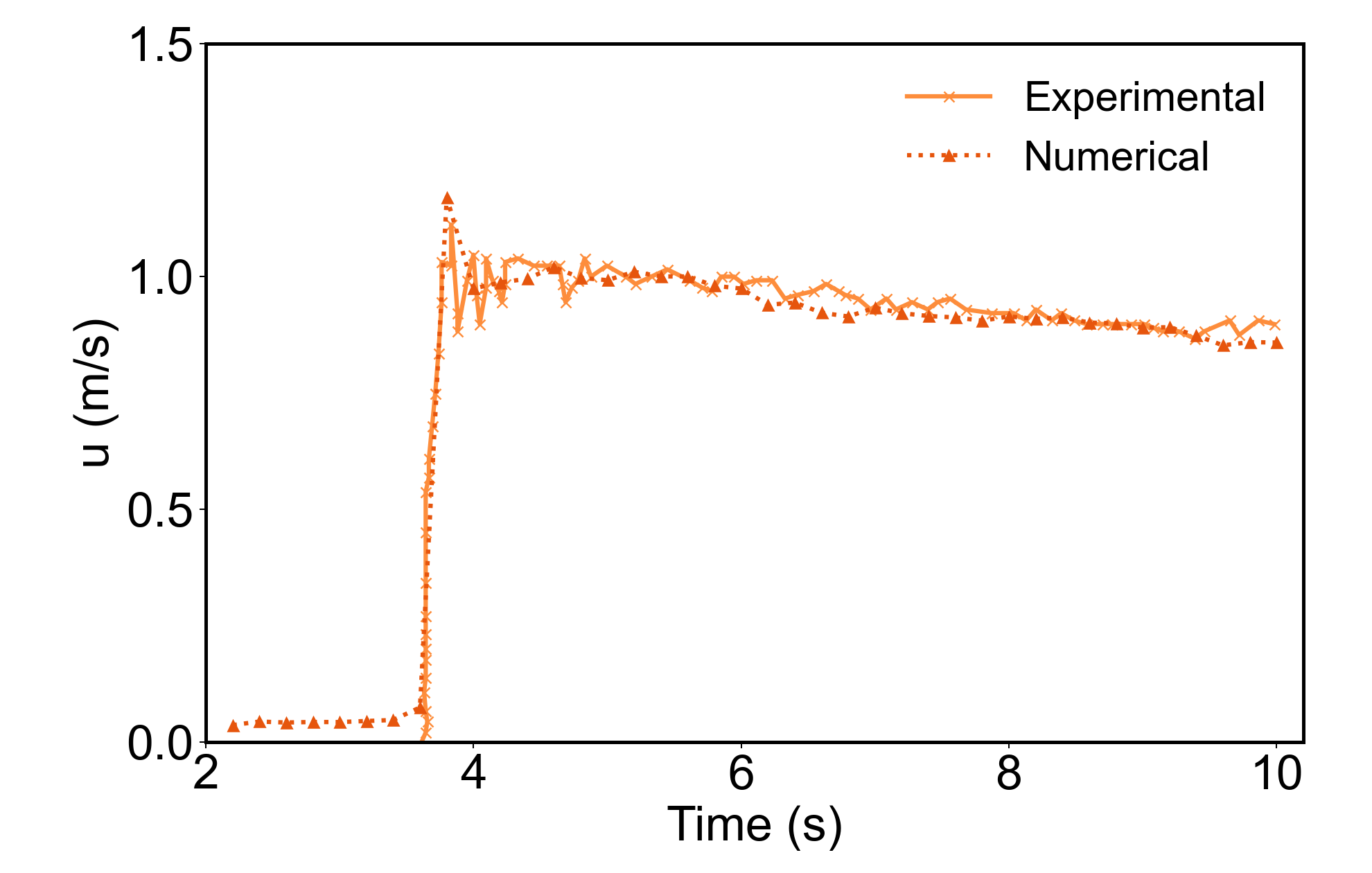}
 	\caption{Comparison of the time evolution of velocity magnitude at cylinder upstream edge location}
\label{arnason-vel}
\end{figure}

In contrast, the fine grid resolution perfectly captures the maximum water level after the bore impact on the cylinder surface. Once the bore hits the cylinder at  $3\,{\rm s}$, the water level increases at the cylinder leading edge due to reflection, and the bore reaches a maximum of $0.268\,{\rm m}$ at $3.5\,{\rm s}$. The details of the fine mesh grid set up and comparison of maximum water level at the cylinder upstream edge between the experimental and numerical case is reported in table \ref{table1}. We also use the coarse and fine grid simulations to compare the depth-averaged flow velocity against the experimental data at the cylinder location ($x=11.1\,{\rm m}$). As shown in figure \ref{arnason-vel}, before the bore arrival at cylinder upstream edge, a negligible discrepancy of velocity exists between numerical analysis and laboratory experiment due to numerical oscillation ($t=2-3.7\,{\rm s}$), but once the bore reaches the cylinder upstream of the wall, the numerical result approaches to the average position of the velocity-time curve of the experiment.\\
From the comparisons, we conclude that the current numerical method, including the turbulence model, wall function, grid size ($0.0075\,{\rm m}$), and the boundary conditions, correctly predict the water depth and flow velocity evolution and can be employed to simulate the tsunami bore and energy flux analysis for multiple cylinders configuration.
\begin{table}[t]
\centering
\caption{Comparison of maximum water surface elevation at cylinder upstream edge of numerical study against the laboratory data by \cite{arnason2009tsunami}}
\begin{tabular}{cccc}
\toprule
\multicolumn{1}{p{2.0cm}}{\centering Method} & 
\multicolumn{1}{p{2.5cm}}{\centering Number of elements} & 
\multicolumn{1}{p{1.5cm}}{\centering Grid size (m) } &
\multicolumn{1}{p{1.2cm}}{\centering $h_{max}$ (m)} \\
\midrule
Experiment & - & - & 0.27 \\
Numerical: Coarse grid resolution & $2 \times 10^{6}$ & 0.025 & 0.22 \\
Numerical: Fine grid resolution & $12 \times 10^{6}$ & 0.0075 & 0.268 \\
\bottomrule
\end{tabular}
\label{table1}
\end{table}
\section{Numerical experiments}
\label{model setup}
We carry out numerical experiments for different vegetation parameters such as elastic moduli, diameters, and stem spacing to calculate the energy balance. The details of numerical flume setup, non-dimensional parameters, and energy balance calculations are described in the following sections.      
\subsection{Size of the numerical flume}
\label{size-flume}
\begin{figure}[t]
\centering
\includegraphics[scale=0.26]{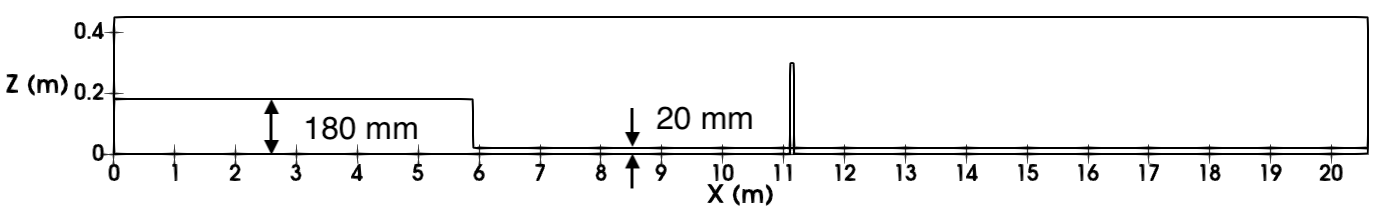}
\includegraphics[scale=0.26]{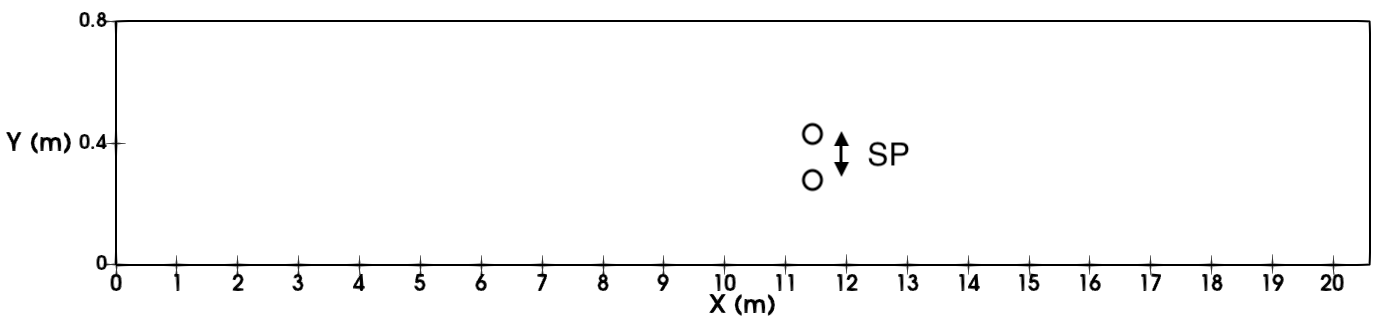}
\begin{minipage}{1.0\textwidth}
\centering
\caption{Schematic illustration of numerical testing flume used for multiple cylinder configurations with cylindrical columns. Top: XZ plane (front view) of the flume showing the dam depth at the upstream and downstream regions. Bottom: XY plane (top view) of the flume; here, $SP$ corresponds to the center-to-center distance between the cylinders.}
\label{fig:case_setup}
\end{minipage}
\end{figure}

We modify our numerical setup used to validate our simulations against the experimental results by \cite{arnason2009tsunami} to accommodate multiple cylinders in the domain along the spanwise direction. The modified testing flume is $20.6\,{\rm m}$ long, $0.8\,{\rm m}$ wide, and $0.45\,{\rm m}$ deep. The dam depth at the upstream reservoir is $0.18\,{\rm m}$, and in the downstream region, the initial water depth is $0.02\,{\rm m}$ water depth. We increase the streamwise length of the domain to avoid wave reflection at the end wall and the domain width to minimize the blockage effect between the lateral walls and cylinders. The sketch of the numerical testing flume is shown in figure \ref{fig:case_setup}. 
\subsection{Scaled parameters of tsunami bore generation}
The onshore height of a tsunami bore can vary from $2\,{\rm m}$ to $15\,{\rm m}$ based on the tsunami intensity scale \cite{imamura1942history,imamura1949list,iida1956earthquakes,iida1970generation,iida1967preliminary}. In this study, we scale down the model to reduce the computational cost. Since the grid size of the numerical model needs to be sufficiently smaller to capture the small-scale eddies, the computational cost will be huge due to a large scale separation between the eddies unless  the problem has been scaled down. Therefore, to simulate the tsunami phenomena, we scale down the computational model using the scale factor $SF = 1:60$. In this study, We choose a possible tsunami bore height of $4.5\,{\rm m}$ for the Indian Ocean tsunami \cite{lavigne2009reconstruction} on a real scale, which corresponds to a bore height of $0.075\,{\rm m}$ in the scaled simulations. Following the method presented by \cite{arnason2009tsunami}, we then estimate the initial dam depth as $0.18\,{\rm m}$ based on the onshore tsunami depth in the scaled simulations. 

Our approach to scaling implies that the Froude number is approximately the same in a field setting as in our scaled simulations. We focus on maintaining scale similarity in the Froude number because it is an important parameter for onshore tsunami propagation and other free-surface flows, particularly where hydraulic jumps occur \cite{chanson2002hydraulics}, \cite{hager1988b}. A tsunami generally flows inland as a supercritical flow characterized by $Fr>1$. Different studies proposed different ranges of Froude numbers. For example, \cite{kawata1999tsunami} suggested a range of $0.7-2$ for the Froude number based on field surveys conducted for past tsunamis. Similarly, a study by \cite{foytong2013analysis} found Froude numbers between $1$ and $1.5$. Informed by these estimates, we use $Fr=1.2$ for all simulations, calculated based on the tsunami inflow condition before impact on the cylinders. We keep the inflow condition the same for all simulations, implying that $Fr$ is the same for all cases (see table \ref{table_scaled}). 

The main drawback of ensuring scale-similarity in the Froude number is that the Reynolds in our simulations is lower than it would typically be in the field. Since the flow structure changes in the vicinity of the tree trunks due to dissipation, the Reynolds number ($Re_{h}$) is calculated for the scaled model using the hydraulic radius ($R_{h}$) as a characteristic length as employed by \cite{arnason2009tsunami}, where the hydraulic radius is defined as a ratio of flow area ($A$) and wetted perimeter ($P$),
which are defined as \[A = h_{avg}B\] 
\[P = 2(h_{avg}+B)
\]
where $h_{avg}$ is the average incoming bore depth and $B$ is the flume width.
In this study, the $Re_{h}$ is $2.8\times 10^{4}$ for scaled model, which represents the $Re_{h}$ value of $13\times 10^{6}$ on a real scale. 
In the scaled simulation, the initial bore depth is the same for all configurations. Since the hydraulic radius is calculated based on the flow area before the bore impact on vegetation, the value of $Re_{h}$ is the same for all configurations in scaled simulations. The bore parameters corresponding to the real scale and scaled model are presented in table \ref{table_scaled}. 

\begin{table}[t]
\centering
\caption{Scaled tsunami parameters used in the simulations}\label{tbl-scaled}
\resizebox{\textwidth}{!}{\begin{tabular}{ccc}
\toprule
\multicolumn{1}{p{5.2cm}}{\centering Tsunami \\ parameters} & 
\multicolumn{1}{p{2.2cm}}{\centering Real scale \\ parameters} & 
\multicolumn{1}{p{2.2cm}}{\centering Scaled model \\ parameters } \\
\midrule
Onshore bore depth: $h_{avg}$ (m)  & $4.5$ & 0.075   \\
Onshore bore velocity: $u_{avg}$ (m/s) & $7.9$ & $1.02$  \\
Cylinder diameter: $D$ (m) & $0.4-1$ & 0.14,0.07,0.035  \\
Height of vegetation: $H$ (m) & $15$ & $0.4$ \\
Modulus of elasticity: $E$ (Pa) & $10^{8}-10^{10}$ & $10^{4}-10^{6}$ \\
Non-dimensional modulus of elasticity: $E^{*}=E/\rho_{f} u_{avg}^{2} $ & $3 \times 10^{3}-3 \times 10^{5}$ & $20-2 \times 10^{3}$ \\
Froude Number: $Fr=v_{avg}/\sqrt{gh_{avg}}$ &$1.2$&$1.2$ \\
Reynolds Number: $Re_{h}=\frac{v_{avg} R_{h}}{\nu}$ & $13 \times 10^{6}$ &  $2.8 \times 10^{4}$ \\
\bottomrule
\end{tabular}}
\label{table_scaled}
\end{table}

\begin{table}[t]
\centering
\caption{Non-dimensional parameters used in the present study}
\begin{tabular}{cc}
\toprule
\multicolumn{1}{p{3cm}}{\centering Parameters} & 
\multicolumn{1}{p{3cm}}{\centering Values \\ considered } \\
\midrule
Vegetation flow parameter: $VFP = \frac{E^{*}I BD}{\rho_{r} h_{avg} H^{3}SP^{2}}$  & $90-225$ \\
Vegetal parameter: $VP = \frac{B D}{SP^{2}}$ & $1.4-5.7$   \\
Roughness density: $\lambda_{f}=aH$ & $0.28-2$ \\
Solid volume fraction: $\phi=\frac{\pi}{4}aD$ & $0.2-0.35$ \\
\bottomrule
\end{tabular}
\label{table_non_dim}
\end{table}

\subsection{Different parameters of multiple cylinder configurations}
\label{cyl-config}
In these simulations, we have modeled trees as cylinders since emergent plants with rounded stems can be approximated as circular base cylinders \cite{nepf2012flow}. To incorporate the vegetation parameters into the cylinders, we have chosen the properties of hardwoods such as pine, oak, and maple trees.  The diameters of vegetation in the coastal forest can vary depending on the canopy height and age. Based on the study by \cite{bingham1992canopy} and  \cite{tanaka2012effectiveness}, we have chosen a range of the diameters of vegetation which is $0.4 - 1\,{\rm m}$. The average height  ($H$) is chosen as $15\,{\rm m}$. According to the field investigation conducted by \cite{pasha2018tsunami}, \cite{rodriguez2016field} coastal forests can be dense, intermediate, and sparse based on the $SP/D$ ratio ($SP$ is the gap between the vegetation along a cross-stream direction and $D$ is the diameter), which can vary from $1.1-5$. In this study, we have mainly used dense and intermediate arrangements where $SP/D$ is $1.5$ and $2$. To study the effect of bendability on energy balance, a suitable parameter should be identified to represent flexible vegetation in the real scenario. For studying the deformation of elastic bodies, one of the common parameters is Young's modulus ($E$) which is approximately around $10^{8}-10^{10}\,{\rm Pa}$ \cite{green1999mechanical} for hardwoods. All the vegetation parameters are reported in table \ref{table_scaled}. 

Since we have scaled the tsunami flow parameters, the vegetation parameters must be scaled down as well. The main challenge is to scale the modulus of elasticity so that the deflection of cylinders will be correctly captured. Scaling the modulus of elasticity using the scale factor $SF = 1:60$ may not be correct because, apart from the modulus of elasticity, rigidity also depends on the moment of inertia, which can not be captured unless a proper scaling parameter is used. Hence, to properly scale the modulus of elasticity, the parameter $EI$, known as flexural rigidity, is scaled down. Here, the parameter $I$ is the moment of inertia ($I=\pi D^{4}/4$). A similar approach was implemented by \cite{lakshmanan2012}, where rigidity was scaled using a non-dimensional vegetation flow parameter ($VFP$) expressed as
\begin{equation}
   VFP \equiv \bigg(\frac{E^{*}I BD}{\rho_{r} h_{avg} H^{3}SP^{2}}\bigg)
\end{equation}
where $H$ is the vegetation height, $u_{avg}$ is the flow velocity of the bore, $B$ is the bed width, $SP$ is the center to center distance between the cylinders, $\rho_{r} = \rho_{s}/\rho_{f}$, $\rho_{s}$ is the vegetation density which is $500\,{\rm kg/m^{3}}$, $\rho_{f}$ is the fluid density ($1000\,{\rm kg/m^{3}}$) and $E^{*} = E/\rho_{f}u_{avg}^{2}$ is the non-dimensional modulus of elasticity. Here, we choose hardwood trees such as pine as a reference point for our scaling. First, we calculate $VFP$ based on the parameters on the real scale reported in table \ref{table_scaled}. The range of $VFP$ used in this study is shown in table \ref{table_non_dim}. Then, the scaled dimensional modulus of elasticity of the model can be calculated by the following
\begin{equation}
 E_{s} = \frac{h_{s_{avg}} H^{3}_{s} u^{2}_{s_{avg}} SP_{s}^{2}}{I_{s} B_{s} D_{s}} VFP_{real-scale}
 \label{eq:scale}
\end{equation}

\begin{table}[t] 
\centering
\caption{Configuration details of cylinder parameters used in scaled simulations}
\resizebox{\textwidth}{!}{\begin{tabular}{cccc}
\toprule
\multicolumn{1}{p{4.5cm}}{\centering Vegetation configurations} &
\multicolumn{1}{p{3.5cm}}{\centering One cylinder} & 
\multicolumn{1}{p{3.5cm}}{\centering Two-cylinder} & 
\multicolumn{1}{p{3.5cm}}{\centering Four-cylinder} \\
\midrule
Diameter (D) (m) &  $0.14$ & $0.07$ & $0.035$\\
$SP/D=1.5$ & - & $0.105$ & $0.0525$ \\
$\phi_{SP/D=1.5}$ & - & $0.35$ & $0.35$ \\
$\lambda_{f_{SP/D=1.5}}$ & - & $1$ & $2$ \\
$SP/D=2$ & - & $0.14$ & $0.07$ \\
$\phi_{SP/D=2}$ & - & $0.2$ & $0.2$ \\
$\lambda_{f_{SP/D=2}}$ & - & $0.56$ & $1.12$ \\
Total flow blockage area ($\rm m^{2}$) & $0.0154$ & $0.00384$ & $0.0019$ \\
Scaled modulus of elasticity (Pa) & $ 1 \times 10^{5}, 5 \times 10^{4}, 1 \times 10^{4}$ & $5 \times 10^{5}, 1 \times 10^{5}, 5 \times 10^{4}$ & $5 \times 10^{6}, 1 \times 10^{6}, 5 \times 10^{5}$ \\
Maximum deflection (m) & $0.01$ & $0.02$ & $0.05$\\ 
\bottomrule
\end{tabular}}
\label{table_cylinder}
\end{table}
where subscript $s$ refers to the scaled parameters. From equation \ref{eq:scale} the modulus of elasticity of the scaled model is found to be in the range of $0.01-1\,{\rm GPa}$ and the diameters are in the range of $0.035-0.1\,{\rm m}$ for the three different configurations shown in table \ref{table_cylinder}. Since the rigidity ($EI$) changes with the diameter, we choose three sets of scaled moduli of elasticity and diameters for each case, totaling nine cases reported in table \ref{table_cylinder}. The diameter of the cylinders is set in such a way that the effective channel width along the spanwise direction (Y direction) will be the same for all cases. Therefore, three cases with the same cylinder-perimeter are considered, one cylinder with a diameter of $0.14\,{\rm m}$, two cylinders with a diameter of $0.07\,{\rm m}$, and four cylinders with a diameter of $0.035\,{\rm m}$. These configurations are chosen in such a way so that the rigidity ($EI$) is properly captured based on the vegetation parameters. Because once the diameter increases, rigidity also increases, and therefore to maintain the rigidity in the proper range, elastic modulus needs to be modified to model the cylinders properly. In this study, the range of diameters is chosen based on the rigidity of the scaled models used in the simulation.   These configurations will provide more meaningful insight into what the best choice for tsunami mitigation should be, whether by increasing the elastic moduli or the number of cylinders with a smaller diameter.
We also vary the gap between the cylinders to investigate how the spacing between the cylinders will affect the energy reflection and dissipation. Table \ref{table_cylinder} shows all the parameters of cylinders used in the scaled simulations.

\subsection{Non-dimensional parameters}
As stated in \ref{cyl-config}, $VFP$ is used to scale down the modulus of elasticity. However, additional non-dimensional parameters need to be calculated to verify the scaled model to capture the tsunami bore propagation through vegetation. This set of non-dimensional parameters can either characterize the flow or vegetation configurations.
In the review by Nepf \cite{nepf2012flow}, the non-dimensional parameters such as the roughness density and solid volume fraction are introduced to measure the non-dimensional canopy density. 
As a starting point, and considering fully emergent vegetation, the non-dimensional measure of the vegetation density is the solid volume fraction \cite{wooding1973drag}.
\[
\phi = \frac{\pi}{4}aD,
\]
where $a= D/SP^2$ is the frontal area per canopy volume for a given diameter $D$, and  $SP$ is the average distance between two consecutive stems. The parameters $a$ and $\phi$ vary as a function of the tree diameter and spacing between the trunks. If the vegetation height is considered instead of the diameter, the non-dimensional canopy density is represented by the roughness density, which is defined as $\lambda_{f} = aH$,
where $H$ is the canopy height.
The range of both parameters is presented in table \ref{table_non_dim}. For each set of cylinder gaps, the solid volume fraction is constant in multi-cylinder configurations, whereas the roughness density varies as the diameter of the cylinders changes. Table \ref{table_cylinder} shows the solid volume fraction and roughness density for each configuration.
As reported by Mazda et al. \cite{mazda1997drag}, $\phi$ can reach up to $0.45$ for dense mangrove forests. In the current configurations, since we are using pine forests as reference vegetation parameters, the value of $\phi$ should be lower than the solid volume fraction of dense mangrove forests. In our study, the solid volume fraction ranges from $0.2$ to $0.35$. 

Since the tsunami becomes supercritical while propagating through the onshore vegetation, we calculate the Froude number before the bore impacts cylinders. As described by Arnason et al. \cite{arnason2009tsunami}, we calculate the surge Froude number, which is defined as 
$Fr_{s}=u/\sqrt{gh_{0}}$ where $h_{0}$ is the downstream water depth which is $0.02\,{\rm m}$ at $t=0$ and $u$ is the bore front celerity.
The stationary Froude number at the cylinder upstream edge location is also calculated as $v_{avg}/\sqrt{gh_{avg}}$ where $h_{avg}$ is the average onshore bore depth and $v_{avg}$ is average flow velocity in scaled simulations (see table \ref{table_scaled}). Both parameters match well with the results reported by Arnason et al. \cite{arnason2009tsunami}, indicating that the flow becomes supercritical before hitting the cylinders.   
The Froude number, average water depth, and bore velocity for the individual cylinder configuration are compiled in table \ref{table_froude}.

\begin{table}[t]
\centering
\caption{Bore attributes for different cylinder configurations: Calculation of the flow velocity ($v_{avg}$) and the average bore depth $h_{avg}$ and the non-dimensional parameters such as surge Froude number ($Fr_{s}$), stationary Froude number ($Fr$) and flow Reynolds number ($Re_{D}$). For all scenarios, the downstream water depth ($h_{0}$) is $0.02\,{\rm m}$ (see figure \ref{fig:case_setup})}
\begin{tabular}{ccccccc}
\toprule
\multicolumn{1}{p{1.0cm}}{\centering Cylinder \\ configuration} & 
\multicolumn{1}{p{1.4cm}}{\centering Diameter (D) (m)} & 
\multicolumn{1}{p{1.4cm}}{\centering Flow velocity \\ $v_{avg}$(m/s) } &
\multicolumn{1}{p{1.6cm}}{\centering Bore front celerity \\ $u$(m/s) } &
\multicolumn{1}{p{1.3cm}}{\centering Bore height \\ $h_{avg}$(m)} &
\multicolumn{1}{p{1.8cm}}{\centering $Fr=v_{avg}/\sqrt{gh_{avg}}$} &
\multicolumn{1}{p{1.5cm}}{\centering $Fr_{s}=u/\sqrt{gh_{0}}$} \\
\midrule
One-cylinder & $0.14$ & 1.02 & 1.3 & 0.075 & 1.2 & 2.93 \\ 
Two-cylinder & $0.07$ & 1.02 & 1.3 & 0.075 & 1.2 & 2.93 \\ 
Four-cylinder & $0.035$ & 1.02 & 1.3 & 0.075 & 1.2 & 2.93 \\
\bottomrule
\end{tabular}
\label{table_froude}
\end{table}

\subsection{Energy Flux} \label{energy flux}
The energy flux at a particular location gives information about how much energy is available for mixing and transport \cite{mackinnon2003mixing}. 
To quantify the protective benefits of tsunami mitigation, we calculate the energy flux at different streamwise locations as computed by \cite{venayagamoorthy2006numerical,lunghino2020protective}. The energy flux is defined as
    \begin{equation}
        \mathbf{f} = \mathbf{u} (p/\rho + g z + q),
        \label{flux}
    \end{equation}
where $q = \mathbf{u} \cdot \mathbf{u}/2$ is the kinetic energy per unit mass.
Since, in this study, we aim to investigate the energy budget along the flow or streamwise direction, we integrate the energy flux at a fixed location $x_{i}$ along the water depth (along the Z axis) and spanwise direction (along the Y axis). We then normalize the depth and spanwise integrated energy flux by the width of the domain so that the energy flux will only vary along the horizontal, streamwise direction.
    \begin{equation}
        \mathbf{F}(t,x=x_{i}) = \frac{\int_{0}^{Y} \int_{0}^{H} \mathbf{f}(t,x=x_{i},y,z)\, dy \, dz}{\int_{0}^{Y} \, dy}
        \label{total flux}
    \end{equation}
where $H = $ water depth.
The kinetic and potential components of the horizontal energy flux are, respectively
    \begin{equation}
        \mathbf{F_{k}} = \frac{\int_{0}^{Y} \int_{0}^{H} \mathbf{u}q\, dy \, dz}{{{\int_{0}^{Y} \, dy}}}
        \label{kinetic flux}
    \end{equation}
    \begin{equation}
        \mathbf{F_{p}} = \frac{\int_{0}^{Y} \int_{0}^{H} \mathbf{u} gz \, dy \, dz}{{\int_{0}^{Y} \, dy}}
        \label{potential flux}
    \end{equation}
The cumulative time-integrated horizontal energy flux is defined as
    \begin{equation}
    \begin{aligned}
        \mathbf{F_{k_{t}}} = \int_{0}^{t} \mathbf{F_{k}}(t,x=x_{i}) \, dt\\
        \mathbf{F_{p_{t}}} = \int_{0}^{t} \mathbf{F_{p}}(t,x=x_{i}) \, dt\\
        \mathbf{F_{t}} = \mathbf{F_{k_{t}}} + \mathbf{F_{p_{t}}}
        \end{aligned}
        \label{cumulative flux}
    \end{equation}
where $\mathbf{F_{t}}$ is cumulative time-integrated total energy flux.

\begin{figure}[t]
\centering
\includegraphics[scale=0.26]{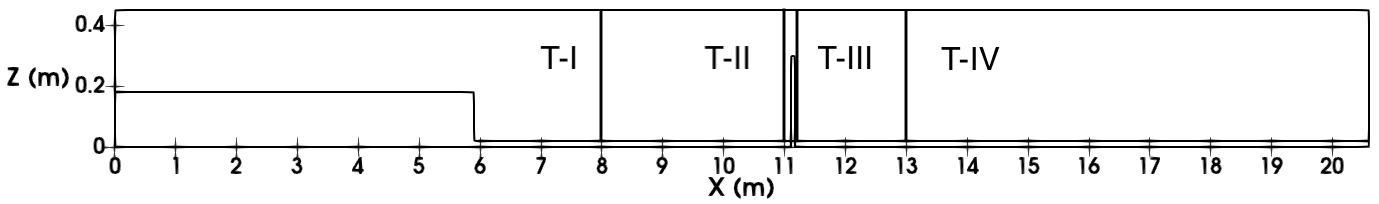}
\begin{minipage}{1.0\textwidth}
\centering
\caption{Location of transects at different streamwise locations for calculation of energy flux}
\label{fig:dam-break-energy-flux}
\end{minipage}
\end{figure}
We take two transects along the streamwise direction to calculate the reflected horizontal energy flux. Transect T-I is taken in the upstream region ($x= 8\,{\rm m}$) and another transect T-II is near to the cylinder leading edge ($x= 11\,{\rm m}$) as shown in figure \ref{fig:dam-break-energy-flux}. If there is any reflection, the energy flux will be drastically reduced at T-I due to the reflected wave arrival. We integrate the horizontal energy flux over the flow period (see equation \ref{cumulative flux}) at T-I and T-II to identify reflection. Finally, we calculate the reflected cumulative horizontal energy flux by taking the difference between the cumulative horizontal energy flux at T-I and T-II. 
  \begin{equation}
    \begin{aligned}
        \mathbf{F^{k}_{R}} = \mathbf{F_{k_{t_{I}}}} - \mathbf{F_{k_{t_{II}}}}\\
        \mathbf{F^{p}_{R}} = \mathbf{F_{p_{t_{I}}}} - \mathbf{F_{p_{t_{II}}}},
    \end{aligned}
  \end{equation}
where $\mathbf{F^{k}_{R}}$ and $\mathbf{F^{k}_{R}}$ are the reflected cumulative horizontal kinetic and potential energy flux, respectively. Finally, we obtain the reflected energy flux coefficient by normalizing kinetic and potential energy flux by the time-integrated total energy flux at T-I, which is calculated as:
    \begin{equation}
    \begin{aligned}
        \mathbf{\phi^{k}_{R_{t}}} = \frac{\mathbf{F^{k}_{R}}}{\mathbf{F}_{t_{I}}}\\
        \mathbf{\phi^{p}_{R_{t}}} = \frac{\mathbf{F^{p}_{R}}}{\mathbf{F}_{t_{I}}}
        \end{aligned}
    \end{equation}
We calculate the cumulative energy flux at T-III, which is located behind the cylinder trailing edge, and at T-IV, located at $2\,{\rm m}$ from the cylinder trailing edge (see figure \ref{fig:dam-break-energy-flux}). Then, the dissipated cumulative horizontal energy flux is calculated by subtracting the cumulative energy flux at the T-III and T-IV, as shown below.

\begin{table}[t]
\centering
\caption{Energy flux parameters calculated in the study}\label{efluxtbl}
\begin{tabular}{cccc}
\toprule
\multicolumn{1}{p{2.5cm}}{\centering Energy flux coefficients} & 
\multicolumn{1}{p{3.5cm}}{\centering Reflected and dissipated energy flux} & 
\multicolumn{1}{p{3.5cm}}{\centering Kinetic and potential energy flux} \\
\midrule
$\mathbf{\phi^{k}_{R_{t}}} = \frac{\mathbf{F^{k}_{R}}}{\mathbf{F}_{t_{I}}}$ & $\mathbf{F^{k}_{R}} = \mathbf{F_{k_{t_{I}}}} - \mathbf{F_{k_{t_{II}}}}$ & $\mathbf{F_{k_{t}}} = \int_{0}^{t} \frac{\int_{0}^{Y} \int_{0}^{H} \mathbf{u}q\, dy \, dz}{{{\int_{0}^{Y} \, dy}}} \, dt$ \\
$\mathbf{\phi^{p}_{R_{t}}} = \frac{\mathbf{F^{p}_{R}}}{\mathbf{F}_{t_{I}}}$ & $\mathbf{F^{p}_{R}} = \mathbf{F_{p_{t_{I}}}} - \mathbf{F_{p_{t_{II}}}}$ & $\mathbf{F_{p_{t}}} = \int_{0}^{t} \frac{\int_{0}^{Y} \int_{0}^{H} \mathbf{u} gz \, dy \, dz}{{\int_{0}^{Y} \, dy}} \, dt$ \\
$\mathbf{\phi^{k}_{D_{t}}} = \frac{\mathbf{F^{k}_{D}}}{\mathbf{F}_{t_{III}}}$ & $\mathbf{F^{k}_{D}} = \mathbf{F_{k_{t_{III}}}} - \mathbf{F_{k_{t_{IV}}}}$ & $\mathbf{F_{k_{t}}} = \int_{0}^{t} \frac{\int_{0}^{Y} \int_{0}^{H} \mathbf{u}q\, dy \, dz}{{{\int_{0}^{Y} \, dy}}} \, dt$ \\
$\mathbf{\phi^{p}_{D_{t}}} = \frac{\mathbf{F^{p}_{D}}}{\mathbf{F}_{t_{III}}}$ & $\mathbf{F^{p}_{D}} = \mathbf{F_{p_{t_{III}}}} - \mathbf{F_{p_{t_{IV}}}}$ &  $\mathbf{F_{p_{t}}} = \int_{0}^{t} \frac{\int_{0}^{Y} \int_{0}^{H} \mathbf{u} gz \, dy \, dz}{{\int_{0}^{Y} \, dy}} \, dt$ \\
\bottomrule
\end{tabular}
\label{table_eflux}
\end{table}
\begin{equation}
 \begin{aligned}
        \mathbf{F^{k}_{D}} = \mathbf{F_{k_{t_{III}}}} - \mathbf{F_{k_{t_{IV}}}}\\
        \mathbf{F^{p}_{D}} = \mathbf{F_{p_{t_{III}}}} - \mathbf{F_{p_{t_{IV}}}}
    \end{aligned}
  \end{equation}
where $\mathbf{F^{k}_{D}}$ and $\mathbf{F^{k}_{D}}$ are the cumulative dissipated kinetic and potential energy flux, respectively.
Similarly, the energy flux dissipation coefficient is defined as:
    \begin{equation}
    \begin{aligned}
        \mathbf{\phi^{k}_{D_{t}}} = \frac{\mathbf{F^{k}_{D}}}{\mathbf{F}_{t_{III}}}\\
        \mathbf{\phi^{p}_{D_{t}}} = \frac{\mathbf{F^{p}_{D}}}{\mathbf{F}_{t_{III}}}
        \end{aligned}
    \end{equation}
Since the bore generated in the wet bed reaches the transects (T-I, T-II, T-III, T-IV) located at different streamwise locations at different times, it is important to apply the time shift while calculating the energy flux difference between the transects. In the present simulation, the bore passes through the T-I (located at $x= 8\,{\rm m}$) at $t= 3.4\,{\rm s}$ and reaches at T-II at $t= 4\,{\rm s}$. To compare and calculate the cumulative energy flux difference, we apply the time shift to the T-I to synchronize the bore arrival time with T-II.
We implement the time shift in T-III and T-IV as well to calculate the energy dissipation. Table \ref{table_eflux} summarizes the energy flux parameters calculated in this study. 

\graphicspath{{Figure_new_2021}}

\section{Results}
\label{results}
\subsection{Rigidity enhances energy reflection for a single cylinder}\label{sec5.1}
\begin{figure}[!t]
    \centering
    \begin{subfigure}{\textwidth}
    \centering
    \includegraphics[trim={100 350 60 20},clip,scale=0.2]{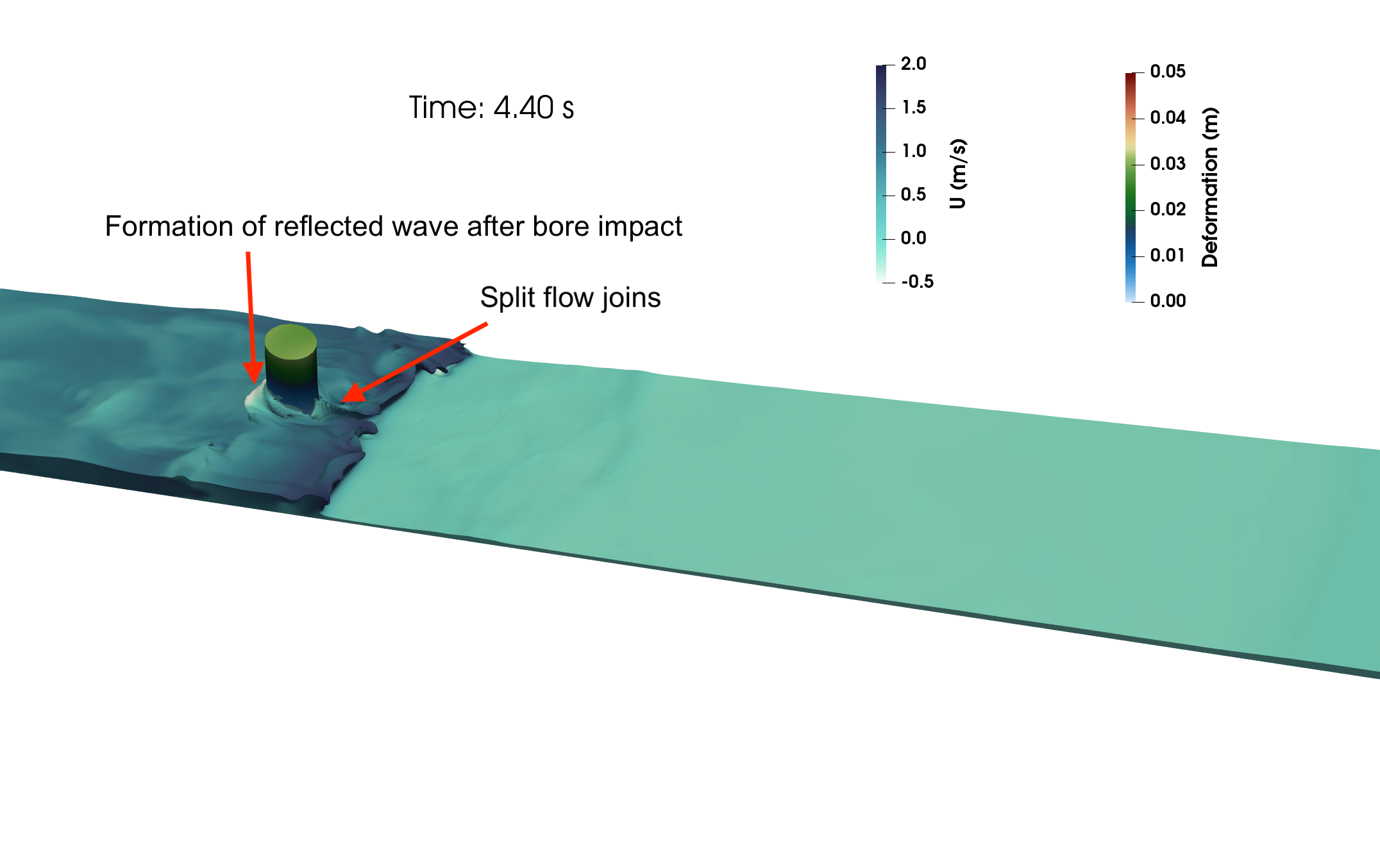}
    \caption{One cylinder: $D=0.14\,{\rm m}$, $E=1 \times 10^{4}\,{\rm Pa}$}\label{fig:8a}		
	\end{subfigure}\\
	\begin{subfigure}[t]{\textwidth}
    \centering
    \includegraphics[trim={100 350 60 20},clip,scale=0.2]{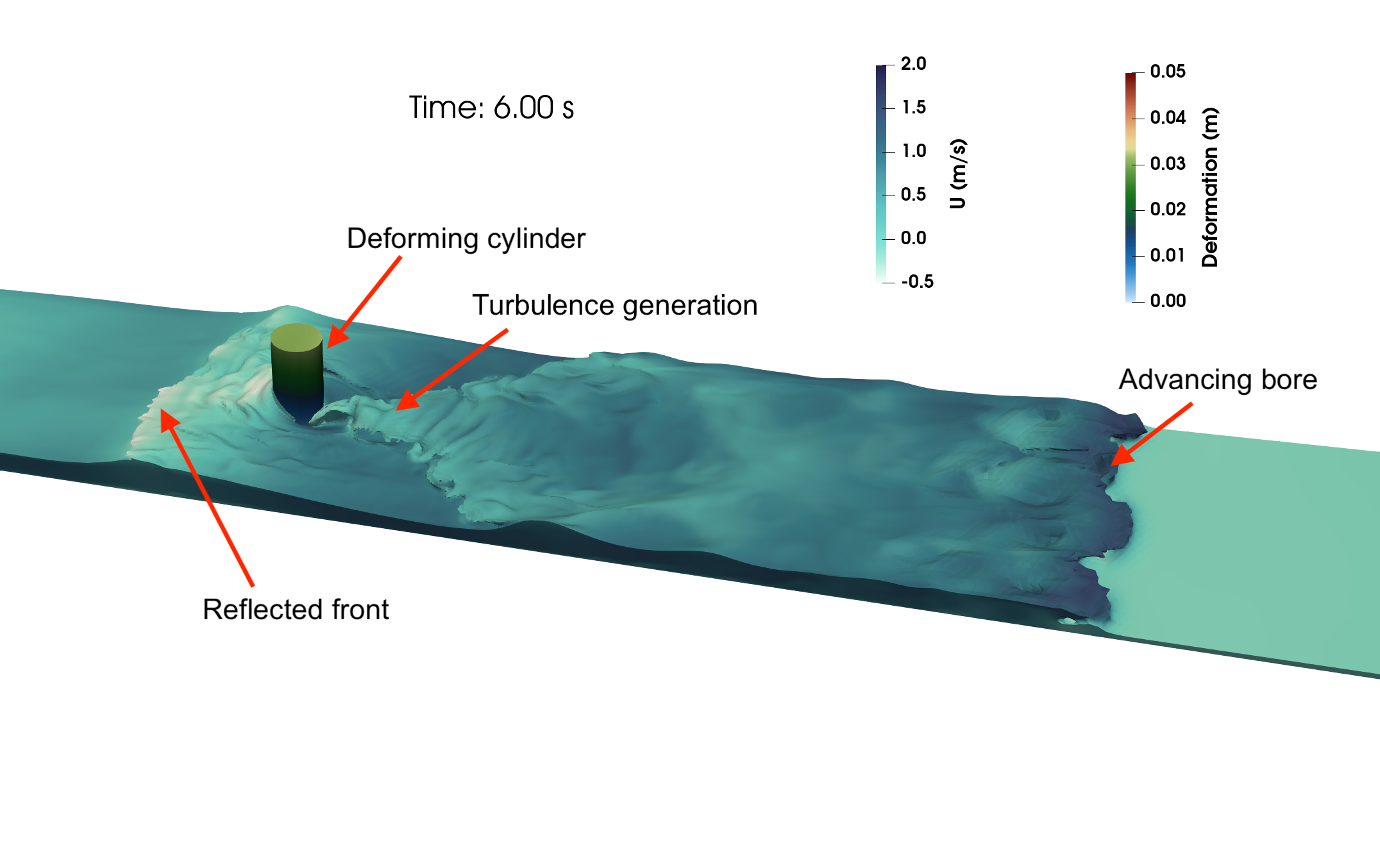}
    \caption{One cylinder: $D=0.14\,{\rm m}$, $E=1 \times 10^{4}\,{\rm Pa}$}\label{fig:8b}
	\end{subfigure}
\caption{Velocity profile and cylinder displacement of one-cylinder configuration for $E=1 \times 10^{4}\,{\rm Pa}$ at different flow time. Figure (a): The flow structures when the cylinders show the maximum deflection, figure (b): The velocity profile and cylinder displacement when the bore wave flows further downstream.}
\label{fig:3d-vel-profile-one-cyl}
\end{figure}
Energy reflection is one of the primary means of attenuating tsunami waves \cite{lunghino2020protective}. When the bore hits the cylinder, part of its energy is reflected. To identify the reflection, we show the free surface profile at different flow times for one cylinder in figure \ref{fig:3d-vel-profile-one-cyl}. At $t=4\,{\rm s}$ the bore front reaches the cylinder upstream face with a bore depth of approximately $0.075\,{\rm m}$. As the bore continues to propagate, the bore depth at the cylinder's upstream face continues to increase due to flow blockage and reaches a maximum of $0.14\,{\rm m}$ before reflection begins. At the downstream side, the split flow joins and begins to form a turbulent wake, whereas the reflected wave flattens out towards the sides of the flume and continues to propagate back upstream (see figure \ref{fig:8b}).

\begin{figure}[t]
        \centering
    \includegraphics[trim={20 10 10 10},clip,scale=0.32]{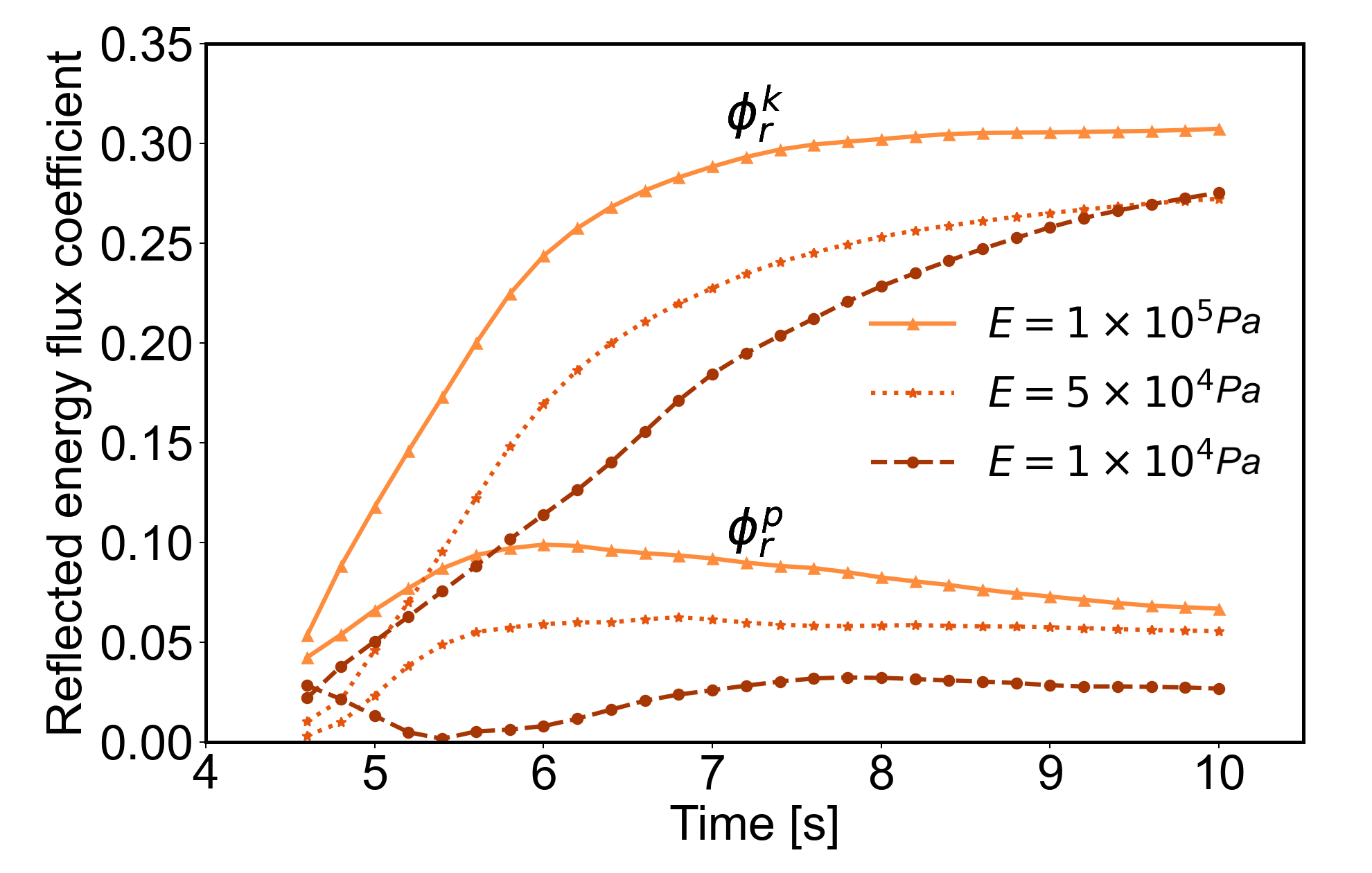}%
\caption{The energy reflection flux coefficient as a function of time of one cylinder for different elastic moduli and constant $Re_{h}=2.8\times 10^{4}$. $\phi^{k}_{R}$: Reflected kinetic energy flux coefficient. $\phi^{p}_{R}$: Reflected potential energy flux coefficient}
\label{fig:energy-flux-ref-one}
\end{figure}

To quantify energy reflection, we calculate the energy flux reflection coefficients as described in section \ref{energy flux}. To study how  reflection varies with the vegetation properties, we compute the reflected energy flux coefficients for different elastic moduli. Figure \ref{fig:energy-flux-ref-one} shows the comparison of the reflected energy flux coefficient of cylinders with three different moduli of elasticity defined in Section \ref{cyl-config}. Notice that deformation increases for decreasing values of the elasticity modulus. As shown in figure \ref{fig:energy-flux-ref-one}, the energy reflection flux coefficient increases sharply after the bore impact ($t= 4.4 - 7\,{\rm s}$). During the later stages of bore propagation ($t= 8 - 10\,{\rm s}$), the reflection flux coefficient becomes almost constant. 
For all three moduli of elasticity, the reflected potential flux is $4$ to $5$ times smaller than the reflected kinetic energy flux, indicating that the reflection of kinetic energy mainly contributes to the total energy reflection.

As stated in section \ref{energy flux}, the reflected energy flux coefficient is calculated based on the cumulative time-integrated reflected energy flux, which means that the slope of the reflected energy flux coefficient shown in figures \ref{fig:energy-flux-ref-one} 
indicates the rate of change of reflected energy over the flow period. 
During the initial bore impact and propagation ($t= 4.4 - 7\,{\rm s}$), the slope of the curve of reflected kinetic energy flux coefficient becomes steeper for the cylinder with the highest modulus of elasticity than with the lowest modulus of elasticity, indicating that the amount of reflected energy increases as rigidity increases.  
As the bore propagates further downstream, the gradient of reflected kinetic energy flux coefficient becomes constant at higher rigidity. However, at the lowest modulus of elasticity, the gradient still increases (see figure \ref{fig:energy-flux-ref-one}) for $t = 8-10\,{\rm s}$. Since the rigid cylinder deforms less after the bore impact but also rebounds less with a negligible deflection after passage of the bore front, the reflected energy is higher during the initial bore impact, and  becomes constant quickly. On the other hand, flexible cylinder deforms more by the hydrodynamic forces exerted on them and hence reflects less energy. However, the flexible cylinder also rebounds more compared to the rigid one, resulting in a continuous increase of reflected energy during the later phase of flow propagation ($t = 8-10\,{\rm s}$).
 
\begin{figure}[t]
        \centering
    \includegraphics[trim={20 10 10 15},clip,scale=0.27]{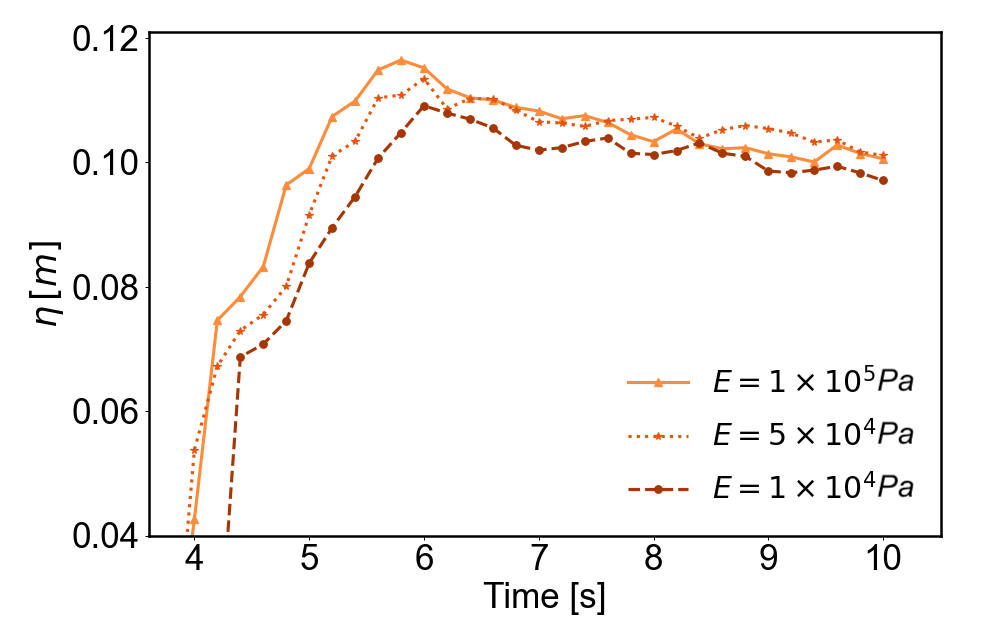}
	\caption{
	 The time evolution of water surface at transect II ($x=11\,{\rm m}$) for different elastic moduli in one cylinder}
\label{fig:eta-flex-one-cyl}
\end{figure}

The reflected potential energy flux coefficient is consistent with the evolution of the reflected kinetic flux coefficient. In figure \ref{fig:energy-flux-ref-one} we find the gradient of reflected potential energy flux coefficient is steeper at the highest rigidity, but during the later phase of flow, the gradient declines for all cases, although the slope declines more rapidly as the rigidity increases. Since rigidity triggers reflection, the bore depth increases at the upstream face of the cylinder. This finding is supported by the average water surface evolution at the cylinder upstream edge location at different elastic moduli shown in figure \ref{fig:eta-flex-one-cyl}. Since the potential flux depends on the water depth, rigidity-induced water elevation increases the potential flux. 

\begin{figure}[t]
        \centering
        \begin{subfigure}[b]{0.49\textwidth}
    \includegraphics[trim={20 10 10 10},clip,scale=0.27]{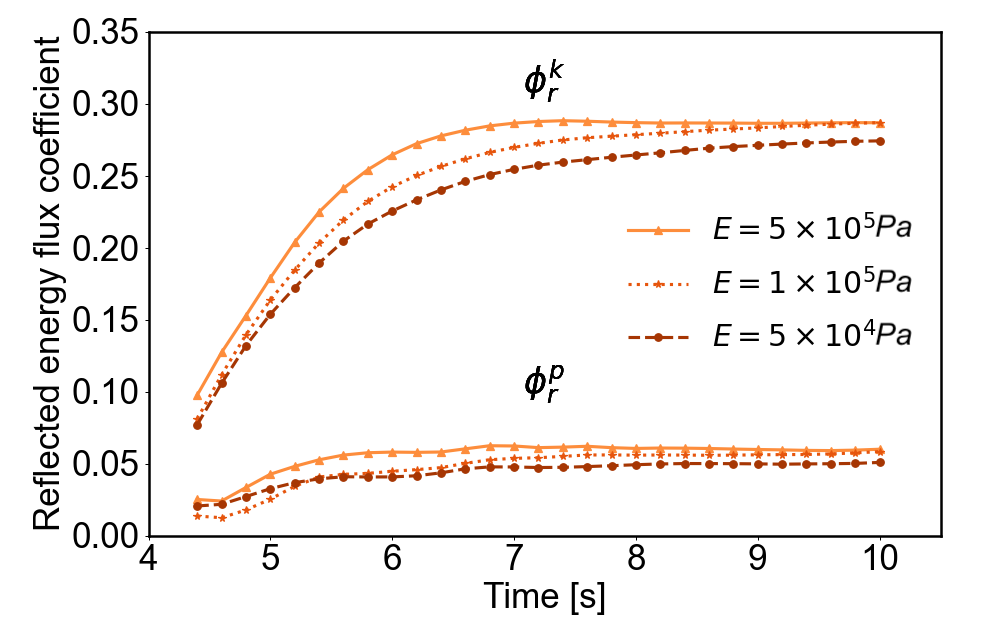}
    \caption{Two cylinders, $SP/D=1.5$}\label{fig:11a}
	\end{subfigure}%
	\begin{subfigure}[b]{0.49\textwidth}
    \includegraphics[trim={20 10 10 15},clip,scale=0.27]{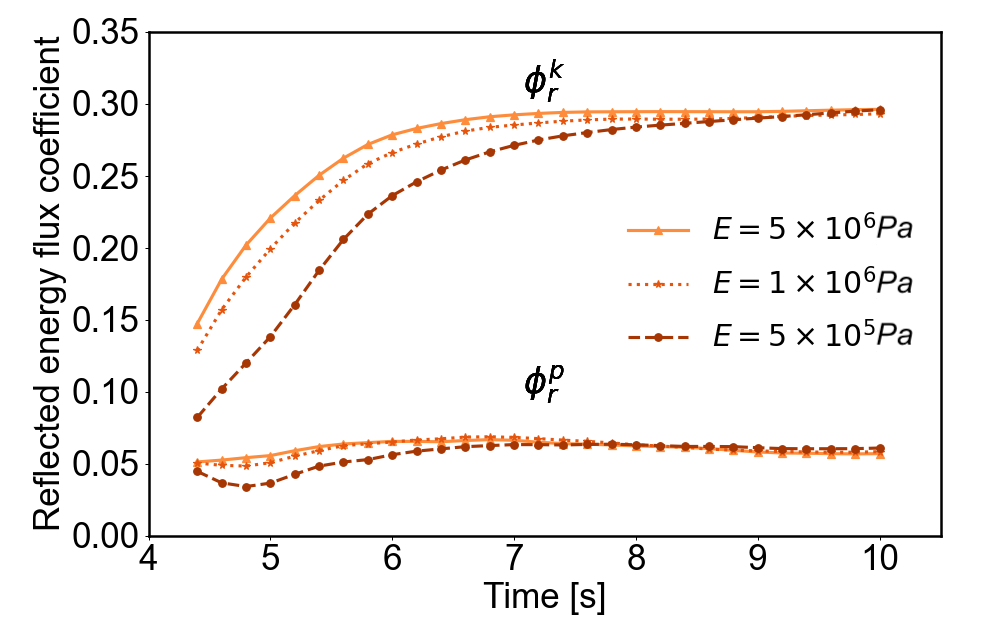}
    \caption{Four cylinders, $SP/D=1.5$}\label{fig:11b}		
	\end{subfigure}%
\caption{ The energy reflection flux coefficient as a function of time of two-cylinder and four-cylinder configuration for different elastic moduli and constant $Re_{h}=2.8\times 10^{4}$. Left: Reflected kinetic and potential energy flux coefficients for two-cylinder arrangement. Right: Reflected kinetic and potential energy flux coefficients for four-cylinder configuration}
\label{fig:energy-flux-ref-multi}
\end{figure}

\begin{figure}[!t]
   \centering
   \begin{subfigure}[t]{\textwidth}
    \centering
    \includegraphics[trim={100 350 60 20},clip,scale=0.2]{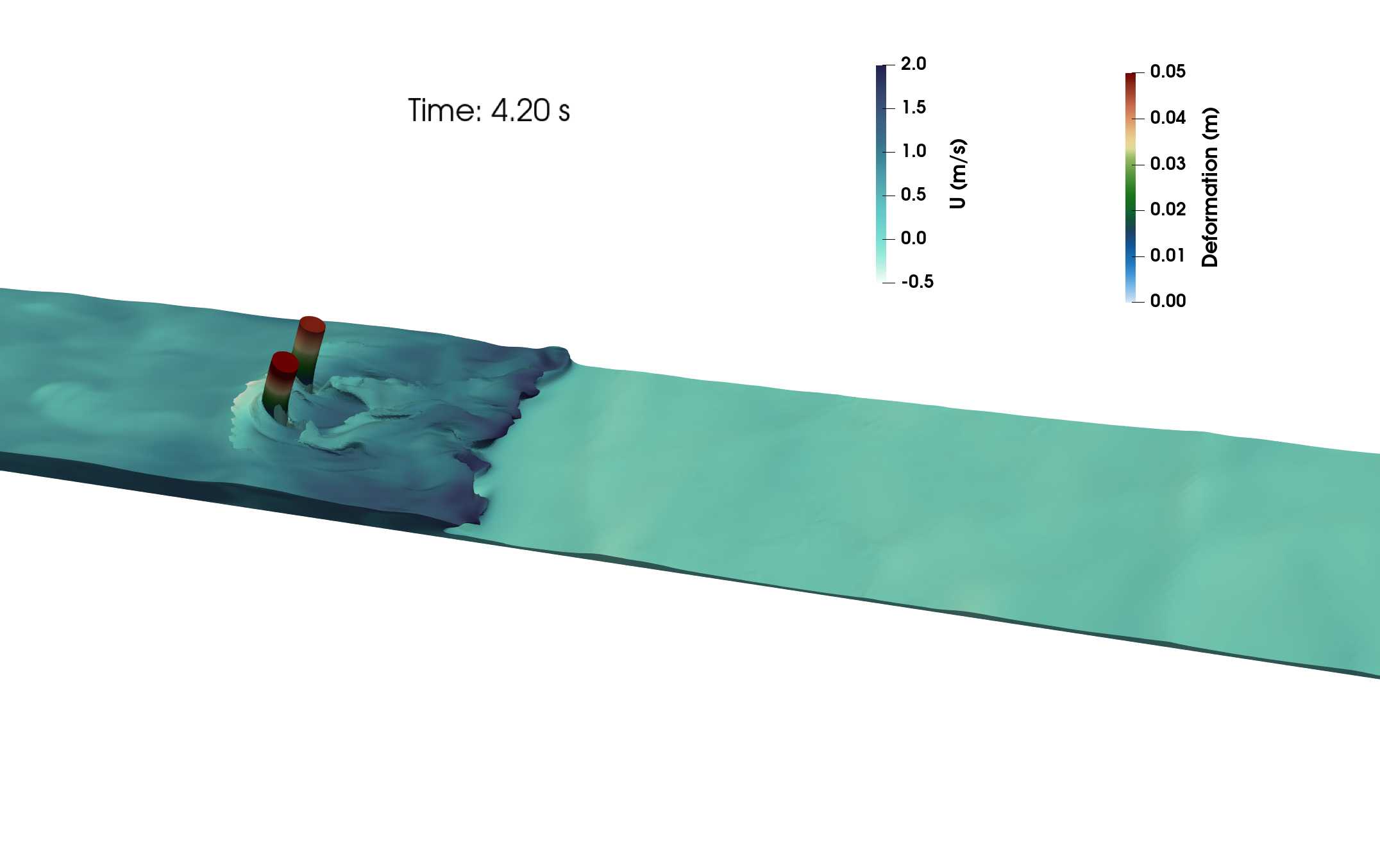}
    \caption{Two cylinders: $D=0.07\,{}\rm m$, $E=5 \times 10^{4}\,{\rm Pa}$}\label{fig:12a}		
	\end{subfigure}\\
	\begin{subfigure}[t]{\textwidth}
    \centering
    \includegraphics[trim={100 350 60 20},clip,scale=0.2]{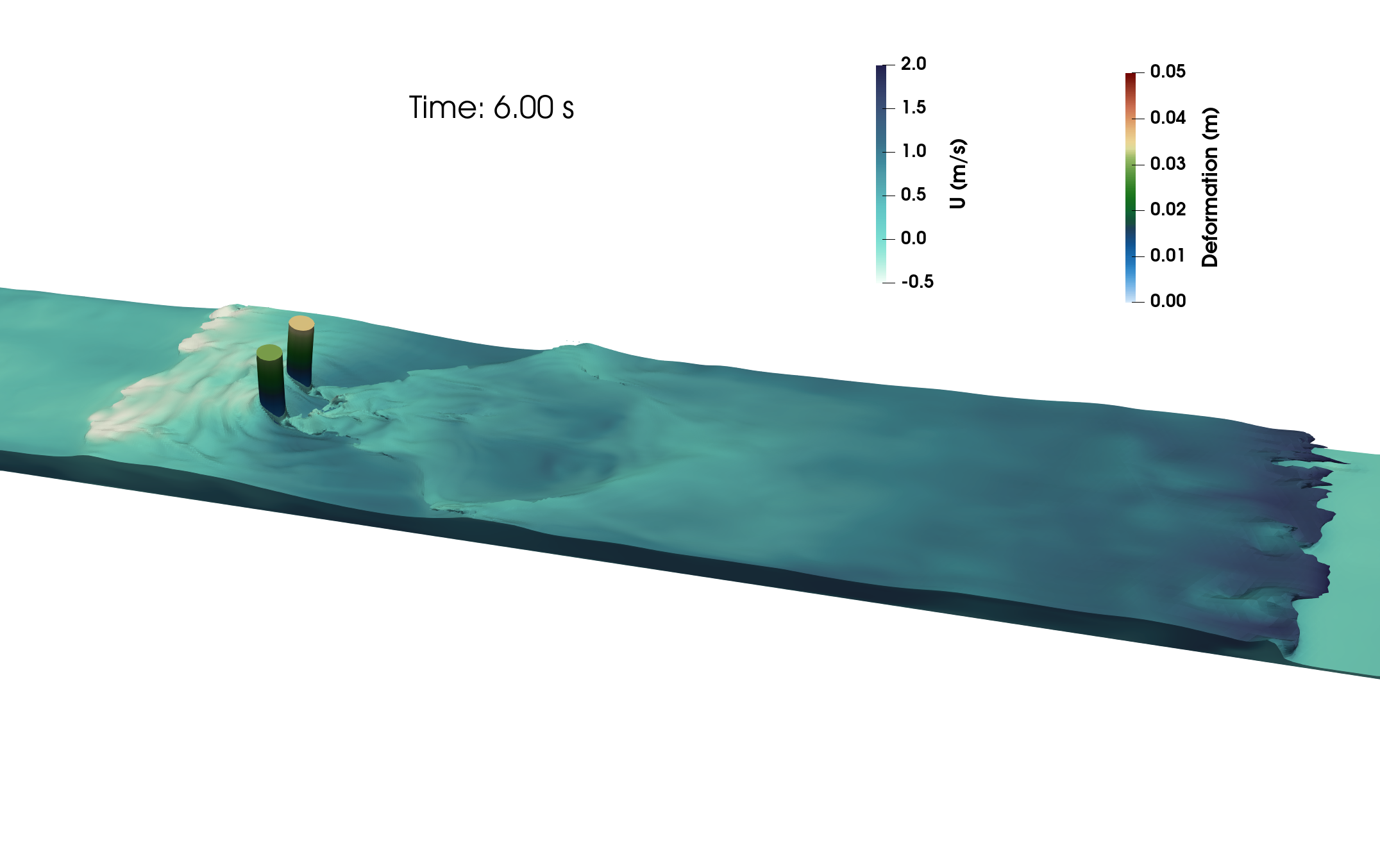}
    \caption{Two cylinders: $D=0.07\,{}\rm m$, $E=5 \times 10^{4}\,{\rm Pa}$, $SP/D=2$}\label{fig:12b}
	\end{subfigure}
\caption{Velocity profile and cylinder displacement of two-cylinder configuration at $SP/D=2$ at different flow times. Figure (a): The flow structures when the cylinders show the maximum deflection, figure (b): The velocity profile and cylinder displacement when the bore wave flows further downstream.}
\label{fig:3d-vel-profile-two-cyl}
\end{figure}

\begin{figure}[!t]
   \centering
   \begin{subfigure}[t]{\textwidth}
    \centering
    \includegraphics[trim={100 350 60 20},clip,scale=0.2]{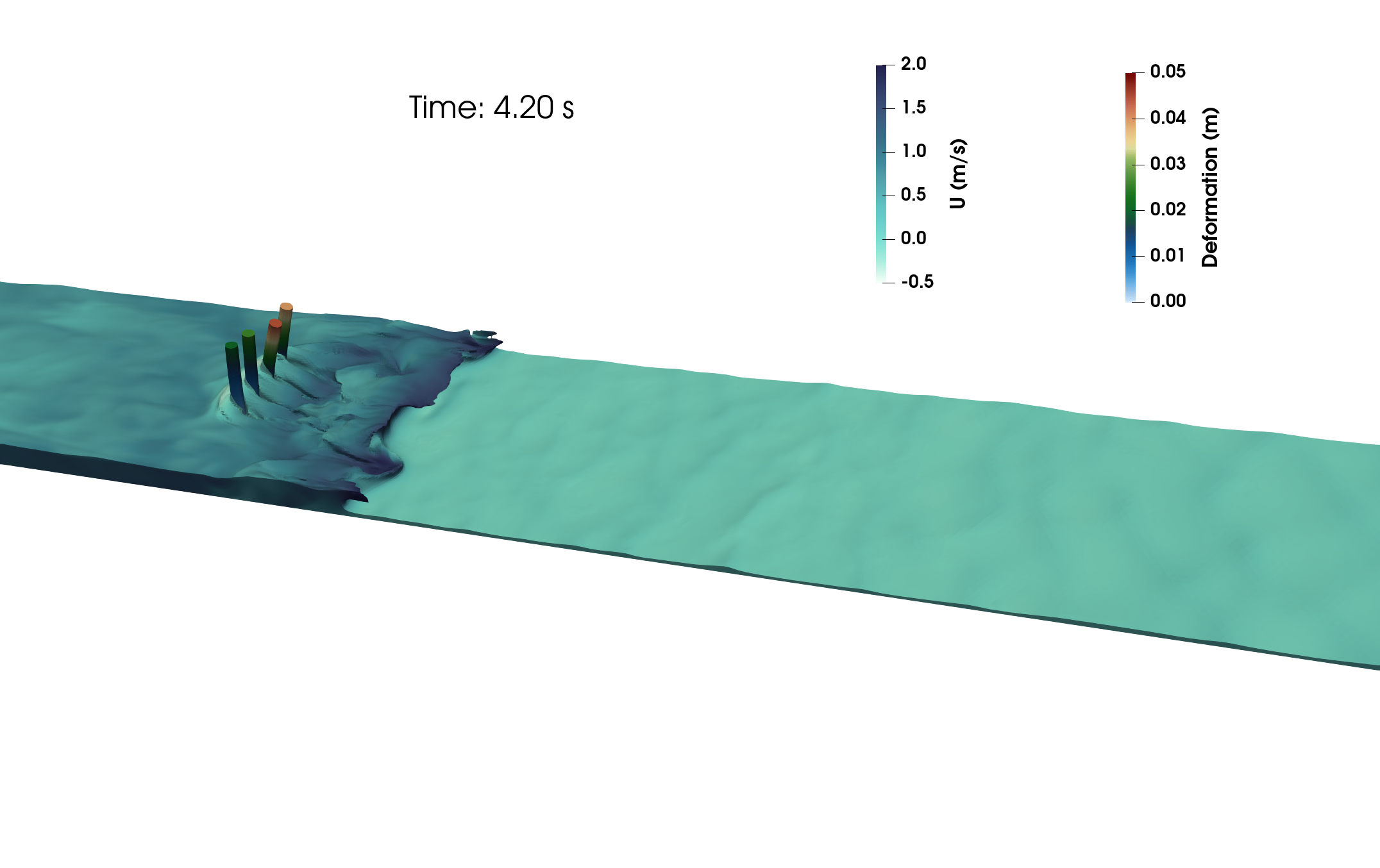}
    \caption{Four cylinders: $D=0.035\,{}\rm m$, $E=5 \times 10^{5}\,{\rm Pa}$}
    \label{fig:13a}		
	\end{subfigure}\\
	\begin{subfigure}[t]{\textwidth}
    \centering
    \includegraphics[trim={100 350 60 20},clip,scale=0.2]{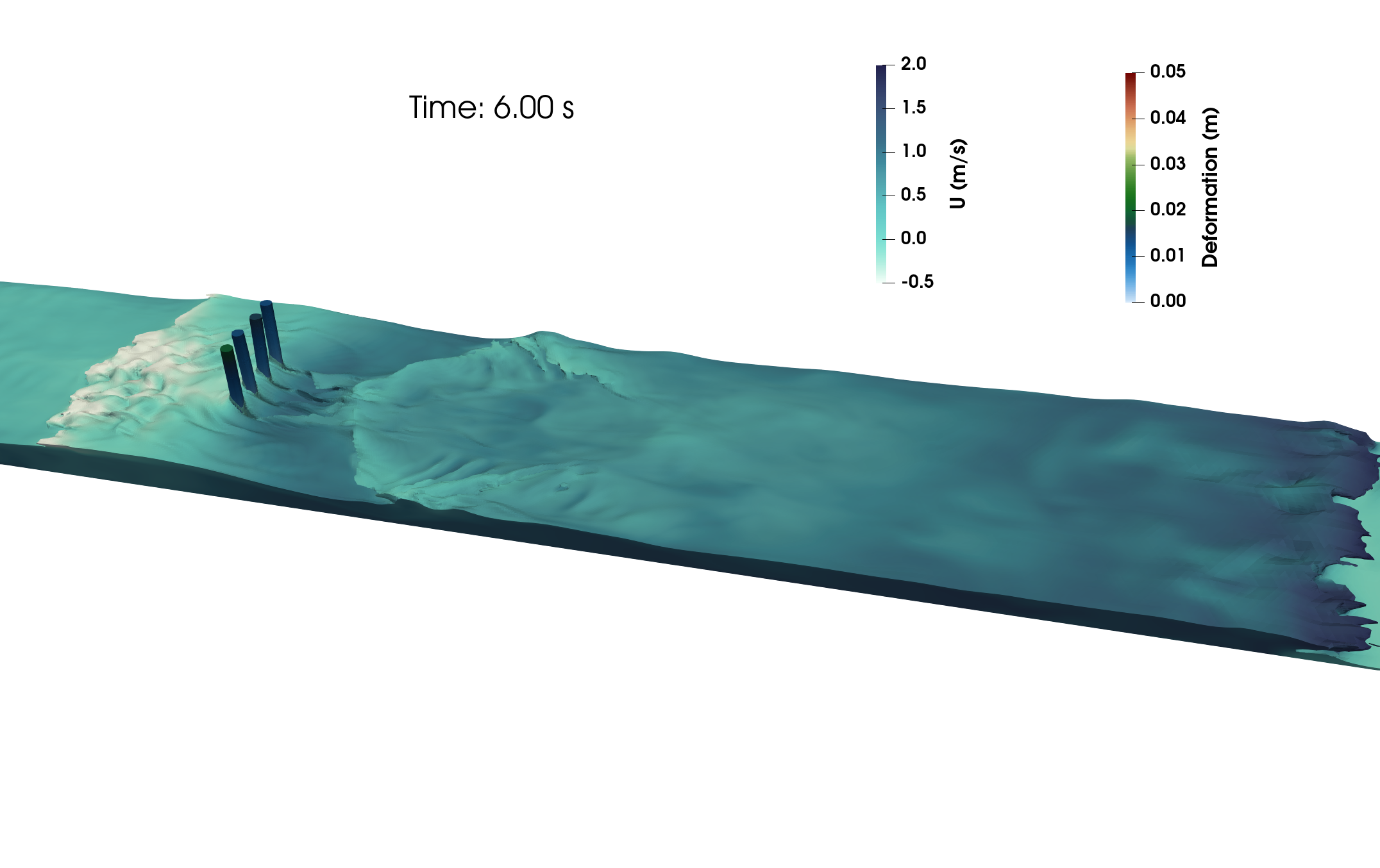}
    \caption{Four cylinders: $D=0.035\,{}\rm m$, $E=5 \times 10^{5}\,{\rm Pa}$}\label{fig:13b}
	\end{subfigure}
\caption{The velocity profile and cylinder displacement of four-cylinder at $SP/D=2$ configuration at different flow times. Figure (a): The flow structures when the cylinders show the maximum deflection, figure (b): The velocity profile and cylinder displacement when the bore wave flows further downstream.}
\label{fig:3d-vel-profile-four-cyl}
\end{figure}

\subsection{Reflective effect of rigidity is lost for multiple cylinders}\label{sec5.2}
Rigidity is not the only parameter affecting onshore energy flux; the number of cylinders and the cylinder spacing also alter the flow field sensitively and hence the energy flux. Interestingly, our simulations show that the dependency of reflected energy on the rigidity is most pronounced for the single cylinder shown in figure \ref{fig:energy-flux-ref-one}. The equivalent results for multiple cylinders are summarized in figures \ref{fig:11a} and \ref{fig:11b}. In table \ref{table_kf_multcyl}, we compare the percentage of maximum reflected kinetic energy flux during the time period $4$ - $8\,{\rm s}$ for one cylinder to the two- and four-cylinder configurations at two different spacings. The maximum difference in reflected kinetic energy flux for single cylinders with differing rigidity is around $4.6\%$. In a two-cylinder arrangement, this difference decreases to $2\%$. In a four-cylinder arrangement, this difference further decreases to $1\%$, suggesting that the effect of rigidity on the reflected kinetic energy flux becomes negligible as the number of cylinders increases. 
 
In a multiple-cylinder configuration, the blockage area along the span-wise direction is split into smaller areas. The flow can propagate through the gap between the cylinders, leading to a different flow field with multiple wakes. Figures \ref{fig:3d-vel-profile-two-cyl} and  \ref{fig:3d-vel-profile-four-cyl} show two snapshots of the flow around two cylinders and four cylinders at the time of maximum deflection and after passage of the bore front, respectively. At $t=4.2\,{\rm s}$ after bore impact, the increase in water depth at the cylinders' upstream faces is clearly visible for two cylinders. The interactions between the four cylinders are apparent, particularly in figure \ref{fig:3d-vel-profile-four-cyl}, where the split flow around each one of the cylinders joins into one wake that is reminiscent of the wake behind a single cylinder in figure \ref{fig:3d-vel-profile-one-cyl} (b).

For four cylinders, these flow-field interactions become even more pronounced, as shown in figure \ref{fig:3d-vel-profile-four-cyl}. In this configuration, the cylinders no longer deform in unison, because of the dynamic feedback between deflection and flow. In figure \ref{fig:13a} only the second cylinder from the upper side deflects primarily in the flow direction while the cylinder closest to the lower side of the domain bends partially outwards to accommodate flow in the gap between the cylinders. As a consequence. the reflected wave front in the four-cylinder configuration does not increase in depth, but instead flattens out along a spanwise direction. However, the total blockage area of two cylinders is twice of four cylinders, which explains why increasing the number of split regions between the cylinders reduces the energy reflection. 

The dynamic feedback between cylinder deflection and flow explains why the relationship between gap and energy flux is not a simple one. Our simulations demonstrate that the gap between the cylinders alters flow restriction and modifies the energy flux reflection, but the difference between different configurations is relatively small. The blockage area of cylinders are the same at $SP/D=1.5$ and $SP/D=2$; as the gaps are widened, the cylinders are shifting towards the sidewall, which explains why increasing the gap does not change the energy reflection much. As shown in table \ref{table_kf_multcyl}, the difference between $SP/D=1.5$ and $SP/D=2$ in energy flux reflection is $0.3\%$ for two-cylinder and $0.2\%$ for four-cylinder configurations and hence negligible for practical purposes. Since the blockage area of a two-cylinder configuration is two times of that in a four-cylinder configuration, the cylinders with a higher blockage area cause slightly higher reflection in closely spaced cylinders.

\begin{figure}[t]
        \centering
	\begin{subfigure}[b]{0.49\textwidth}
    \includegraphics[trim={20 10 10 15},clip,scale=0.27]{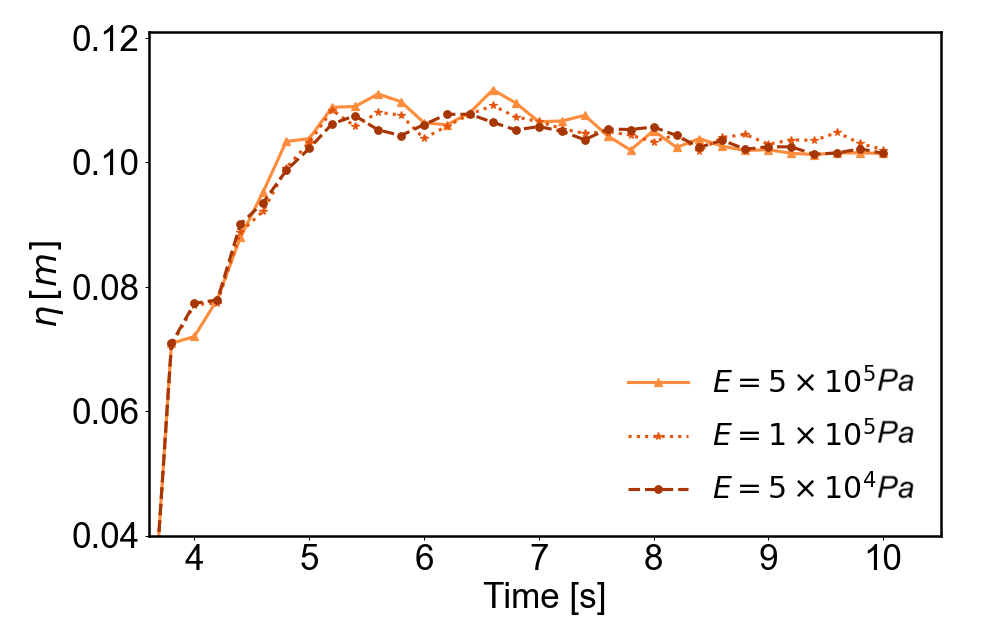}
    \caption{Two cylinders, $SP/D=1.5$}\label{fig:14a}
	\end{subfigure}
	\begin{subfigure}[b]{0.49\textwidth}
    \includegraphics[trim={20 10 10 15},clip,scale=0.27]{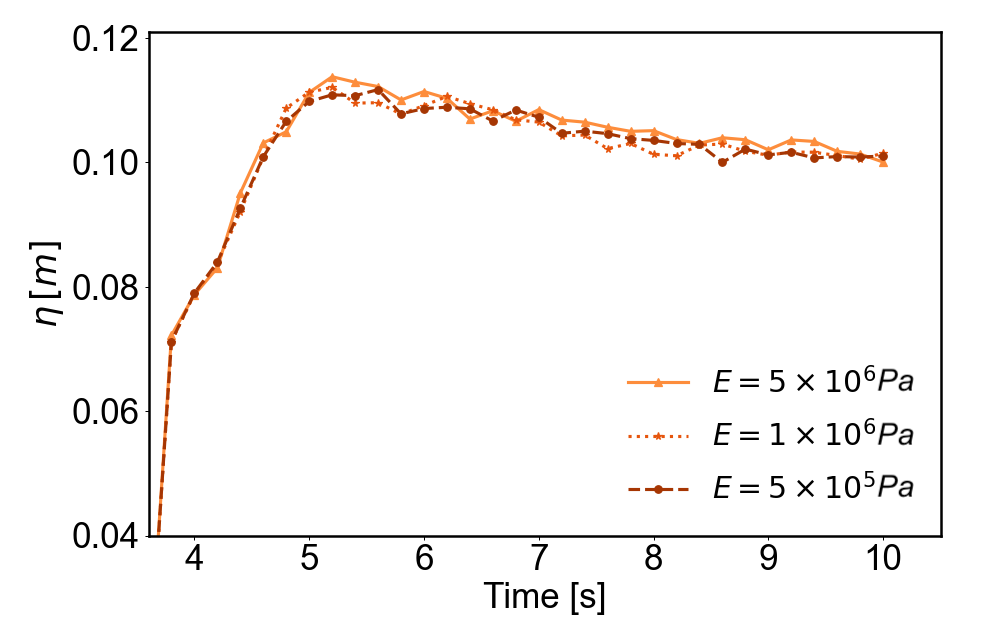}
    \caption{Four cylinders, $SP/D=1.5$}\label{fig:14b}
	\end{subfigure}
	\caption{
	 The time evolution of water surface at transect II ($x=11\,{\rm m}$) for different cylinder arrangements. Left: Comparison of water surface evolution for different elastic moduli in two-cylinder at $SP/D=1.5$. Right: Four-cylinder at $SP/D=1.5$}
\label{fig:eta-flex}
\end{figure}

The percentage of reflected kinetic energy flux lies between $25-30\%$ for all configurations, whereas the amount of maximum potential energy flux reflection is $4-5$ times smaller than the kinetic energy flux reflection. The reflected potential flux shows the same pattern as kinetic flux, where the importance of rigidity is lost when the number of cylinders increases. Since potential flux depends on the water depth at the cylinder upstream edge (transect II, see figure \ref{fig:dam-break-energy-flux}), the comparison of water depth evolution for different elastic moduli is shown for multi-cylinder 
arrangement in figure \ref{fig:eta-flex} to verify out findings. Since the difference in energy flux reflection at $SP/D=1.5$ and at $SP/D=2$ is minimal (see table \ref{table_kf_multcyl}), the water surface evolution for different flexibility is shown only at $SP/D=1.5$. Also, the difference in wave height for different rigidities at the cylinder upstream edge becomes minimal in a two and four-cylinder arrangement, indicating that the flexibility-induced wave reflection is negligible in multi-cylinder configurations. 

\begin{table}[t] \centering
\caption{Percentage of maximum reflected kinetic energy flux calculated for different cylinder configurations}
\begin{tabular}{cccccc}
\toprule \toprule
\addlinespace
 & \multicolumn{1}{c}{\makecell{one-cylinder \\($D=0.14\,{\rm m}$)}} &
 \multicolumn{2}{c}{\makecell{Two-cylinder configuration \\($D=0.07\,{\rm m}$)}} & \multicolumn{2}{c}{\makecell{Four-cylinder configuration \\($D=0.035\,{\rm m}$)}} \\
 \cmidrule(lr){2-2}
 \cmidrule(lr){3-4}
 \cmidrule(lr){5-6}
{$\phi^{k}_{R_{t\,max}} \%$}&{}&{$SP/D=1.5$}&{$SP/D=2$}&{$SP/D=1.5$}&{$SP/D=2$}
\tabularnewline
\cmidrule[\lightrulewidth](lr){1-6}\addlinespace[1ex]
E=$1\times10^{4}\,{\rm Pa}$&27.2\%&-&-&- \tabularnewline
E=$5\times10^{4}\,{\rm Pa}$&27.6\%&27.4\%&27\%&- \tabularnewline
E=$1\times 10^{5}\,{\rm Pa}$&30.8\%&28.31\%&28.3\%&-&- \tabularnewline
E=$5\times10^{5}\,{\rm Pa}$&-&29.1\%&28.8\%&28.9\%&27.5\% \tabularnewline
E=$1\times10^{6}\,{\rm Pa}$&-&-&-&28.95\%&28.12\% \tabularnewline
E=$5\times10^{6}\,{\rm Pa}$&-&-&-&29\%&28.85\% \tabularnewline
\addlinespace
\bottomrule
\end{tabular}
\label{table_kf_multcyl}
\end{table}

\subsection{Flow blockage area enhances the wave reflection}\label{sec5.3}

 \begin{figure}[t]
        \centering
    \includegraphics[trim={20 0 20 20},clip,scale=0.27]{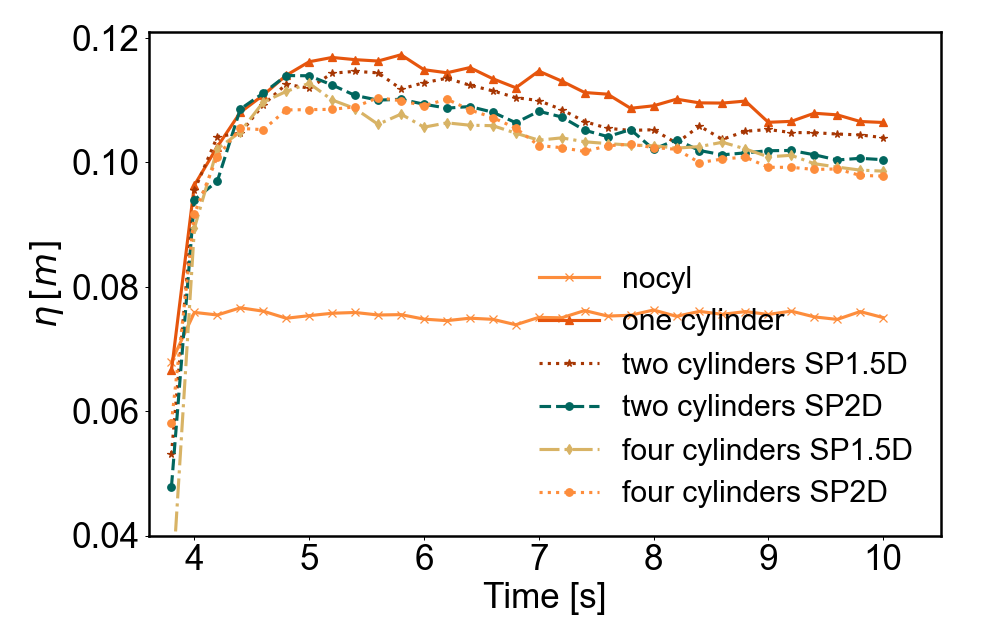}%
    \includegraphics[trim={20 0 20 20},clip,scale=0.27]{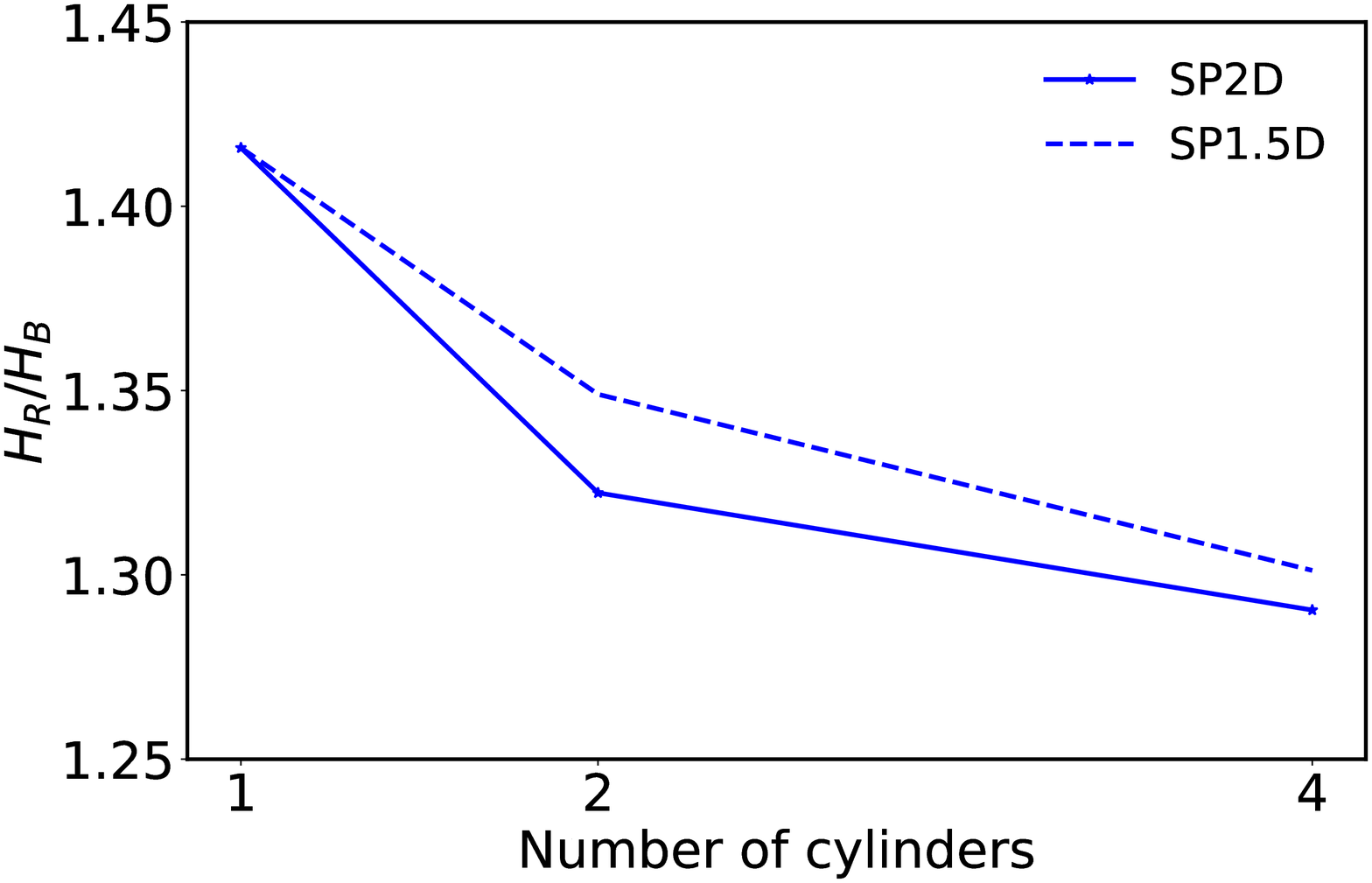}
\caption{
The evolution of water surface profile and reflection index for different cylinder arrangements at $x=11\,{\rm m}$. Left: comparison of water surface profile, Right: Reflection index for different cylinder parameters}
\label{fig:eta-rigid}
\end{figure}

In this study, we define three different cylinder configurations, each cylinder configuration has a different flow blockage area. The flow blockage area in a one-cylinder is two times higher than in a two-cylinder and four times higher than in a four-cylinder configuration. As shown in table \ref{table_kf_multcyl}, the one-cylinder with the highest flow blockage causes more flow restriction than multiple cylinders, resulting in the maximum amount of energy reflection. To compare the wave reflection for all cylinder arrangements, the free-surface evolution for different cylinder arrangements at transect T-II ($x=11\,{\rm m})$ is shown in figure \ref{fig:eta-rigid}. Without the cylinders, the bore maintains a constant depth over the flow period. The bore depth increases when cylinders are present, indicating the water level rise due to flow blockage, which causes the reflection of the wave. The reflected bore depth is highest in one cylinder with a higher flow blockage area and higher in closely spaced cylinders. To compare the wave reflection for different cylinder configurations, we measure the maximum reflected bore depth ($H_{R}$) by taking the maximum value of wave height over the flow period. We define the incident bore depth in the absence of cylinders as the maximum bore depth (see figure \ref{fig:eta-rigid}).

Finally, we obtain the reflection index by taking the ratio of $H_{R}$ and $H_{B}$ and comparing it for different cylinder configurations. A similar approach was implemented by \cite{oshnack2009effectiveness} to calculate the reflection index of tsunami bore to study the influence of different onshore seawalls on tsunami wave attenuation. From the reflection index shown in figure \ref{fig:eta-rigid}, we conclude that the reflection of the wave follows the same pattern as the energy flux reflection variation in different cylinder arrangements.
 As expected, the reflection index in one cylinder with a larger diameter is highest among the other configurations, whereas the four-cylinder with a larger gap ($SP/D=2$) has the minimum reflection index. The percentage of reflection for different cylinder configurations is also shown in table \ref{table_reflection}. The smaller gap between the cylinders is beneficial in terms of the reflection because cylinders with a smaller gap produce more flow blockage. Nevertheless, when the number of cylinders is increased to four, the effect of closely spaced cylinders in flow blockage becomes negligible. The percentage of wave reflection for the four-cylinder with a smaller gap ($SP/D=1.5$) is only $1\%$ higher than the four-cylinder at $SP/D=2$.

\begin{table}[t]
\centering
\caption{Percentage wave reflection for different cylinder configurations}
\begin{tabular}{cccccc}
\toprule
\multicolumn{1}{p{2.0cm}}{\centering Percentage of \\ reflection } & 
\multicolumn{1}{p{1.5cm}}{\centering One cylinder} & 
\multicolumn{1}{p{2cm}}{\centering Two-cylinder \\ $SP = 1.5D$} &
\multicolumn{1}{p{2cm}}{\centering Two-cylinder \\ $SP = 2D$} &
\multicolumn{1}{p{2.0cm}}{\centering Four-cylinder \\ $SP = 1.5D$} &  
\multicolumn{1}{p{2.0cm}}{\centering Four-cylinder \\ $SP = 2D$}\\
\midrule
$(\frac{H_{R}-H_{B}}{H_{B}})\%$  &  $41.5\%$   & $35.5\%$  & $32\%$ & $31\%$ & $30\%$\\
\bottomrule
\end{tabular}
\label{table_reflection}
\end{table}

\begin{figure}[t]
        \centering
	\begin{subfigure}[b]{0.49\textwidth}
    \includegraphics[trim={20 10 10 15},clip,scale=0.27]{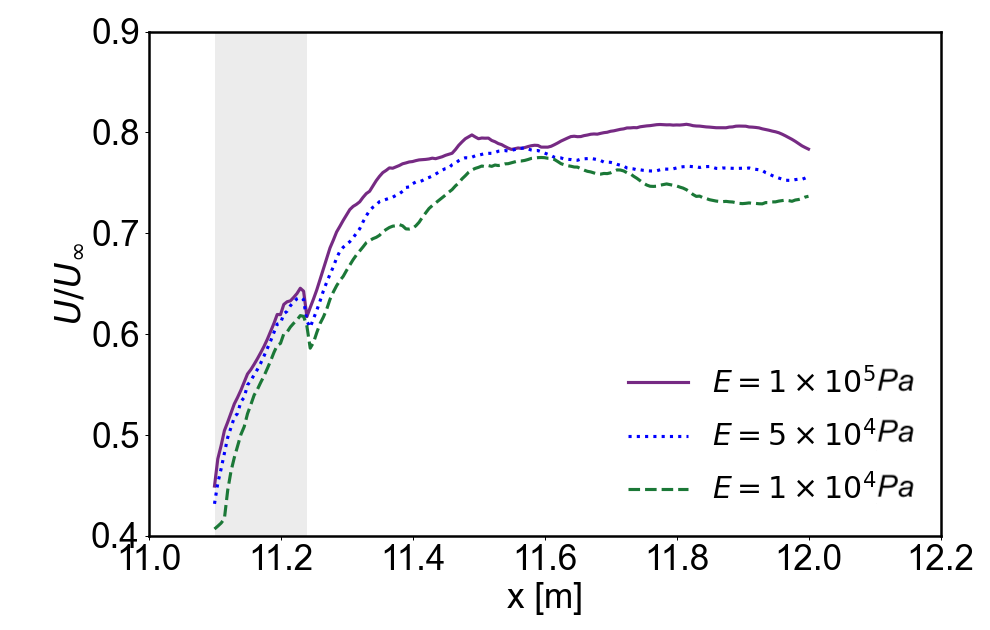}
    \caption{One cylinder}\label{fig:16a}
	\end{subfigure}%
	\begin{subfigure}[b]{0.49\textwidth}
    \includegraphics[trim={20 10 10 15},clip,scale=0.27]{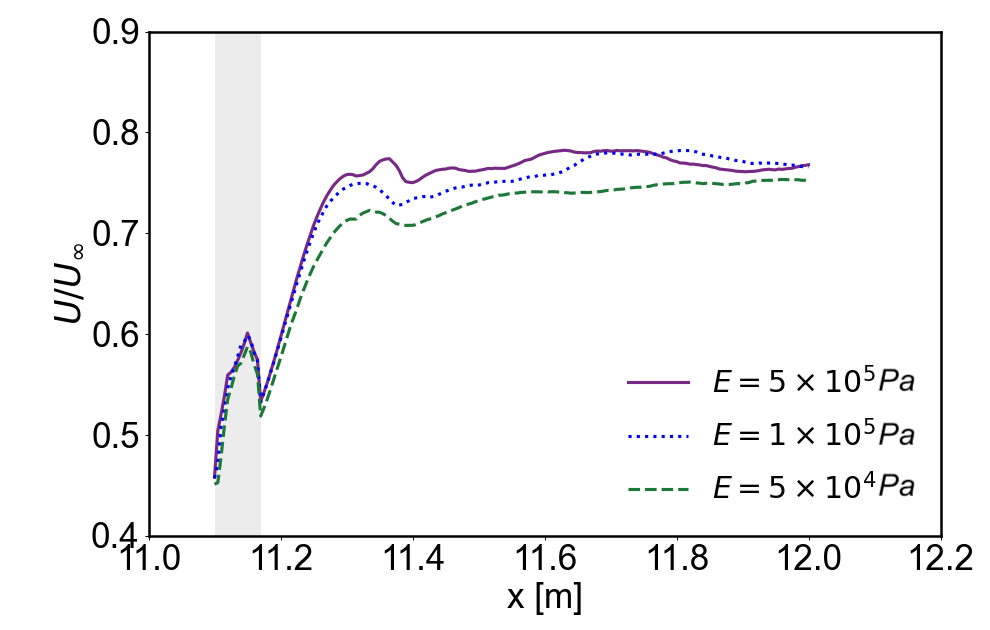}
    \caption{Two cylinders, $SP/D=2$}\label{fig:16b}
	\end{subfigure}
	\begin{subfigure}[b]{0.49\textwidth}
        \centering
    \includegraphics[trim={20 10 10 15},clip,scale=0.27]{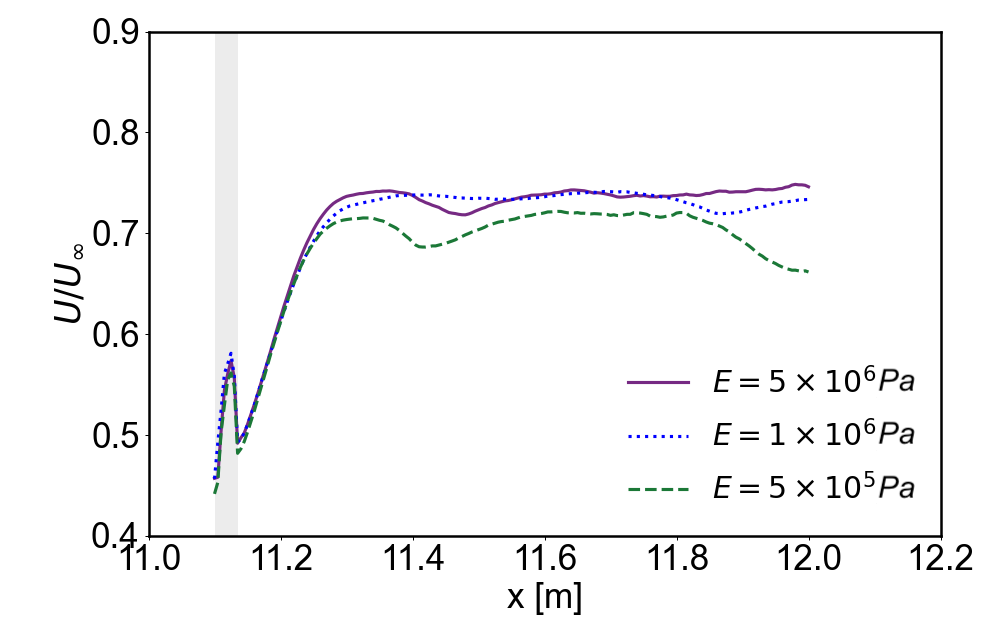}
    \caption{Four cylinders, $SP/D=2$}\label{fig:16c}		
	\end{subfigure}
	\caption{The longitudinal velocity profile along the streamwise direction for multi-cylinder configuration. Top row: Normalized longitudinal velocity profile for one-cylinder and two-cylinder configurations. Bottom Row: Normalized longitudinal velocity profile for four-cylinder configuration.}
\label{fig:longitudinal-vel-multi-cyl}
\end{figure} 

\subsection{Rigidity reduces flow damping and turbulent kinetic energy at downstream region}\label{sec5.5}
Understanding the flow characteristics in the downstream region is important because it gives us an idea of how the flow field changes in multi-cylinder configuration and at different rigidities. In the wake region, the bendability not only controls the motion of the cylinder body but also flow structures around the. Since not all cylinders move uniformly, the dynamic response between the flow and the cylinder bodies makes the flow structure more complex, which affects the flow velocity and hence can alter dissipation.
 To understand the effect of flexibility in the wake region, we show the longitudinal average velocity profile along the streamwise direction in figure \ref{fig:longitudinal-vel-multi-cyl}. The velocity is averaged along the depth and spanwise direction. The field averaged velocity is then further averaged over the flow period after the bore has reached the downstream region. Here, the flow period is $4\,{\rm s}$ which is the time interval between the bore arrival at the downstream region and the end wall (see figures \ref{fig:3d-vel-profile-one-cyl}, \ref{fig:3d-vel-profile-two-cyl} and \ref{fig:3d-vel-profile-four-cyl}). The velocity is normalized by the approach velocity ($U_{\infty}$), which is the velocity of the bore before hitting the cylinders to compare how the longitudinal velocity changes along the wake region with respect to the approach velocity. As shown in figure \ref{fig:longitudinal-vel-multi-cyl}, the averaged velocity is lower than the approach velocity at the upstream edge of the cylinders. Once the bore flows past the cylinders, the flow decelerates, and the flow energy is reduced due to reflection. At the flow separation region, the normalized velocity steadily increases from $0.42$ and maintains a constant value at a further downstream region. As expected, flexibility induces more velocity reduction in the wake region. The normalized velocity is reduced by $15\%$ at the streamwise location $x= 11.4-12\,{\rm m}$. At the farthest downstream region ($x= 11.8-12\,{\rm m}$), the normalized velocity is reduced by around $20\%$ at modulus of elasticity $E= 5 \times 10^{5}\,{\rm Pa}$ in four-cylinder arrangement indicating that flexibility alters the flow structure and the change of the flow structure at the downstream region increases as the number of cylinder increases. Figure \ref{fig:longitudinal-vel-multi-cyl} shows the longitudinal velocity profiles at  $SP/D=2$ only because we have not observed any significant changes in normalized velocity as the gap between the cylinder changes.
 
 \begin{figure}[t]
        \centering
        \begin{subfigure}[b]{0.49\textwidth}
    \includegraphics[trim={20 10 10 15},clip,scale=0.27]{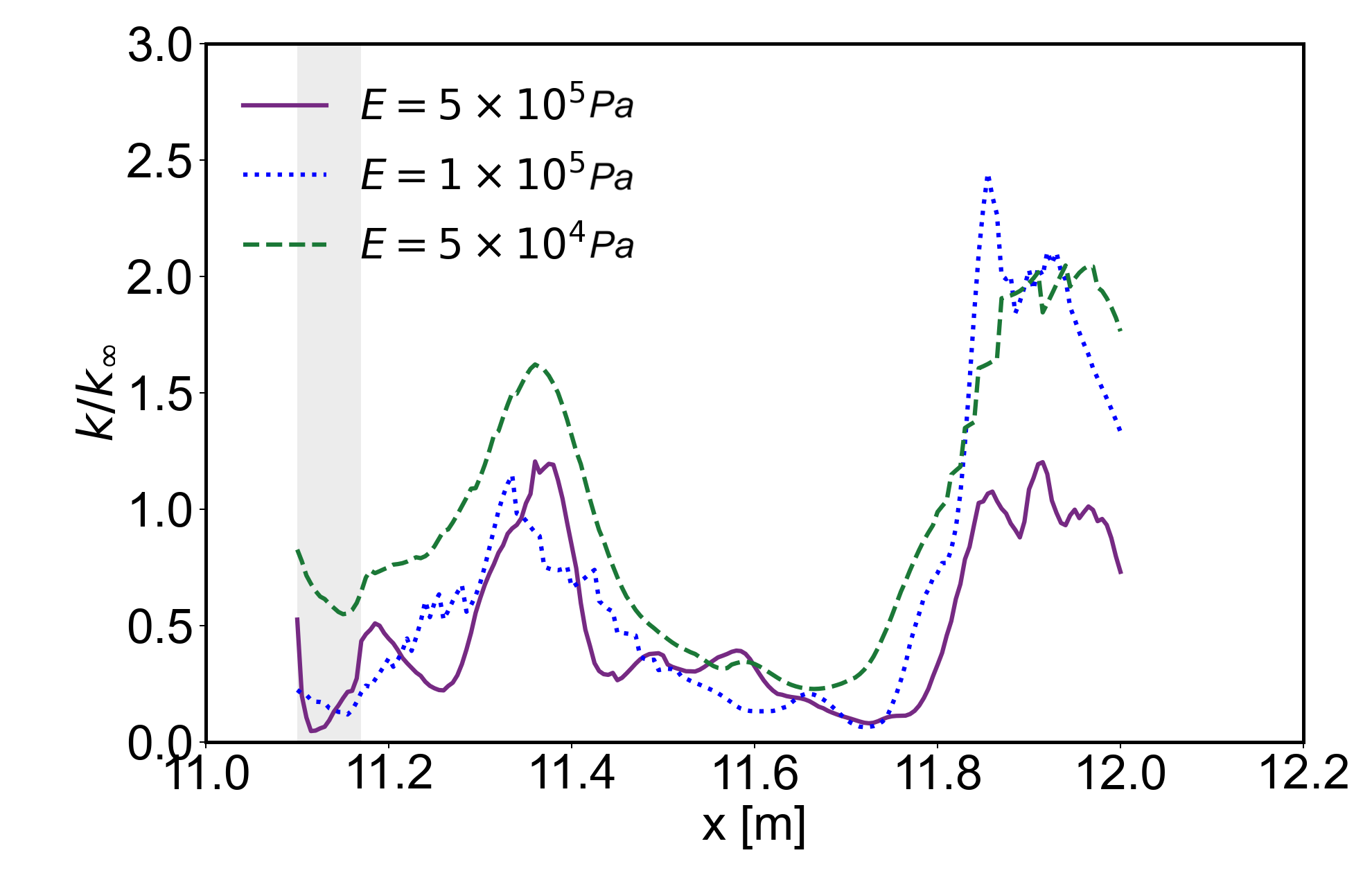}
    \caption{Two cylinders, $SP/D=1.5$}\label{fig:17a}		
	\end{subfigure}%
	\begin{subfigure}[b]{0.49\textwidth}
        \centering
    \includegraphics[trim={20 10 10 15},clip,scale=0.27]{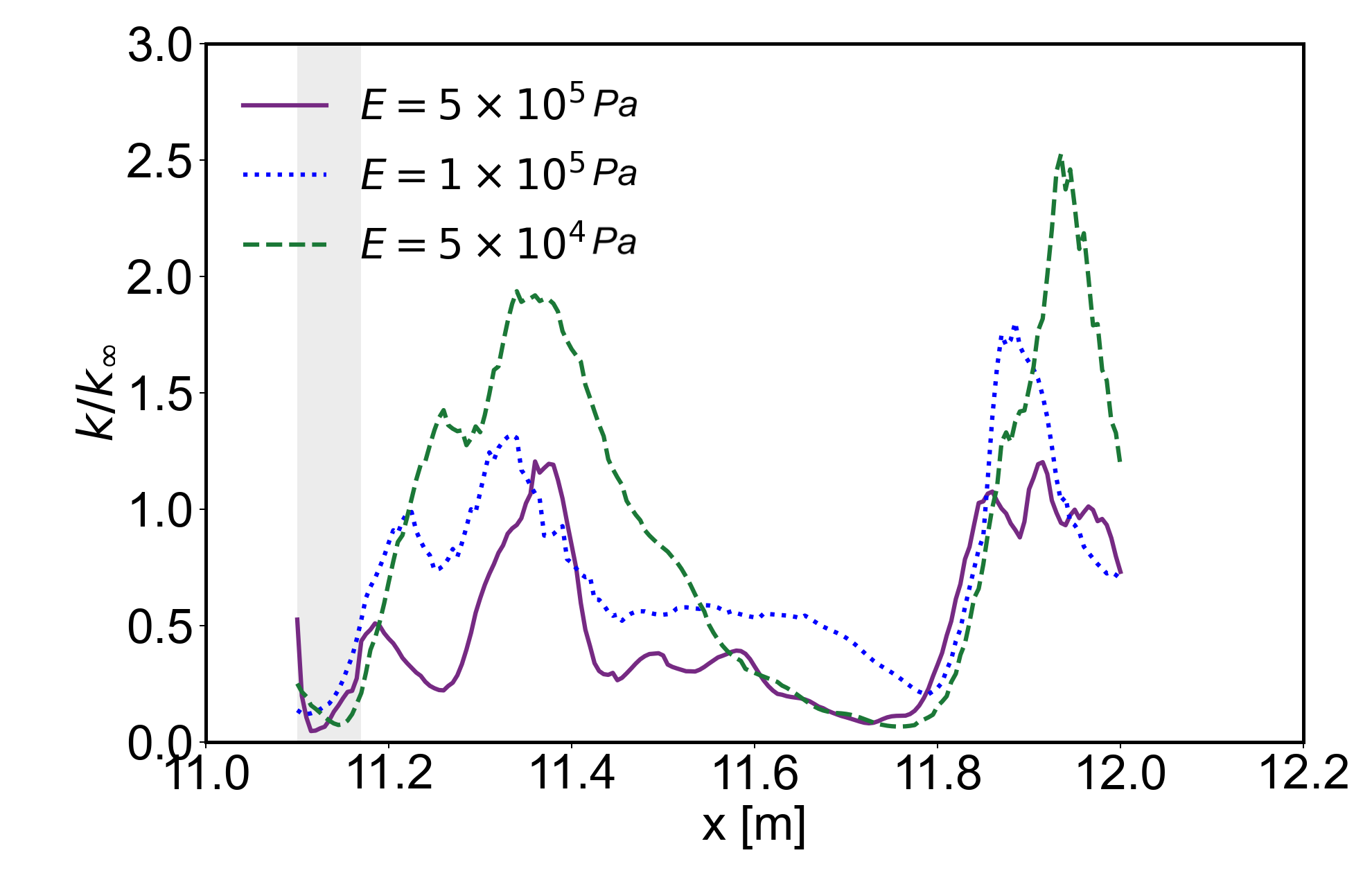}
    \caption{Two cylinders, $SP/D=2$}\label{fig:17b}		
	\end{subfigure}
	\caption{The instantaneous longitudinal distributions of normalized turbulent kinetic energy along the streamwise direction for two-cylinder configuration at $t=6\,{\rm s}$. The turbulent kinetic energy has two peaks, one at the behind of cylinders due to flow separation and at the further downstream region where split flow joins and hydraulic jumps occur.}
\label{fig:longitudinal-tke-double-cyl}
\end{figure}

\begin{figure}[t]
        \centering
	\begin{subfigure}[b]{0.49\textwidth}
    \includegraphics[trim={20 10 10 15},clip,scale=0.27]{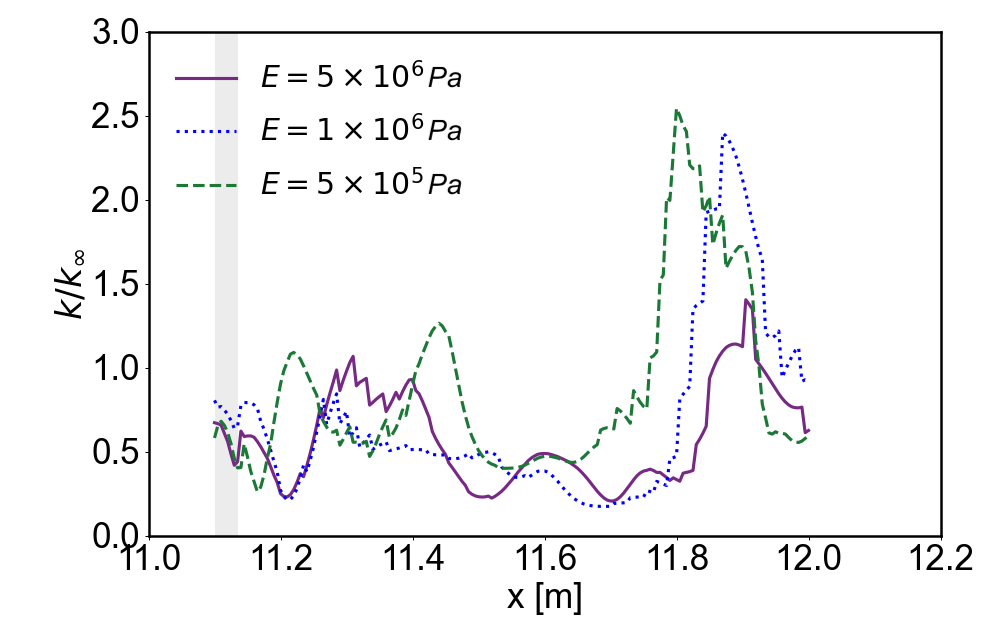}
    \caption{Four cylinders, $SP/D=1.5$}\label{fig:18a}
	\end{subfigure}%
	\begin{subfigure}[b]{0.49\textwidth}
    \includegraphics[trim={20 10 10 15},clip,scale=0.27]{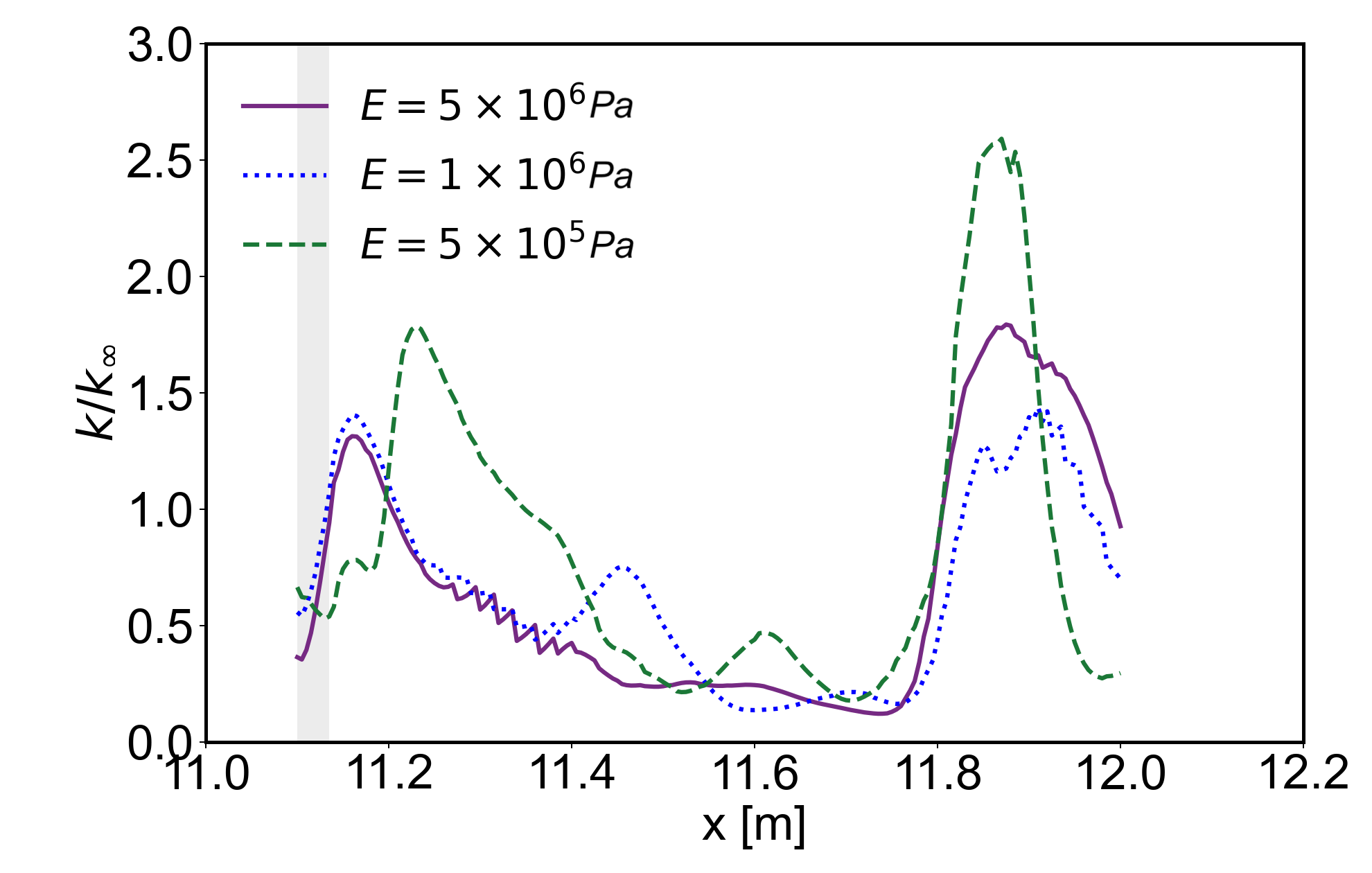}
    \caption{Four cylinders, $SP/D=2$}\label{fig:18b}
	\end{subfigure}
	\caption{The longitudinal distributions of turbulent kinetic energy along the streamwise direction for four-cylinder configuration at $t=6\,{\rm s}$}
\label{fig:longitudinal-tke-four-cyl}
\end{figure}

The longitudinal profiles of instantaneous turbulent kinetic energy (TKE) in figures \ref{fig:longitudinal-tke-double-cyl} and \ref{fig:longitudinal-tke-four-cyl} show the effect of flexibility and gap between cylinders on the reconfiguration of turbulent flow structures in the wake region of cylinders. 
The turbulent kinetic energy per unit mass is calculated by taking a spanwise and depth average over the velocity fluctuation fields as:
\begin{equation}
    k = \bigg\langle\frac{1}{2}(\langle u^{\prime^2} \rangle + \langle v^{\prime^2} \rangle + \langle w^{\prime^2} \rangle)\bigg\rangle
    \label{eq:k}
\end{equation}
where $u^{\prime}= u - \langle u \rangle$ , $v^{\prime}=v-\langle v \rangle$ and $w^{\prime}=w-\langle w \rangle$ are the velocity fluctuations 
and $\langle \rangle$ represents the spanwise average operator. 
Finally, the turbulent kinetic energy is normalized by $k_{\infty}$ measured at $x=11.1\,{\rm m}$ which is the location of the upstream cylinder edge. For all the investigated flows, the normalized instantaneous field-averaged turbulent kinetic energy profile ($k/k_{\infty}=1$) is extracted at $t=6\,{\rm s}$, representing the scenario when the dam front wave has reached the downstream region (see figures \ref{fig:3d-vel-profile-four-cyl} and \ref{fig:3d-vel-profile-two-cyl}).

Figures \ref{fig:longitudinal-tke-double-cyl} and \ref{fig:longitudinal-tke-four-cyl} show two peaks of turbulent kinetic energy: the first peak is located behind the cylinders and is a consequence of flow separation. The second peak occurs further downstream, where the split flow joins, and a hydraulic jump occurs. For both the two-cylinder and the four-cylinder arrangements, the first peak of normalized TKE becomes higher as flexibility increases at the trailing edge of cylinders. We detect the highest first peak at $SP/D=2$, where the normalized TKE increases by $30\%$ at the lowest modulus of elasticity. After the first peak, a region of reduced TKE extends up to $x=11.8\,{\rm m}$ for all scenarios. Between $x=11.8\,{\rm m}$ and $x=12\,{\rm m}$, a second peak forms, initiating a von Ka\'rma\'n vortex street. The highest second peak arises at $SP/D=2$ for both two and four-cylinder arrangements at the lowest modulus of elasticity. For four cylinders, the difference between TKE at $E= 5 \times 10^{5}\,{\rm Pa}$ and $E= 1 \times 10^{6}\,{\rm Pa}$ is higher at $SP/D=2$ than at $SP/D=1.5$. A similar pattern emerges in the two-cylinder arrangement. The reconfiguration of the cylinders alters the flow field and enhances the turbulence generation in the downstream region. Also, the flexibility alters the turbulent structures in the wake of cylinders, particularly at $SP/D=2$. 

\begin{figure}[t]
        \centering
    \includegraphics[trim={20 10 10 15},clip,scale=0.32]{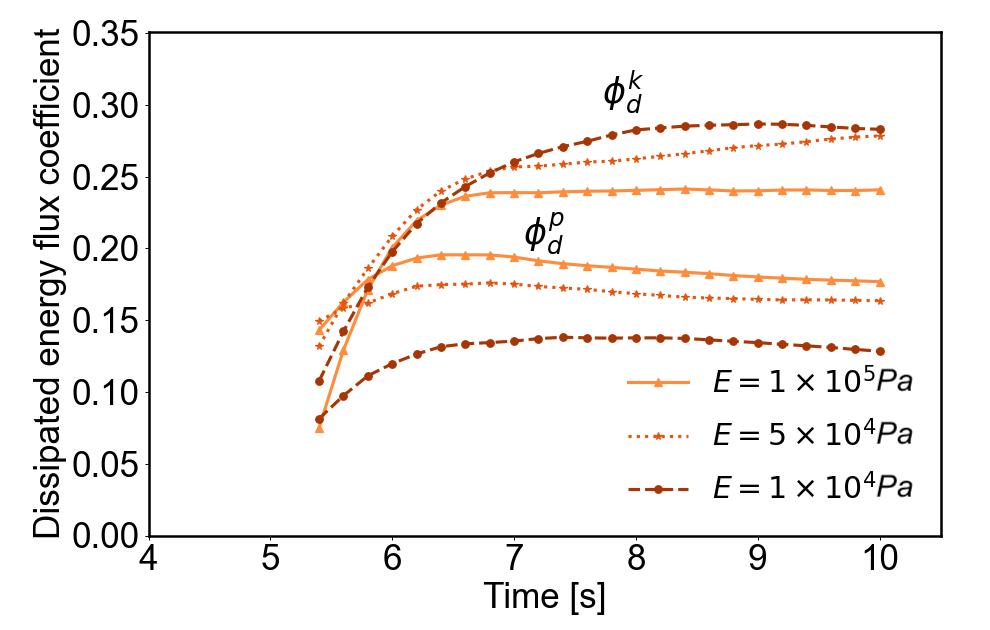}
\caption{Energy dissipation flux coefficient as a function of time of one cylinder for different elastic moduli. $\phi^{k}_{d}$ dissipated kinetic energy flux coefficient. $\phi^{p}_{d}$: dissipated potential energy flux coefficient}
\label{fig:energy-flux-dis-one}
\end{figure}

\subsection{Rigidity reduces kinetic energy dissipation for all configurations}\label{sec5.4}
Energy reflection is not the only physical mechanism that contributes to reducing the energy carried onshore by a tsunami. Energy dissipation can help reduce the energy flux created by the tsunami and is hence an important parameter to assess the protective benefit afforded by different vegetation configurations in tsunami mitigation. Figure \ref{fig:energy-flux-dis-one} is the equivalent of figure \ref{fig:energy-flux-ref-one}, showing the dissipated rather than the reflected energy flux coefficient for a single cylinder. It demonstrates that increasing rigidity has opposing effects on the dissipation of kinetic and potential energy fluxes: More rigidity dissipates less kinetic energy but more potential energy. 
Once the bore hits the cylinders, the flexible cylinder deforms more than the rigid cylinder, dissipating more kinetic energy. After the impact of the bore front, the energy of the incoming flow is not sufficient to cause further deformation of the cylinder. Therefore, the dissipation rate of kinetic energy becomes almost constant at the later phase of the flow period ($t=8-10\,{\rm s}$) in figure \ref{fig:energy-flux-dis-one}. The dissipated potential energy flux is around $80\%$ of the dissipated kinetic energy flux coefficient in the one-cylinder configuration. The effect of rigidity in the potential flux dissipation is exactly the opposite of what we observe in kinetic energy dissipation. The highest potential flux is observed in a less deforming cylinder. The more pronounced deformation of the flexible as compared to the rigid cylinder also reduces build-up of the free surface around the cylinder and hence dissipates less potential energy flux. As a consequence, the effect of rigidity on the total dissipation, including both the kinetic and the potential component, is small.

\begin{figure}[t]
        \centering
        \begin{subfigure}[b]{0.49\textwidth}
    \includegraphics[trim={20 10 10 15},clip,scale=0.142]{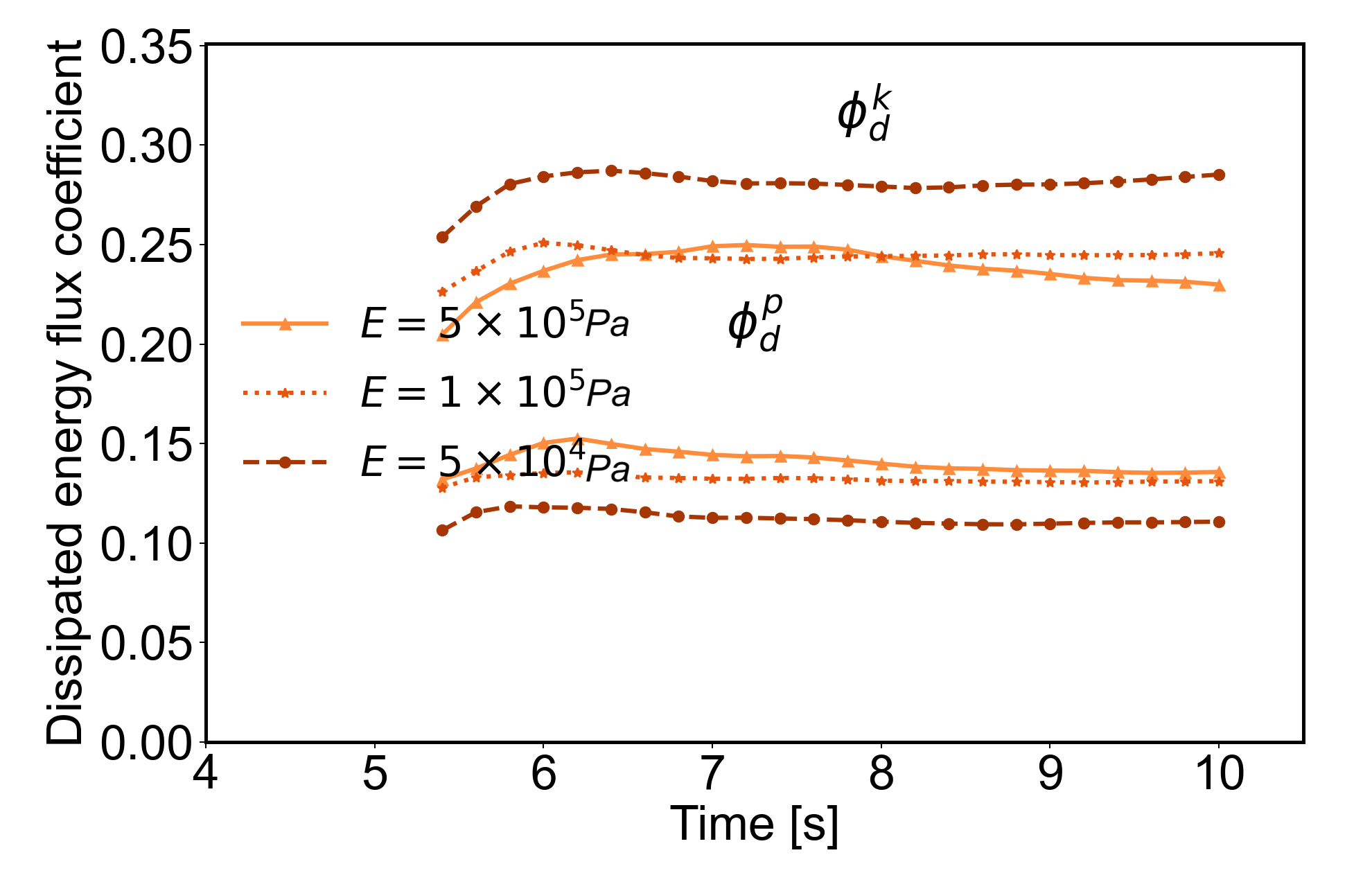}
    \caption{Two cylinders, $SP/D=1.5$}\label{fig:20a}		
	\end{subfigure}%
	\begin{subfigure}[b]{0.49\textwidth}
    \includegraphics[trim={20 10 10 15},clip,scale=0.142]{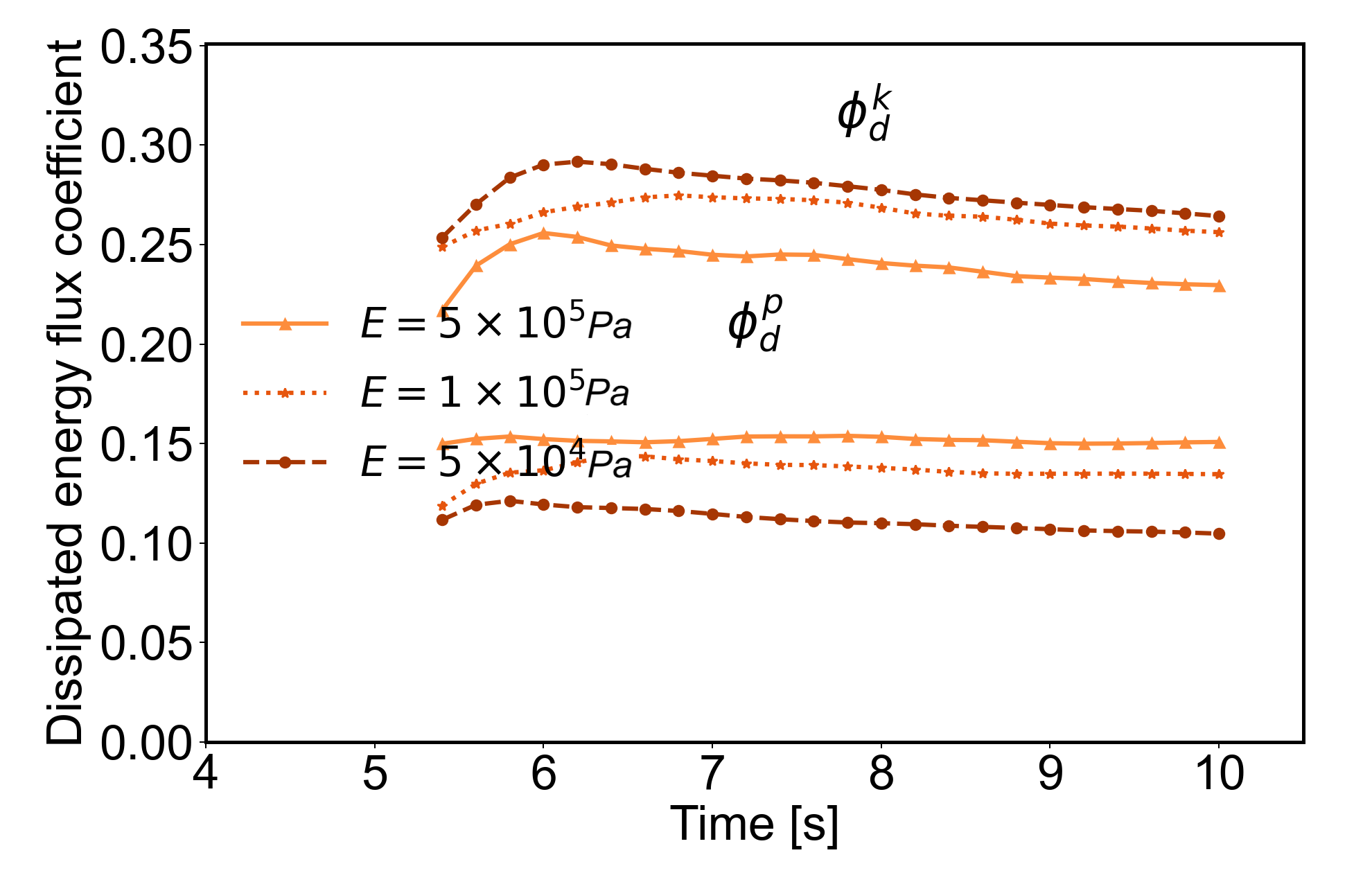}
    \caption{Two cylinders, $SP/D=2$}\label{fig:20b}
	\end{subfigure}
	\begin{subfigure}[b]{0.49\textwidth}
        \centering
    \includegraphics[trim={20 10 10 15},clip,scale=0.27]{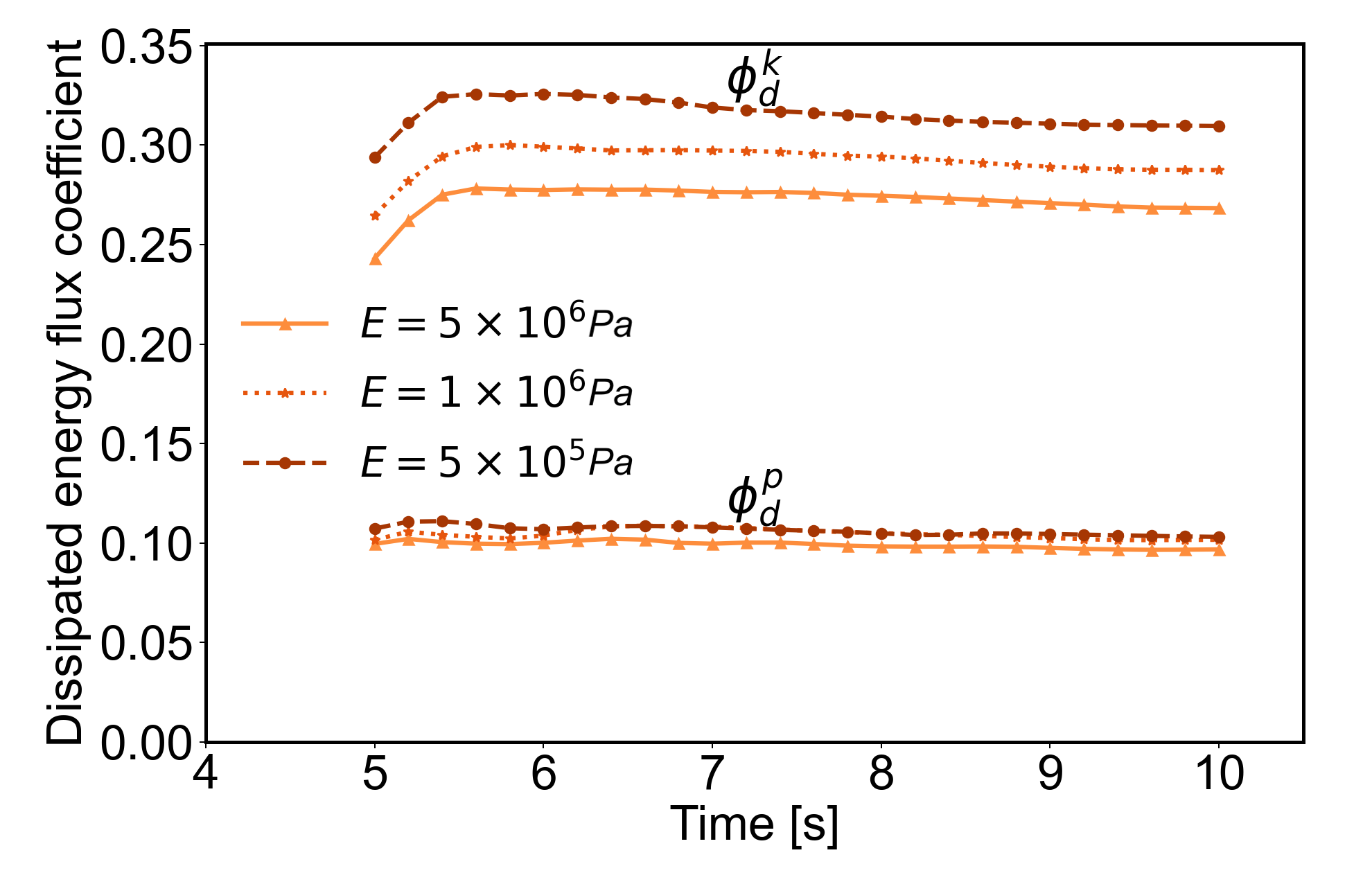}
    \caption{Four cylinders, $SP/D=1.5$}\label{fig:20c}		
	\end{subfigure}%
	\begin{subfigure}[b]{0.49\textwidth}
    \includegraphics[trim={20 10 10 15},clip,scale=0.27]{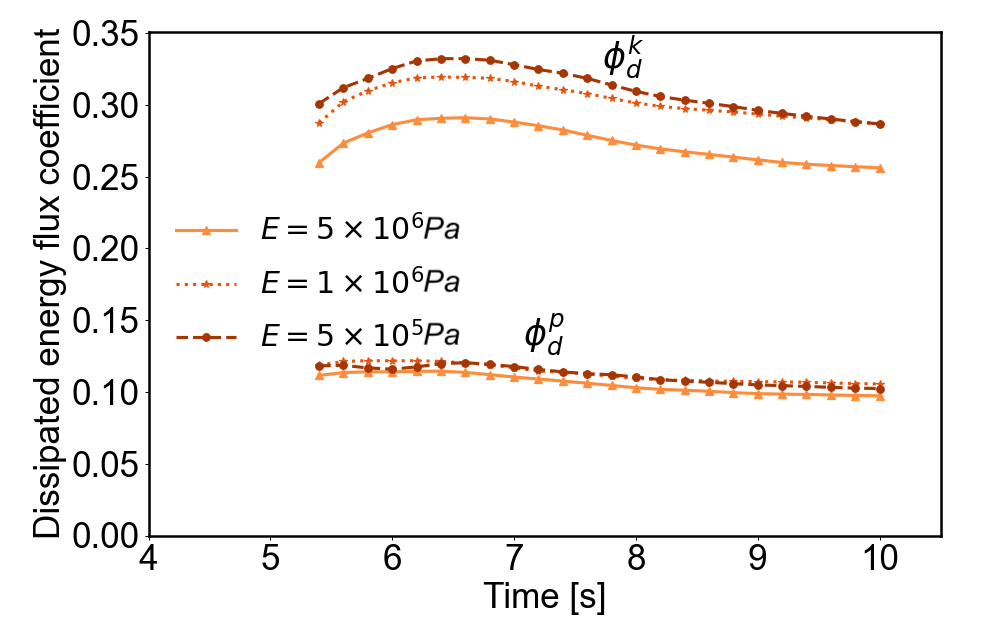}
    \caption{Four cylinders, $SP/D=2$}\label{fig:20d}
	\end{subfigure}

\caption{The energy dissipation flux coefficient as a function of time of two-cylinder and four-cylinder arrangement for different elastic moduli and cylinder gaps. Top Row: dissipated kinetic and potential energy flux coefficients for two-cylinder configuration. Bottom Row: dissipated kinetic and potential energy flux coefficients for four-cylinder arrangement}
\label{fig:energy-flux-dis-multi}
\end{figure}

We observe the same pattern of energy dissipation in multi-cylinder configurations. Figure \ref{fig:energy-flux-dis-multi} shows that rigidity continues to decrease the dissipation of kinetic energy but increases the potential flux dissipation even in the presence of multiple cylinders. The tendency for rigidity to increase the dissipation of potential energy flux remains notable for two-cylinders in figures \ref{fig:20a} and \ref{fig:20b}, but disappears in the four-cylinder configuration (\ref{fig:20c}, \ref{fig:20d}). Since we do not observe any noticeable build-up of the wave surface around the cylinder, the variability of wave depth along the downstream region is small in multi-cylinder as compared to one-cylinder. The dissipated potential energy flux becomes $50\%$ of dissipated kinetic flux in the two-cylinder configuration. In the four-cylinder configuration, the dissipated potential flux is further reduced to $30\%$ of kinetic flux. Also, with increasing rigidity, the dissipated potential flux increases. Since flexibility alters the wave reflection, the difference in wave height between the upstream and downstream regions becomes smaller. However, as the number of cylinders increases, flexibility has less impact on dissipated potential energy flux. In the two-cylinder configuration, the dissipated potential flux decreases by $5\%$ as flexibility increases. In a four-cylinder configuration, the changes of potential flux are less than $1\%$ since the changes in the depth of the transmitted wave are negligible as flexibility changes.  
\begin{table}[t] \centering
\caption{Percentage of maximum dissipated kinetic energy flux in different cylinder configurations}
\resizebox{\textwidth}{!}{\begin{tabular}{cccccc}
\toprule \toprule
\addlinespace
 & \multicolumn{1}{c}{\makecell{one-cylinder \\($D=0.14\,{\rm m}$)}} &
 \multicolumn{2}{c}{\makecell{Two-cylinder configuration \\($D=0.07\,{\rm m}$)}} & \multicolumn{2}{c}{\makecell{Four-cylinder configuration \\($D=0.035\,{\rm m}$)}} \\
 \cmidrule(lr){2-2}
 \cmidrule(lr){3-4}
 \cmidrule(lr){5-6}
{$\phi^{k}_{D_{t\,max}} \%$}&{}&{$SP/D=1.5$}&{$SP/D=2$}&{$SP/D=1.5$}&{$SP/D=2$}
\tabularnewline
\cmidrule[\lightrulewidth](lr){1-6}\addlinespace[1ex]
E=$1\times10^{4}\,{\rm Pa}$&28.7\%&-&-&- \tabularnewline
E=$5\times10^{4}\,{\rm Pa}$&27.9\%&28.7\%&29.2\%&- \tabularnewline
E=$1\times 10^{5}\,{\rm Pa}$&24.1\%&25.1\%&27.5\%&-&- \tabularnewline
E=$5\times10^{5}\,{\rm Pa}$&-&25\%&25.6\%&31\%&33.2\% \tabularnewline
E=$1\times10^{6}\,{\rm Pa}$&-&-&-&28.4\%&32\% \tabularnewline
E=$5\times10^{6}\,{\rm Pa}$&-&-&-&26.2\%&29\% \tabularnewline
\addlinespace
\bottomrule
\end{tabular}}
\label{table_dis_kf_multcyl}
\end{table}

We report the percentage of maximum dissipated kinetic energy flux for different cylinder parameters in table \ref{table_dis_kf_multcyl}. For one cylinder, the percentage of the maximum dissipated kinetic flux coefficient is around $24\%$ at an elastic modulus of $1\times 10^{5}\,{\rm Pa}$, and the dissipated kinetic energy increases by $4\%$ as the structure becomes more flexible. In the two-cylinder and four-cylinder configurations, the percentage of dissipated kinetic energy flux coefficient for the lowest modulus of elasticity steadily goes up to $33\%$. For both scenarios, higher flexibility induces more energy dissipation. Moreover, the difference in kinetic energy flux dissipation increases as the number of cylinders increases. In a two-cylinder configuration, the difference is $4\%$, whereas, in a four-cylinder, the difference becomes $5\%$. Our finding is also supported by the turbulent kinetic energy shown in figures \ref{fig:longitudinal-tke-four-cyl} and \ref{fig:longitudinal-tke-four-cyl}, which prove that flexibility-induced dissipation enhances the turbulent kinetic energy due to dissipation. The comparison shows that the energy dissipation increases with an increasing number of cylinders, suggesting that the structure with a larger diameter contributes to energy reduction by reflection primarily. 
The contribution of potential flux dissipation is smaller in multi-cylinder configuration than in one-cylinder. The total energy flux dissipation is maximum in the four-cylinder configuration. However, the difference in total dissipated energy flux is not significant as the number of cylinders increases.
Apart from flexibility, the gap between the cylinders also modifies the energy dissipation. However,  the gap between the cylinders has a minimal effect on energy dissipation, the same phenomenon we notice in energy reflection. The difference in kinetic energy dissipation is $2\%-3\%$for all configurations. At the highest modulus of elasticity, where cylinders are almost rigid, the maximum energy dissipation obtained is approximately $2.2\%$ higher in the four-cylinder at $SP/D=2$ than at $SP/D=1.5$. Whereas, In the two-cylinder arrangement, the difference is small ($0.6\%$) for the rigid body, which indicates that increasing the cylinders alters the energy dissipation.


\section{Discussion}
\label{sct:discussion}
The hardening of the shoreline by way of concrete sea walls comes at a staggering price, not only in terms of the construction costs, but also in terms of long-term negative impact on coastal ecosystems \cite{petersonLowe2009, duganHubbard2010, bulleriChapman2010} and shoreline stability \cite{deanDalrymple2002, komar1998}. Less obvious than the drawbacks of lining the coast with concrete, often meter-high sea walls is the alternative, particularly along high-impact coasts like Japan or Indonesia, where tsunamis are frequent and destructive. The idea that vegetation may act as a bioshield against flooding was first proposed in the context of 
marshlands attenuating storm surges in southern Louisiana (e.g., \cite{USACE1963,fosberg1971mangroves}), but the wrath of a tsunami differs fundamentally from the comparatively mild tidal or seasonal flows for which vegetation is thought to be an effective bioshield \cite{shephardEtAl2012}.

An important aspect of tsunami runup is its profound dependence on both off-shore bathymetry and on-shore topography \cite{titov1997extreme,matsuyama1999effect,hentry2010influence,lekkas2011critical,selvakumar2013revealing,dilmen2018role}. While off-shore bathymetry is difficult to modify, on-shore topography could be altered in a way that reduces tsunami impacts, essentially combining nature-based elements and traditional engineering elements into a hybrid approach to mitigating tsunami risk. An example of this approach is a vegetated coastal mitigation park: A landscape unit on the shoreline built specifically to protect communities or critical infrastructure and provide vertical evacuation space while strategically integrating trees or other vegetated elements to enhance protective benefit. 

Prior work has characterized the protective benefit of the onshore topography created by the hillscape of coastal mitigation parks \cite{lunghino2020protective}, but did not explicitly consider the role of vegetation. As communities across the world increasingly consider or adopt coastal mitigation parks, it remains an open question whether the presence of forests of large trees such as the maritime pines of Japan could be used to enhance the protective benefits of coastal mitigation parks and, if so, whether some species of vegetation might be more effective than others. The latter question has immediate relevance for ongoing international efforts to mitigate tsunami-risks. For example, field assessments of damage associated with the 2004 Indian Ocean earthquake and tsunami have posited that Casuarina forests mitigated tsunami impacts more effectively than native ecosystems like sand dunes \cite{mascarenhas2008environmental}. As a consequence, natural sand-dune ecosystems along the Coromandel Coast, India, were destroyed and replaced with Casuarina Equisetifolia, an exotic timber with potentially adverse ecological effects \cite{bhalla2007bio}.

Several field assessments have argued that different species vary significantly in the protective benefits they provide, as reviewed in \cite{cochard2008}. However, many of these assessments are based on observed correlations between tsunami damage and the presence of different ecosystems \cite{danielsen2005asian,kathiresan2005coastal, dahdouh2005effective,tanaka2007coastal,tanaka2007effects,chatenoux2007impacts,iverson2007using,olwig2007using,tanaka2009vegetation,bayas2011influence} and stop short of identifying a possible causal relationship between vegetation and tsunami damage. A key challenge in comparing the effect of different ecosystems is that they thrive in different environments. For example, after the Indian Ocean tsunami Chatenoux and Peduzzi found that tsunami damage was reduced behind seagrass beds but increased behind coral reefs \cite{chatenoux2007impacts}. As pointed out by the authors, an important confounding factor is the difference in the morphological settings where these ecosystems are located with seagrass meadows growing primarily on shallow, gently-sloping continental shelves and barrier reefs lining a break in the continental slope.  

Here, we intentionally remove the complexities associated with the geomorphological setting and consider a highly simplified flat topography to isolate the effect of vegetation properties. Since we are interested primarily in the role of onshore vegetation in tsunami-risk reduction, we focus specifically on coastal forests and the rigidity of tree trunks. In \ref{sec5.1}, we show that rigidity enhances energy reflection for a single cylinder, but the case of a single cylinder is primarily of abstract value since tree trunks rarely grow in isolation. In the practically more relevant case of multiple cylinders in \ref{sec5.2}, the difference created by variable rigidity all but disappears. Moreover, in \ref{sec5.3}, we observe that the gap between the cylinders has less effect on energy reflection as the number of cylinders increases. The additional complexity introduced into the flow field by the presence of multiple cylinders dominates over the rigidity effect, even in the idealized setup of our numerical experiments. There would be significantly more noise in an actual field setting, further reducing the difference in response. 

Taken together, our simulations from sections \ref{sec5.1} and \ref{sec5.2} suggest that the rigidity of tree trunks is less relevant than the arrangement and density of the tree trunks within the coastal forest. An additional factor that reduces the role of rigidity is that rigidity reduces the dissipation of kinetic energy flux afforded by the presence of trees. At least for a single cylinder, rigidity hence affects the reflection and the dissipation of the onshore energy flux in opposite ways. Compared to the effect of rigidity on the reflection of kinetic energy flux, the reduction in dissipation of kinetic energy appears less sensitive to the presence of other tree trunks as demonstrated in section \ref{sec5.4}. However, the effect itself is on the order of a few percent and may hence not generalize to a field setting where many other factors confound flow behavior.

An interesting aspect of our findings in sections \ref{sec5.1} - \ref{sec5.5} is that tree trunks reduce the on-shore energy flux to a comparable degree through both reflection and dissipation. In contrast, the protective benefit of coastal mitigation parks lies primarily in the reflection of incoming wave energy \cite{lunghino2020protective}. This finding suggests that the protective benefit of tree trunks and hillscapes could complement the protective benefit of coastal mitigation parks by augmenting reflection and adding dissipation. An important caveat to this interpretation is that we only consider reversible, elastic deformation of the tree trunks and do not include tree breakage or uprooting, which tend to be the main cause for trees increasing damage in the field (e.g., \cite{tanaka2013breaking}). The breaking patterns of trees do tend to vary among different species \cite{tanaka2018effective} and also depend on trunk diameter \cite{torita2021assessment}. Particularly the dependence on diameter highlights the value of working with existing, native vegetation, and potentially integrating it into mitigation park, rather than planting a mono-culture of trees of a particular species from scratch.

We emphasize that our work adopts a highly idealized view and does not capture sedimentation and erosion that could significantly affect the stability of tree trunks. Neither do we consider the role of the canopy or other vegetation that would undoubtedly grow around trees. By reducing trees to cylinders with a constant elastic modulus, we neglect possible species-specific differences in the breaking pattern of the trunk, the branching structure emerging from the trunk or the root structure. Our work is hence merely a small step towards improving our understanding of the intricate interactions between onshore vegetation and tsunami impacts. Further work would be needed to assess the protective benefit of vegetation with an inherently three-dimensional structure like mangroves that can not be meaningfully reduced to a single trunk or stem.

\section{Conclusions}
\label{sct:conclusions}

In this study, we quantify how the reflection and dissipation of onshore tsunami flow depends on the rigidity and arrangement of trees in a coastal forest. We use a three-dimensional LES turbulent model to study the effect of varying rigidity, vegetation gap, and  the number of trees on tsunami energy attenuation. To represent the dynamic interplay between the tsunami bore and the trees, we employ a two-way FSI model to capture the coupling between fluid and solid body interaction. 
We compute energy flux at different streamwise locations to quantify the energy reflection and dissipation for three sets of moduli of elasticity and different cylinder configurations. 
Our simulations suggest that the rigidity of tree trunks only increase the reflection of kinetic energy flux in the somewhat abstract case of a single cylinder. Even small deflections of the tree trunk alter the flow structures downstream, impacting flow velocity and enhancing turbulent kinetic energy. This variability in velocity distribution and turbulent kinetic energy is more pronounced when the number of cylinders increases, highlighting that rigidity is less important than the number and arrangement of the trees. An important caveat to this conclusion is that we did not capture shear-driven erosion, sediment transport or tree breakage, all of which could limit the ability of the tree trunks to withstand tsunami impact. 
\clearpage

\bibliographystyle{unsrt} 
\bibliography{main}

\begin{thebibliography}{100}

\bibitem{cochard2008}
Roland Cochard, Senaratne~L Ranamukhaarachchi, Ganesh~P Shivakoti, Oleg~V
  Shipin, Peter~J Edwards, and Klaus~T Seeland.
\newblock The 2004 tsunami in aceh and southern thailand: a review on coastal
  ecosystems, wave hazards and vulnerability.
\newblock {\em Perspectives in Plant Ecology, Evolution and Systematics},
  10(1):3--40, 2008.

\bibitem{behrensEtAl2021}
Jörn Behrens, Finn Løvholt, Fatemeh Jalayer, Stefano Lorito, Mario~A.
  Salgado-Gálvez, Mathilde Sørensen, Stephane Abadie, Ignacio Aguirre-Ayerbe,
  Iñigo Aniel-Quiroga, Andrey Babeyko, Marco Baiguera, Roberto Basili, Stefano
  Belliazzi, Anita Grezio, Kendra Johnson, Shane Murphy, Raphaël Paris, Irina
  Rafliana, Raffaele De~Risi, Tiziana Rossetto, Jacopo Selva, Matteo Taroni,
  Marta Del~Zoppo, Alberto Armigliato, Vladimír Bureš, Pavel Cech, Claudia
  Cecioni, Paul Christodoulides, Gareth Davies, Frédéric Dias, Hafize~Başak
  Bayraktar, Mauricio González, Maria Gritsevich, Serge Guillas, Carl~Bonnevie
  Harbitz, Utku Kânoǧlu, Jorge Macías, Gerassimos~A. Papadopoulos, Jascha
  Polet, Fabrizio Romano, Amos Salamon, Antonio Scala, Mislav Stepinac,
  David~R. Tappin, Hong~Kie Thio, Roberto Tonini, Ioanna Triantafyllou, Thomas
  Ulrich, Elisa Varini, Manuela Volpe, and Eduardo Vyhmeister.
\newblock Probabilistic tsunami hazard and risk analysis: A review of research
  gaps.
\newblock {\em Frontiers in Earth Science}, 9, 2021.

\bibitem{rabyEtal2015}
A.~Raby, J.~Macabuag, A.~Pomonis, S.~Wilkinson, and T.~Rossetto.
\newblock {Implications of the 2011 Great East Japan Tsunami on sea defence
  design}.
\newblock {\em International Journal of Disaster Risk Reduction}, 14:332--346,
  2015.

\bibitem{danielsen2005asian}
Finn Danielsen, Mikael~K S{\o}rensen, Mette~F Olwig, Vaithilingam Selvam,
  Faizal Parish, Neil~D Burgess, Tetsuya Hiraishi, Vagarappa~M Karunagaran,
  Michael~S Rasmussen, Lars~B Hansen, et~al.
\newblock The asian tsunami: a protective role for coastal vegetation.
\newblock {\em Science(Washington)}, 310(5748):643, 2005.

\bibitem{kathiresan2005coastal}
Kandasamy Kathiresan and Narayanasamy Rajendran.
\newblock Coastal mangrove forests mitigated tsunami.
\newblock {\em Estuarine, Coastal and shelf science}, 65(3):601--606, 2005.

\bibitem{dahdouh2005effective}
Farid Dahdouh-Guebas, Loku~Pulukkuttige Jayatissa, Diana Di~Nitto, Jared~O
  Bosire, D~Lo Seen, and Nico Koedam.
\newblock How effective were mangroves as a defence against the recent tsunami?
\newblock {\em Current biology}, 15(12):R443--R447, 2005.

\bibitem{tanaka2007coastal}
Norio Tanaka, Yasushi Sasaki, MIM Mowjood, KBSN Jinadasa, and Samang Homchuen.
\newblock Coastal vegetation structures and their functions in tsunami
  protection: experience of the recent indian ocean tsunami.
\newblock {\em Landscape and Ecological Engineering}, 3(1):33--45, 2007.

\bibitem{tanaka2007effects}
Norio Tanaka, Yasushi Sasaki, and MIM Mowjood.
\newblock Effects of sand dune and vegetation in the coastal area of sri lanka
  at the indian ocean tsunami.
\newblock In {\em Advances in Geosciences: Volume 6: Hydrological Science
  (HS)}, pages 149--159. World Scientific, 2007.

\bibitem{chatenoux2007impacts}
Bruno Chatenoux and Pascal Peduzzi.
\newblock Impacts from the 2004 indian ocean tsunami: analysing the potential
  protecting role of environmental features.
\newblock {\em Natural Hazards}, 40(2):289--304, 2007.

\bibitem{iverson2007using}
Louis~R Iverson and Anantha~M Prasad.
\newblock Using landscape analysis to assess and model tsunami damage in aceh
  province, sumatra.
\newblock {\em Landscape Ecology}, 22(3):323--331, 2007.

\bibitem{olwig2007using}
MF~Olwig, MK~S{\o}rensen, MS~Rasmussen, F~Danielsen, V~Selvam, LB~Hansen,
  L~Nyborg, KB~Vestergaard, F~Parish, and VM~Karunagaran.
\newblock Using remote sensing to assess the protective role of coastal woody
  vegetation against tsunami waves.
\newblock {\em International Journal of Remote Sensing}, 28(13-14):3153--3169,
  2007.

\bibitem{bayas2011influence}
Juan Carlos~Laso Bayas, Carsten Marohn, Gerd Dercon, Sonya Dewi, Hans~Peter
  Piepho, Laxman Joshi, Meine van Noordwijk, and Georg Cadisch.
\newblock Influence of coastal vegetation on the 2004 tsunami wave impact in
  west aceh.
\newblock {\em Proceedings of the National Academy of Sciences},
  108(46):18612--18617, 2011.

\bibitem{kerr2006comments}
Alexander~M Kerr, Andrew~H Baird, and Stuart~J Campbell.
\newblock Comments on" coastal mangrove forests mitigated tsunami" by k.
  kathiresan and n. rajendran [estuar. coast. shelf sci. 65 (2005) 601-606].
\newblock {\em Estuarine, Coastal and Shelf Science}, 67:539--541, 2006.

\bibitem{kerr2007natural}
Alexander~M Kerr and Andrew~H Baird.
\newblock Natural barriers to natural disasters.
\newblock {\em BioScience}, 57(2):102--103, 2007.

\bibitem{kerr2009reply}
Alexander~M Kerr, Andrew~H Baird, Ravi~S Bhalla, and V~Srinivas.
\newblock Reply to ``using remote sensing to assess the protective role of
  coastal woody vegetation against tsunami waves''.
\newblock {\em International Journal of Remote Sensing}, 30(14):3817--3820,
  2009.

\bibitem{liu2013}
Haijiang Liu, Takenori Shimozono, Tomohiro Takagawa, Akio Okayasu, Hermann~M
  Fritz, Shinji Sato, and Yoshimitsu Tajima.
\newblock The 11 march 2011 tohoku tsunami survey in rikuzentakata and
  comparison with historical events.
\newblock {\em Pure and Applied Geophysics}, 170(6):1033--1046, 2013.

\bibitem{mori2013overview}
Nobuhito Mori, Daniel~T Cox, Tomohiro Yasuda, and Hajime Mase.
\newblock Overview of the 2011 tohoku earthquake tsunami damage and its
  relation to coastal protection along the sanriku coast.
\newblock {\em Earthquake Spectra}, 29(1\_suppl):127--143, 2013.

\bibitem{tanaka2009vegetation}
Norio Tanaka.
\newblock Vegetation bioshields for tsunami mitigation: review of
  effectiveness, limitations, construction, and sustainable management.
\newblock {\em Landscape and Ecological Engineering}, 5(1):71--79, 2009.

\bibitem{dengler2003mitigation}
Lori Dengler and Jane Preuss.
\newblock Mitigation lessons from the july 17, 1998 papua new guinea tsunami.
\newblock In {\em Landslide Tsunamis: Recent Findings and Research Directions},
  pages 2001--2031. Springer, 2003.

\bibitem{houzeaux2009}
G.~Houzeaux, M.~V{\'a}zquez, R.~Aubry, and J.~Cela.
\newblock A massively parallel fractional step solver for incompressible flows.
\newblock {\em Journal of Computational Physics}, 228(17):6316--6332, 2009.

\bibitem{houzeauxAubryVazquez20111}
G.~{Houzeaux}, R.~{Aubry}, and M.~{V\'azquez}.
\newblock Extension of fractional step techniques for incompressible flows: The
  preconditioned orthomin(1) for the pressure schur complement.
\newblock {\em Comput. Fluids}, 44:297--313, 2011.

\bibitem{green1999mechanical}
D.~W. {Green}, E.~W. {Jerrold}, and E.~K. {David}.
\newblock Mechanical properties of wood.
\newblock {\em Journal of Wood handbook: US Department of Agriculture, Forest
  Service, Products Laboratory}, pages 4--1, 1999.

\bibitem{lunghino2020protective}
B.~Lunghino, A.~F.~S. Tate, M.~Mazereeuw, A.~Muhari, F.~X Giraldo, S.~Marras,
  and J.~Suckale.
\newblock The protective benefits of tsunami mitigation parks and ramifications
  for their strategic design.
\newblock {\em Proceedings of the National Academy of Sciences},
  117(20):10740--10745, 2020.

\bibitem{okalSynolakis2015}
E.A. Okal and C.E. Synolakis.
\newblock {Sequencing of tsunami waves: why the first wave is not always the
  largest}.
\newblock {\em Geophys. J. Int.}, 204, 2015.

\bibitem{borreroEtAl2015b}
J.C. Borrero, P.~Lynett, and N.~Kalligeris.
\newblock {Tsunami currents in ports}.
\newblock {\em Phil. Trans. R. Soc. A}, 373:20140372, 2015.

\bibitem{apotsos2011wave}
Alex Apotsos, Bruce Jaffe, and Guy Gelfenbaum.
\newblock Wave characteristic and morphologic effects on the onshore
  hydrodynamic response of tsunamis.
\newblock {\em Coastal Engineering}, 58(11):1034--1048, 2011.

\bibitem{lin2004numerical}
P.~Lin.
\newblock A numerical study of solitary wave interaction with rectangular
  obstacles.
\newblock {\em Coast. Eng.}, 51(1):35--51, 2004.

\bibitem{marrasMandli2021}
S.~Marras and K.~T. Mandli.
\newblock Modeling and simulation of tsunami impact: A short review of recent
  advances and future challenges.
\newblock {\em Geosciences}, 11:5--, 2021.

\bibitem{qin2018comparison}
X.~{Qin}, M.~{Motley}, R.~{LeVeque}, F.~{Gonzalez}, and K.~{Mueller}.
\newblock A comparison of a two-dimensional depth-averaged flow model and a
  three-dimensional rans model for predicting tsunami inundation and fluid
  forces.
\newblock {\em Natural Hazards and Earth System Sciences}, 18(9):2489--2506,
  2018.

\bibitem{mazaEtAl2015}
M.~Maza, J.L. Lara, and I.J. Losada.
\newblock {Tsunami wave interaction with mangrove forests: a 3-D numerical
  approach,}.
\newblock {\em Coast. Eng.}, 98, 2015.

\bibitem{huangEtAl2011}
Z.~Huang, Y.~Yao, S.~Y. Sim, and Y.~Yao.
\newblock {Interaction of solitary waves with emergent stationary vegetation}.
\newblock {\em Ocean Eng.}, 38, 2011.

\bibitem{mazaEtAl2013}
M.~Maza, J.L. Lara, and I.J. Losada.
\newblock {A coupled model of submerged vegetation under oscillatory flow using
  Navier–Stokes equations}.
\newblock {\em Coast. Eng.}, 80, 2013.

\bibitem{tsaiEtAl2017}
C.-P. Tsai, Y.-C. Chen, T.~Octaviani~Sihombing, and C.~Lin.
\newblock {Simulations of moving effect of coastal vegetation on tsunami
  damping}.
\newblock {\em Nat. Hazards Earth Sys. Sci.}, 17, 2017.

\bibitem{larsen2019full1}
B.~E. Larsen and D.~R Fuhrman.
\newblock Full-scale cfd simulation of tsunamis. part 1: Model validation and
  run-up.
\newblock {\em Coastal engineering}, 151:22--41, 2019.

\bibitem{williamsFuhrman2016}
I.A. Williams and D.R. Fuhrman.
\newblock {Numerical simulation of tsunami-scale wave boundary layers}.
\newblock {\em Coastal Engineering}, 110, 2016.

\bibitem{larsen2019full2}
B.~E. Larsen and D.~R Fuhrman.
\newblock Full-scale cfd simulation of tsunamis. part 2: Boundary layers and
  bed shear stresses.
\newblock {\em Coastal engineering}, 151:42--57, 2019.

\bibitem{larsenFuhrman2018}
B.E. Larsen and D.R. Fuhrman.
\newblock {On the over-production of turbulence beneath surface waves in
  Reynold-averaged Navier-Stokes models}.
\newblock {\em J. Fluid Mech.}, 853, 2018.

\bibitem{tonkin2003tsunami}
Susan Tonkin, Harry Yeh, Fuminori Kato, and Shinji Sato.
\newblock Tsunami scour around a cylinder.
\newblock {\em Journal of Fluid Mechanics}, 496:165--192, 2003.

\bibitem{lakshmanan2012}
N.~{Lakshmanan}, M.~{Kantharaj}, and V.~{Sundar}.
\newblock The effects of flexible vegetation on forces with a
  keulegan-carpenter number in relation to structures due to long waves.
\newblock {\em Journal of Marine Science and Application}, 11(1):24--33, 2012.

\bibitem{irish2014laboratory}
J.L {Irish}, R.~{Weiss}, Y.~{Yang}, Y.K. {Song}, A.~{Zainali}, and
  R.~{Marivela-Colmenarejo}.
\newblock Laboratory experiments of tsunami run-up and withdrawal in patchy
  coastal forest on a steep beach.
\newblock {\em Natural hazards}, 74(3):1933--1949, 2014.

\bibitem{ali2019energy}
A.~H.~M. Rashedunnabi and N.~Tanaka.
\newblock Energy reduction of a tsunami current through a hybrid defense system
  comprising a sea embankment followed by a coastal forest.
\newblock {\em Geosciences}, 9(6):247, 2019.

\bibitem{behrensDias2015}
J.~Behrens and F.~Dias.
\newblock New computational methods in tsunami science.
\newblock {\em Phil. Trans. R. Soc. A}, 373:20140382, 2015.

\bibitem{synolakis2006tsunami}
C.~E. Synolakis and E.~N. Bernard.
\newblock Tsunami science before and beyond boxing day 2004.
\newblock {\em Philosophical Transactions of the Royal Society A: Mathematical,
  Physical and Engineering Sciences}, 364(1845):2231--2265, 2006.

\bibitem{watanabeEtAl2022}
M.~Watanabe, T.~Arikawa, N.~Kihara, C.~Tsurudome, K.~Hosaka, K.~Kimura,
  T.~Hashimoto, F.~Ishihara, T.~Shikata, T.~Morikawa, D.~Makino, M.~Asai,
  Y.~Chida, Y.~Ohnishi, S.~Marras, A.~Mukherjee, J.C. Cajas, G.~Houzeaux,
  B.~Di~Paolo, J.L. Lara, G.~Barajas, I.J. Losada, M.~Hasebe, Y.~Shigihara,
  T.~Asai, T.~Ikeya, S.~Inoue, H.~Matsutomi, Y.~Nakano, Y.~Okuda, S.~Okuno,
  T.~Ooie, G.~Shoji, and T.~Tateno.
\newblock Validation of tsunami numerical simulation models for an idealized
  coastal industrial site.
\newblock {\em Coastal Engineering J.}, 64(2), 2022.

\bibitem{ghani2019numerical}
Usman Ghani, Naveed Anjum, Ghufran~Ahmed Pasha, and Muhammad Ahmad.
\newblock Numerical investigation of the flow characteristics through
  discontinuous and layered vegetation patches of finite width in an open
  channel.
\newblock {\em Environmental Fluid Mechanics}, 19(6):1469--1495, 2019.

\bibitem{yang2017impact}
Yongqian Yang, Jennifer~L Irish, and Robert Weiss.
\newblock Impact of patchy vegetation on tsunami dynamics.
\newblock {\em Journal of Waterway, Port, Coastal, and Ocean Engineering},
  143(4):04017005, 2017.

\bibitem{kundu2016numerical}
Partha Kundu, Vimal Kumar, Yannick Hoarau, and Indra~Mani Mishra.
\newblock Numerical simulation and analysis of fluid flow hydrodynamics through
  a structured array of circular cylinders forming porous medium.
\newblock {\em Applied Mathematical Modelling}, 40(23-24):9848--9871, 2016.

\bibitem{mattisEtAl2015}
S.A. Mattis, C.N. Dawson, C.E. Kees, and M.W. Farthing.
\newblock {An immersed structure approach for fluid-vegetation interaction}.
\newblock {\em Adv. Water Res.}, 80, 2015.

\bibitem{mattisEtAl2012}
S.A. Mattis, C.N. Dawson, C.E. Kees, and M.W. Farthing.
\newblock {Numerical modeling of drag for flow through vegetated domains and
  porous structures}.
\newblock {\em Adv. Water Res.}, 39, 2012.

\bibitem{iimura2012numerical}
K.~Iimura and N.~Tanaka.
\newblock Numerical simulation estimating effects of tree density distribution
  in coastal forest on tsunami mitigation.
\newblock {\em Ocean Engineering}, 54:223--232, 2012.

\bibitem{venayagamoorthy2006numerical}
S.~K. Venayagamoorthy and O.~B. Fringer.
\newblock Numerical simulations of the interaction of internal waves with a
  shelf break.
\newblock {\em Physics of Fluids}, 18(7):076603, 2006.

\bibitem{USACE1963}
US~Army Corps of~Engineers USACE.
\newblock Interim survey report, morgan city, louisiana and vicinity.
\newblock Technical Report~63, US Army Engineer District, New Orleans, LA,
  1963.

\bibitem{fosberg1971mangroves}
F.~R. Fosberg and V.J. Chapman.
\newblock Mangroves v. tidal waves.
\newblock {\em Biological Conservation}, 4(1):38--39, 1971.

\bibitem{nepf2012flow}
H.M. Nepf.
\newblock Flow and transport in regions with aquatic vegetation.
\newblock {\em Ann. Rev. Fluid Mech.}, 44:123--142, 2012.

\bibitem{cajasEtAl2018}
J.~Cajas, G.~Houzeaux, M.~V\'azquez, M.~Garc\'ia, E.~Casoni, H.~Calmet,
  A.~Artigues, R.~Borrell, O.~Lehmkuhl, D.~Pastrana, D.~Y\'a\^nez, R.~Pons, and
  J.~Martorell.
\newblock {Fluid-Structure Interaction Based on HPC Multicode Coupling}.
\newblock {\em SIAM Journal on Scientific Computing}, 40(6), 2018.

\bibitem{vreman2004}
A.W. Vreman.
\newblock {An eddy-viscosity subgrid-scale model for turbulent shear flow:
  algebraic theory and applications}.
\newblock {\em Phys. Fluids}, 16:3670--3681, 2004.

\bibitem{sagautBook}
P.~{Sagaut}.
\newblock {\em Large eddy simulation for incompressible flows. An
  introduction}.
\newblock Springer, 2000.

\bibitem{casoni2015}
E.~Casoni, A~J\'erusalem, C.~Samaniego, B.~Eguzkitza, P.~LaFortune, D.D.
  Tjahjanto, X.~S\'aez, G.~Houzeaux, and M.~V\'azquez.
\newblock Alya: Computational solid mechanics for supercomputers.
\newblock {\em Archives of computational methods in engineering},
  22(4):557--576, 2015.

\bibitem{thekkethil2019level}
N.~Thekkethil and A.~Sharma.
\newblock Level set function--based immersed interface method and benchmark
  solutions for fluid flexible-structure interaction.
\newblock {\em International Journal for Numerical Methods in Fluids},
  91(3):134--157, 2019.

\bibitem{vazquezEtAlALYA2016}
M.~V\'azquez and G.~Houzeaux.
\newblock Alya: Multiphysics engineering simulation towards exascale.
\newblock {\em J. Comput. Sci}, doi:10.1016/j.jocs.2015.12.007, 2016.

\bibitem{codina2006some}
Ramon Codina and Santiago Badia.
\newblock On some pressure segregation methods of fractional-step type for the
  finite element approximation of incompressible flow problems.
\newblock {\em Computer Methods in Applied Mechanics and Engineering},
  195(23-24):2900--2918, 2006.

\bibitem{belytschko2000w}
Ted Belytschko.
\newblock W. liu and b. moran.
\newblock {\em Nonlinear finite elements for continua and structures}, 2000.

\bibitem{casoni2015alya}
Eva Casoni, Antoine J{\'e}rusalem, Crist{\'o}bal Samaniego, Beatriz Eguzkitza,
  Pierre Lafortune, DD~Tjahjanto, Xavier S{\'a}ez, Guillaume Houzeaux, and
  Mariano V{\'a}zquez.
\newblock Alya: computational solid mechanics for supercomputers.
\newblock {\em Archives of Computational Methods in Engineering},
  22(4):557--576, 2015.

\bibitem{owenCodina2007}
H.~Owen and R.~Codina.
\newblock A finite element model for free surface flows on fixed meshes.
\newblock {\em Int. J. Numer. Methods Fluids}, 54:1151--1171, 2007.

\bibitem{owen2020wall}
H.~Owen, G.~Chrysokentis, M.~Avila, D.~Mira, G.~Houzeaux, R.~Borrell, J.~C.
  Cajas, and O.~Lehmkuhl.
\newblock Wall-modeled large-eddy simulation in a finite element framework.
\newblock {\em International Journal for Numerical Methods in Fluids},
  92(1):20--37, 2020.

\bibitem{reichardt1951vollstandige}
H.~Reichardt.
\newblock Vollst{\"a}ndige darstellung der turbulenten
  geschwindigkeitsverteilung in glatten leitungen.
\newblock {\em ZAMM-Journal of Applied Mathematics and Mechanics/Zeitschrift
  f{\"u}r Angewandte Mathematik und Mechanik}, 31(7):208--219, 1951.

\bibitem{martinMoyce1952}
J.~C. Martin and W.~J. Moyce.
\newblock {Part IV: an experimental study of the collapse of liquid columns on
  a rigid horizontal plane}.
\newblock {\em Phil. Trans. R. Soc. London. Series A}, 224:312--325, 1952.

\bibitem{arnason2009tsunami}
H.~Arnason, C.~Petroff, and H.~Yeh.
\newblock Tsunami bore impingement onto a vertical column.
\newblock {\em Journal of Disaster Research}, 4(6):391--403, 2009.

\bibitem{imamura1942history}
A.~{Imamura}.
\newblock History of japanese tsunamis.
\newblock {\em Kayo-No-Kagaku (Oceanography)}, 2(2):74--80, 1942.

\bibitem{imamura1949list}
A.~{Imamura}.
\newblock List of tsunamis in japan.
\newblock {\em J. Seismol. Soc. Japan}, 2:23--28, 1949.

\bibitem{iida1956earthquakes}
K.~{Iida}.
\newblock Earthquakes accompanied by tunamis occurring under the sea off the
  islands of japan.
\newblock {\em The Journal of Earth Sciences, Nagoya University}, 4(1):1--43,
  1956.

\bibitem{iida1970generation}
K.~{Iida}.
\newblock The generation of tsunamis and the focal mechanism of earthquakes.
\newblock {\em Tsunamis in the Pacific Ocean}, pages 3--18, 1970.

\bibitem{iida1967preliminary}
Kumizi Iida, Doak~C Cox, and George Pararas-Carayannis.
\newblock {P}reliminary catalog of tsunamis occurring in the pacific ocean.
\newblock Technical report, {H}awaii {I}nst. {O}f {G}eophysics Honolulu, 1967.

\bibitem{lavigne2009reconstruction}
F.~{Lavigne}, R.~{Paris}, D.~Grancher, P.~Wassmer, D.~Brunstein, F.~Vautier,
  F.~Leone, F.~Flohic, B.~De~Coster, T.~Gunawan, et~al.
\newblock Reconstruction of tsunami inland propagation on december 26, 2004 in
  banda aceh, indonesia, through field investigations.
\newblock {\em Pure and Applied Geophysics}, 166(1-2):259--281, 2009.

\bibitem{chanson2002hydraulics}
H.~{Chanson}.
\newblock {\em Hydraulics of stepped chutes and spillways}.
\newblock CRC Press, 2002.

\bibitem{hager1988b}
W.H. {Hager}.
\newblock B-jump in sloping channel.
\newblock {\em Journal of Hydraulic Research}, 26(5):539--558, 1988.

\bibitem{kawata1999tsunami}
Y.~{Kawata}, B.C. {Benson}, J.C. {Borrero}, J.L. {Borrero}, H.L. {Davies},
  W.~P. {de Lange}, F.~{Imamura}, H.~{Letz}, J.~{Nott}, and C.E. {Synolakis}.
\newblock Tsunami in papua new guinea was as intense as first thought.
\newblock {\em Eos, Transactions American Geophysical Union}, 80(9):101--105,
  1999.

\bibitem{foytong2013analysis}
P.~{Foytong}, A.~{Ruangrassamee}, G.~{Shoji}, Y.~{Hiraki}, and Y.~{Ezura}.
\newblock Analysis of tsunami flow velocities during the march 2011 tohoku,
  japan, tsunami.
\newblock {\em Earthquake Spectra}, 29(1\_suppl):161--181, 2013.

\bibitem{bingham1992canopy}
B.~B. {Bingham} and J.~O {Sawyer}.
\newblock Canopy structure and tree condition of young, mature, and old-growth
  douglas-fir/hardwood forests.
\newblock In {\em Pages 141-149 in: Harris, RR; Erman, DE,(Technical
  Coordinators). Proceedings of the Symposium on Biodiversity of Northwestern
  California; 1991 October 28-30, Santa Rosa, CA. Berkeley, CA: University of
  California, Wildland Resources Center Report No. 29}, 1992.

\bibitem{tanaka2012effectiveness}
N.~Tanaka.
\newblock Effectiveness and limitations of coastal forest in large tsunami:
  conditions of japanese pine trees on coastal sand dunes in tsunami caused by
  great east japan earthquake.
\newblock {\em Journal of Japan Society of Civil Engineers, Ser. B1 (Hydraulic
  Engineering)}, 68(4):II\_7--II\_15, 2012.

\bibitem{pasha2018tsunami}
G.A. {Pasha}, N.~{Tanaka}, J.~{Yagisawa}, and F.N. {Achmad}.
\newblock Tsunami mitigation by combination of coastal vegetation and a
  backward-facing step.
\newblock {\em Coastal Engineering Journal}, 60(1):104--125, 2018.

\bibitem{rodriguez2016field}
R.~{Rodr{\'\i}guez}, P.~{Encina}, M.~{Espinosa}, and N.~{Tanaka}.
\newblock Field study on planted forest structures and their role in protecting
  communities against tsunamis: experiences along the coast of the biob{\'\i}o
  region, chile.
\newblock {\em Landscape and Ecological Engineering}, 12(1):1--12, 2016.

\bibitem{wooding1973drag}
RA~Wooding, Edward~F Bradley, and JK~Marshall.
\newblock Drag due to regular arrays of roughness elements of varying geometry.
\newblock {\em Boundary-Layer Meteorology}, 5(3):285--308, 1973.

\bibitem{mazda1997drag}
Y.~{Mazda}, E.~{Wolanski}, B.~{King}, A.~{Sase}, D.~{Ohtsuka}, and M.~{Magi}.
\newblock Drag force due to vegetation in mangrove swamps.
\newblock {\em Mangroves and salt marshes}, 1(3):193--199, 1997.

\bibitem{mackinnon2003mixing}
J.A {MacKinnon} and M.C {Gregg}.
\newblock Mixing on the late-summer new england shelf—solibores, shear, and
  stratification.
\newblock {\em Journal of Physical Oceanography}, 33(7):1476--1492, 2003.

\bibitem{oshnack2009effectiveness}
M.~E. Oshnack, J.~van~de Lindt, R.~Gupta, D.~Cox, and F.~Agu{\'\i}{\~n}iga.
\newblock Effectiveness of small onshore seawall in reducing forces induced by
  tsunami bore: large scale experimental study.
\newblock {\em Journal of Disaster Research}, 4(6):382--390, 2009.

\bibitem{petersonLowe2009}
M.S. Peterson and M.R. Lowe.
\newblock {Implications of Cumulative Impacts to Estuarine and Marine Habitat
  Quality for Fish and Invertebrate Resources}.
\newblock {\em Rev. Fish. Sci.}, 17:505--523, 2009.

\bibitem{duganHubbard2010}
J.E. Dugan and D.M. Hubbard.
\newblock Ecological effects of coastal armoring: A summary of recent results
  for exposed sandy beaches in southern california.
\newblock In H.~Shipman, M.N. Dethier, G.~Gelfenbaum, K.L. Fresh, and R.S.
  Dinicola, editors, {\em Puget Sound Shorelines and the Impacts of Armoring,
  U.S. Geol. Surv. Sci. Invest. Rep.}, 2010.

\bibitem{bulleriChapman2010}
F.~Bulleri and M.G. Chapman.
\newblock {The introduction of coastal infrastructure as a driver of change in
  marine environments}.
\newblock {\em J. Appl. Ecol.}, 47, 2010.

\bibitem{deanDalrymple2002}
R.~G. Dean and R.~A. Dalrymple.
\newblock {\em Coastal processes with engineering applications}.
\newblock Cambridge University Press, 2002.

\bibitem{komar1998}
P.~Komar.
\newblock {\em Beach processes and sedimentation}.
\newblock Prantice Hall, 1998.

\bibitem{shephardEtAl2012}
C.C. Shephard, C.M. Crain, and M.W. Beck.
\newblock {The protective role of coastal marshes: A systematic review and
  meta-analysis}.
\newblock {\em PLoS ONE}, 6-e27374, 2012.

\bibitem{titov1997extreme}
Vasily~V Titov and Costas~Emmanuel Synolakis.
\newblock Extreme inundation flows during the hokkaido-nansei-oki tsunami.
\newblock {\em Geophysical Research Letters}, 24(11):1315--1318, 1997.

\bibitem{matsuyama1999effect}
Masafumi Matsuyama, JP~Walsh, and Harry Yeh.
\newblock The effect of bathymetry on tsunami characteristics at sisano lagoon,
  papua new guinea.
\newblock {\em Geophysical Research Letters}, 26(23):3513--3516, 1999.

\bibitem{hentry2010influence}
C~Hentry, N~Chandrasekar, S~Saravanan, and J~Dajkumar~Sahayam.
\newblock Influence of geomorphology and bathymetry on the effects of the 2004
  tsunami at colachel, south india.
\newblock {\em Bulletin of engineering geology and the environment},
  69(3):431--442, 2010.

\bibitem{lekkas2011critical}
Efthymios Lekkas, Emmanouil Andreadakis, Irene Kostaki, and Eleni Kapourani.
\newblock Critical factors for run-up and impact of the tohoku earthquake
  tsunami.
\newblock {\em International Journal of Geosciences}, 2(3):310, 2011.

\bibitem{selvakumar2013revealing}
R~Selvakumar and SM~Ramasamy.
\newblock Revealing effect of bathymetry over tsunami run-up through factor
  analysis.
\newblock {\em Arabian Journal of Geosciences}, 6(12):4701--4708, 2013.

\bibitem{dilmen2018role}
Derya~I Dilmen, Gerard~H Roe, Yong Wei, and Vasily~V Titov.
\newblock The role of near-shore bathymetry during tsunami inundation in a reef
  island setting: A case study of tutuila island.
\newblock {\em Pure and Applied Geophysics}, 175(4):1239--1256, 2018.

\bibitem{mascarenhas2008environmental}
Antonio Mascarenhas and Seelam Jayakumar.
\newblock An environmental perspective of the post-tsunami scenario along the
  coast of tamil nadu, india: Role of sand dunes and forests.
\newblock {\em Journal of Environmental Management}, 89(1):24--34, 2008.

\bibitem{bhalla2007bio}
RS~Bhalla.
\newblock Do bio-shields affect tsunami inundation?
\newblock {\em Current Science}, pages 831--833, 2007.

\bibitem{tanaka2013breaking}
Norio Tanaka, Junji Yagisawa, and Satoshi Yasuda.
\newblock Breaking pattern and critical breaking condition of japanese pine
  trees on coastal sand dunes in huge tsunami caused by great east japan
  earthquake.
\newblock {\em Natural hazards}, 65(1):423--442, 2013.

\bibitem{tanaka2018effective}
Norio Tanaka, Hajime Sato, Yoshiya Igarashi, Yuya Kimiwada, and Hiroyuki
  Torita.
\newblock Effective tree distribution and stand structures in a forest for
  tsunami mitigation considering the different tree-breaking patterns of tree
  species.
\newblock {\em Journal of environmental management}, 223:925--935, 2018.

\bibitem{torita2021assessment}
Hiroyuki Torita, Kazuhiko Masaka, Norio Tanaka, Kenta Iwasaki, Satosi Hasui,
  Masato Hayamizu, and Yasutaka Nakata.
\newblock Assessment of the effect of thinning on the resistance of pinus
  thunbergii parlat. trees in mature coastal forests to tsunami fluid forces.
\newblock {\em Journal of Environmental Management}, 284:111969, 2021.

\end{thebibliography}

\end{document}